\documentclass[a4paper,11pt]{article}
\usepackage{jcappub} 

\usepackage{graphicx}	
\usepackage{amsmath}	
\usepackage{amssymb}	
\usepackage[british]{babel}             
\usepackage{newtxtext}                  
\usepackage{bm}
\usepackage{physics}
\usepackage{academicons}
\usepackage[dvipsnames]{xcolor}
\usepackage{booktabs}
\usepackage{natbib}
\usepackage{times}
\usepackage{url}
\usepackage{appendix}
\usepackage{hyperref}
\usepackage[nolist]{acronym}
\usepackage{ulem}
\newacro{3D}{three-dimensional}
\newacro{MG}{modified gravity}
\newacro{FFT}{Fast Fourier Transformation}
\newacro{GR}{General Relativity}
\newacro{DGP}{Dvali-Gabadadze-Porrati}
\newacro{RSD}{redshift space distortion}

\newcommand{\Mpch}{\,h^{-1}{\rm Mpc}}
\newcommand{\Gpch}{\,h^{-1}{\rm Gpc}}

\newcommand{\revision}[1]{\textcolor{black}{#1}}

\title{Fast full N-body simulations of generic modified gravity: conformal coupling models}

\author[a]{Cheng-Zong Ruan,}
\author[b,c]{C\'{e}sar Hern\'{a}ndez-Aguayo,}
\author[a]{Baojiu Li,}
\author[a]{Christian Arnold,}
\author[a]{Carlton M. Baugh,}
\author[d]{Anatoly Klypin,}
\author[e]{and Francisco Prada}

\affiliation[a]{Institute for Computational Cosmology, Department of Physics, Durham University, South Road, Durham DH1 3LE, UK}
\affiliation[b]{Max-Planck-Institut fur Astrophysik, Karl-Schwarzschild-Str 1, D-85748 Garching, Germany}
\affiliation[c]{Excellence Cluster ORIGINS, Boltzmannstrasse 2, D-85748 Garching, Germany}
\affiliation[d]{Astronomy Department, New Mexico State University, Las Cruces, NM 88001, USA}
\affiliation[e]{Instituto de Astrof\'{i}sica de Andaluc\'{i}a (CSIC), Glorieta de la Astronom\'{i}a, E-18080 Granada, Spain}

\emailAdd{cheng-zong.ruan@durham.ac.uk}
\emailAdd{cesarhdz@MPA-Garching.MPG.DE}
\emailAdd{baojiu.li@durham.ac.uk}
\emailAdd{christian.arnold@durham.ac.uk}
\emailAdd{c.m.baugh@durham.ac.uk}
\emailAdd{aklypin@nmsu.edu}
\emailAdd{f.prada@csic.es}

\abstract{We present {\sc mg-glam}, a code developed for the very fast production of full $N$-body cosmological simulations in modified gravity (MG) models. We describe the implementation, numerical tests and first results of a large suite of cosmological simulations for three classes of MG models with conformal coupling terms: the $f(R)$ gravity, symmetron and coupled quintessence models.
Derived from the parallel particle-mesh code \textsc{glam}, {\sc mg-glam} incorporates an efficient multigrid relaxation technique to solve the characteristic nonlinear partial differential equations of these models. 
For $f(R)$ gravity, we have included new variants to diversify the model behaviour, and we have tailored the relaxation algorithms to these to maintain high computational efficiency. 
In a companion paper, we describe versions of this code developed for derivative coupling MG models, including the Vainshtein- and K-mouflage-type models. 
\textsc{mg-glam} can model the prototypes for most MG models of interest, and is broad and versatile. 
The code is highly optimised, with a tremendous speedup of a factor of more than a hundred compared with earlier $N$-body codes, while still giving accurate predictions of the matter power spectrum and dark matter halo abundance.
\textsc{mg-glam} is ideal for the generation of large numbers of MG simulations that can be used in the construction of mock galaxy catalogues and the production of accurate emulators for ongoing and future galaxy surveys.}

\begin{document}
\maketitle
\flushbottom

\section{Introduction}
\label{sec:intro}

The accelerated expansion of our Universe \cite{SupernovaCosmologyProject:1998vns,SupernovaSearchTeam:1998fmf} is one of the most challenging problems in modern physics, and after decades of attempts to find its origin, we are still far from reaching a clear conclusion. 
While the current standard cosmological model --- $\Lambda$ Cold Dark Matter ($\Lambda$CDM), which assumes that this accelerated expansion is caused by the cosmological constant, $\Lambda$ --- is in excellent agreement with most observational data to \revision{date}, this model suffers from the well-known coincidence and fine-tuning problems. This suggests that a more fundamental theory is yet to be developed which can naturally explain the small value of $\Lambda$ inferred from observations. The alternative theoretical models proposed so far can be roughly classified into two categories: those that involve some exotic new matter species beyond the standard model of particle physics, the so-called \textit{dark energy} \cite{Copeland:2006wr}, which usually has non-trivial dynamics; and the other which involve modifications to Einstein's \ac{GR} on certain (usually cosmic) scales \citep{Clifton:2011jh,2015PhR...568....1J,Koyama:2020zce}, or introduces new fundamental forces between matter particles\footnote{The two classes of models can not always be clearly distinguished, and some of the modified gravity models studied here can also considered as coupled dark energy.}. Leading examples include: quintessence \cite{Ratra:1988_quintessence,Wetterich:1988_quintessence,Zlatev:1998tr_quintessence,Steinhardt:1999nw_quintessence}, k-essence \cite{Armendariz-Picon:2000nqq:kessence,Armendariz-Picon:2000ulo_kessense}, coupled quintessence \cite{Amendola:1999er}, $f(R)$ gravity \cite{Sotiriou:2008rp,DeFelice:2010aj} and chameleon model  \cite{Khoury:2003aq,Khoury:2003rn,Mota:2006fz,Brax:2008hh}, symmetron model \cite{Hinterbichler:2010es,Hinterbichler:2011ca}, the Dvali-Gabadadze-Porrati braneworld (DGP) model \cite{Davis:2011pj}, scalar \cite{Nicolis:2008in,Deffayet:2009wt} and vector \cite{Heisenberg:2014_Proca,Allys:2015sht_Proca,BeltranJimenez:2016rff_Proca} Galileons, K-mouflage \cite{Babichev:2009ee}, and massive gravity \cite[e.g.,][]{Hinterbichler:2011tt_massive_gravity}.

In \ac{MG} models, in addition to a modified, and accelerated,  expansion rate that could explain observations, often the law of gravity is also different from \ac{GR}, which can further affect the evolution of the large-scale structure (LSS) of the Universe. This suggests that we can use various cosmological observations to constrain and test these models \citep[e.g.,][]{Koyama:2015vza,Ferreira:2019xrr,Baker:2019gxo}.
In this sense, the study of \ac{MG} models can be used as a testbed to verify the validity of \ac{GR} on cosmological scales, hence going beyond the usual small-scale or local tests of \ac{GR} \cite{Will:2014_GR_LRR}.

In the last two decades, there have been substantial progresses in the size and quality of cosmological observations, many of which can be excellent probes of dark energy and modified gravity \citep[e.g.,][]{Albrecht:2006um,Weinberg_2013}. 
Some of the leading probes include cosmic microwave background (CMB) \cite{Hinshaw:WMAP9,Hou:SPT2014,Planck2018,Aiola:ACT2020}, supernovae \cite{SupernovaCosmologyProject:1998vns,SupernovaSearchTeam:1998fmf,Astier:2006_SNLS_SN,Wood-Vasey:2007_ESSENCE_SN,Sullivan:2011_SNLS_SN,Scolnic:2013_Pan_STARRS_SN,Rest:2013_Pan_STARRS_SN,Abbott:2018_DES_SN,Betoule:2014_SDSS_SN,Jones:2018_Pan_STARRS_SN}, galaxy clustering \citep{Percival:2004_2dF_Galaxy_Clustering,Guzzo:2008_Galaxy_Clustering,Blake:2011_wiggleZ_Galaxy_Clustering,Beutler:2012_6dF_Galaxy_Clustering,Pezzotta:2016_VIMOS_Galaxy_Clustering,Alam:2017_BOSS_Galaxy_Clustering,Zarrouk:2018_SDSS_Galaxy_Clustering} and baryonic acoustic oscillations (BAO)  \cite{Cole2005_BAO,Eisenstein2005_BAO,Beutler2011_BAO,Blake2011_BAO,Anderson2012_BAO,eBOSS2020_BAO}, gravitational lensing \cite{Heymans2013:CFHTLenS_WL,Abbott:2020_DES_WL_Clusters,Hamana2020:HSC_WL,Amon2021:DES_WL,Secco2021:DES_WL}, and the properties of galaxy clusters \cite{Vikhlinin:2009_Xray_Cluster_count,Planck:2013_SZ_Cluster,Mantz:2014a_Cluster,Mantz:2014b_Cluster,SPT:2016_SZ_Cluster,SPT:2018_SZ_Cluster,Abbott:2020_DES_WL_Clusters,Giocoli:2021_KiDS_WL_Cluster}.
In the near future, a number of large, Stage-IV, galaxy and cluster surveys, such as DESI \citep{DESI:2016zmz}, Euclid \citep{Laureijs:2011gra,EuclidTheoryWorkingGroup:2012gxx}, Vera Rubin observatory \cite{lsst} and eROSITA \cite{erosita:2012arXiv1209.3114M}, are expected to revolutionise our knowledge about the Universe and our understanding of the cosmic acceleration, by providing cutting-edge observational data with unprecedented volume and much better controlled systematic errors. Further down the line, experiments such as CMB-S4 \cite{CMBS4} and LISA \cite{LISA} will offer other independent tests of models by using  improved CMB observables, such as CMB lensing and the kinetic Sunyaev-Zel'dovich effect, and gravitational waves.

To exploit the next generation of observational data, we need to develop accurate theoretical tools to predict the cosmological implications of various models, in particular their behaviour on small scales which encode a great wealth of information. However, predicting LSS formation on small scales is a non-trivial task because structure evolution is in the highly non-linear regime on these scales, with a lot of complicated physical processes, such as gravitational collapse and baryonic interactions, in play. The only tool that could accurately predict structure formation in this regime is cosmological simulations, which follow the evolution of matter through the cosmic time, from some initial, linear, density field all the way down to the highly-clustered matter distribution on small, sub-galactic, scales at late times. Modern cosmological simulation codes, e.g., \textsc{ramses} \cite{Teyssier:2001_RAMSES_code_paper}, \textsc{gadget} \cite{Springel:2005_Gadget_code_paper,Springel:2020plp}, \textsc{arepo} \cite{Springel:2010_AREPO_code_paper}, \textsc{pkdgrav} \cite{Potter:2016_PKDGRAV_code_paper}, \textsc{swift} \cite{Schaller:2016_SWIFT_code_paper}, \revision{\textsc{co\textsl{n}cept}\cite{Dakin2021arXiv211201508D}, \textit{gevolution}\cite{Adamek2016JCAP...07..053A}}, have been able to employ hundreds of billions or trillions of particles in giga-parsec volumes \cite[e.g.,][]{Angulo:2012_MXXL_sim_paper,Kim:2015_Horizon4_sim_paper,Potter:2016_PKDGRAV_code_paper}, and are nowdays indispensable in the confrontation of theories with observational data. In particular, to achieve the high level of precision required by galaxy surveys, one can generate hundreds or thousands of independent galaxy mocks that cover the expected survey volume, using these simulations. However, this has so far been impossible for \ac{MG} models, which usually involve highly non-linear partial differential equations that govern the new physics, solving which has proven to be very expensive even with the latest codes, e.g., \textsc{ecosmog} \cite{Li:2011_ECOSMOG_code_paper,Li:2013_ECOSMOGV_code_paper,2012JCAP...10..002B,Brax:2013mua}, \textsc{mg-gadget} \cite{Puchwein:2013_MGGADGET_code_paper}, \textsc{isis} \cite{Llinares:2013_ISIS_code_paper} and \textsc{mg-arepo} \cite{Arnold:2019_MGAREPO_code_paper,Hernandez-Aguayo:2020_MGAREPO_code_paper} (see \cite{Winther:2015_MG_code_comparison} for a comparison of several MG codes). For example, current \ac{MG} simulations can take between $2$ to $\mathcal{O}(10)$ times longer than standard $\Lambda$CDM simulations with the same specifications. Obviously, to best explore future observations for testing \ac{MG} models, we need a new simulation code for these models with greatly improved efficiency compared with the current generation of codes.

Here, we present such a code, {\sc mg-glam}, which is an extension of the parallel particle-mesh (PPM) $N$-body code {\sc glam}\footnote{{\sc glam} stands for GaLAxy Mocks, which is a pipeline for massive production of galaxy catalogues in the $\Lambda$CDM (GR) model.} \citep{Klypin:2017iwu}, in which various important classes of modified gravity models have been implemented. Efficiency is the main feature of \textsc{mg-glam}, which is partly thanks to the efficiency and optimisations it inherits from its base code,  \textsc{glam}\footnote{The \textsc{glam} code has been shown to be $1.6$--$4$ times faster than similar codes such as {\sc cola} \cite{Koda_15}, {\sc icecola} \cite{Izard_15} and {\sc fastpm} \cite{Feng_16}, while still achieving high resolution and accuracy.}, partly due to optimised numerical algorithms tailored to solve the nonlinear equations of motion in these modified gravity models, and partly thanks to a careful design of the code and data structures to reduce the memory footprint of the simulations.

Modified gravity models can be classified according to the fundamental properties of their new dynamical degrees of freedom, and the interactions the latter have. Here, we study three classes of MG models which introduce scalar-type degrees of freedom that have conformal-coupling interactions: coupled quintessence \cite{Amendola:1999er}, chameleon \cite{Khoury:2003aq,Khoury:2003rn} $f(R)$ gravity \citep{Hu:2007nk}, and symmetron models \citep{Hinterbichler:2010es,Hinterbichler:2011ca}. These models generally introduce a new force (\textit{fifth force}) between matter particles, and the latter two can be considered as special examples of the former, but differ in that they can both employ screening  mechanisms to evade Solar System constraints on the fifth force. These models have been widely studied in recent years and, as we argue below, the implementation of them can lead to prototype MG codes that can be modified to work with minimal effort for other classes of interesting models. 
In a twin paper \citep{Hernandez-Aguayo:2021_twin_paper}, we will describe the implementation and analysis of two classes of derivative-coupling MG models, including the DGP and K-mouflage models.

As we will demonstrate below, the inclusion of modified gravity solvers in \textsc{mg-glam} adds an overhead to the computational cost of \textsc{glam}, and for the models considered in this paper and its twin paper \cite{Hernandez-Aguayo:2021_twin_paper}, a \textsc{mg-glam} run takes about $3$-$5$ times (depending on the resolution) the computing time of an equivalent $\Lambda$CDM simulation using default \textsc{glam}. All in all, this makes this new code at least $100$ times faster than other modified gravity simulation codes such as \textsc{ecosmog} \cite{Li:2011_ECOSMOG_code_paper,Li:2013_ECOSMOGV_code_paper,2012JCAP...10..002B,Brax:2013mua} and \textsc{mg-arepo} \cite{Arnold:2019_MGAREPO_code_paper,Hernandez-Aguayo:2020_MGAREPO_code_paper} for the same simulation boxsize and particle number. In spite of such a massive improvement in speed over those latter codes, it is worthwhile to note that \textsc{mg-glam} is \textit{not} an approximate code: it solves the full Poisson and MG equations, and its accuracy is only limited by the resolution of the PM grid used, which can be specified by users based on their particular scientific objectives. This makes it different from fast approximate simulation codes such as those  \cite{Winther:2017j_MGCOLA_code_paper,Wright:2017_NUCOLA_code_paper,Valogiannis:2016_MGCOLA_code_paper,Fiorini:2021_MGCOLA_application_haloes} based on the COmoving Lagrangian Acceleration method (\textsc{cola}) \cite{Tassev:2013_COLA_code_paper}.

This paper is organised as follows. In Section~\ref{sec:theories}, we present a brief description of the conformally coupled \ac{MG} models covered in this work, which aims at offering a self-contained overview of the key theoretical properties which are relevant for the numerical code.  In Section~\ref{sect:numerics}, we present the details of our numerical implementations to solve the \ac{MG} scalar field equations, including the code and data structure, the implementation of the multigrid relaxation method to solve the MG equations, and the tailored relaxation alogorithms for each model. In Section~\ref{sect:code_tests}, we show various code test results, which help us to verify the accuracy and reliability of the code. Section~\ref{sect:cos_runs} shows the cosmological simulation results for a large suite of MG models, which serve to showcase the potential power of the new code. Finally we summarise and conclude in Section~\ref{sect:discuz}.

Throughout this paper, we adopt the usual conventions that Greek indices label all space-time coordinates ($\mu,\nu,\cdots=0,1,2,3$), while Latin indices label the space coordinates only ($i,j,k,\cdots=1,2,3$). Our metric signature is $(-,+,+,+)$. We will strive to include the speed of light $c$ explicitly in relevant equations, rather than setting it to $1$, given that in numerical implementations $c$ must be treated carefully. Unless otherwise stated, the symbol $\approx$ means `approximately equal' or `equal under certain approximations as detailed in the text', while the symbol $\simeq$ means that two quantities are of a similar order of magnitude. An overdot denotes the derivative with respect to (wrt) the cosmic time $t$, e.g., $\dot{a} \equiv {\dd{a}}/{\dd{t}}$ and the Hubble expansion rate $H(a)$ is defined as $H=\dot{a}/a$, while a prime ($'$) denotes the derivative wrt the conformal time $\tau$, e.g., $a'=\dd{a}/\dd{\tau}$, $\mathcal{H}(a)\equiv{a}'/a=aH(a)$. Unless otherwise stated, we use a subscript $0$ to denote the present-day value of a physical quantity, an overbar for the background value of a quantity, and a tilde for quantities written in code units.

We note that, since they have a lot in common, including the motivation and the design of code structure and algorithms, this paper has identical or similar texts with its twin paper \cite{Hernandez-Aguayo:2021_twin_paper} in the Introduction section, as well as in Sections~\ref{subsect:glam}, \ref{sec:glam_units}, \ref{subsect:extradof} until \ref{subsubsect:relaxation}, \ref{subsubsect:relaxation}, \ref{subsubsect:code_struc}, the last paragraph of \ref{subsubsect:csf_imp}, and part of \ref{subsect:bg_tests}.

\section{Theories}
\label{sec:theories}

In this section we will describe several classes of theoretical models which will later be implemented in the modified {\sc glam} code. The main purpose of this description is to make this paper self-contained, and so we will keep it concise. Interested readers can find more details in the literature elsewhere. 

Consider a general model where a scalar field, $\phi$, couples to matter, described by the following action
\begin{equation}
    S = \int{\rm d}^4x\sqrt{-g}\left[\frac{M^2_{\rm Pl}}{2}R - \frac{1}{2}\nabla^\mu\phi\nabla_\mu\phi-V(\phi)\right] + \sum_i\int{\rm d}^4x\sqrt{-\hat{g}}\mathcal{L}^m\left[\psi_i,\hat{g}_{\mu\nu}\right].
\end{equation}
Here the first term is the gravitational action, where $g$ is the determinant of the metric tensor $g_{\mu\nu}$, $M_{\rm Pl}$ the reduced Planck mass, $R$ the Ricci scalar, $\nabla^\mu$ the covariant derivative, and $V(\phi)$ the potential energy of the scalar field $\phi$. The second term is the matter action, which sums over all matter species labelled by $i$, with $\psi$ being the matter field and $\hat{g}_{\mu\nu}$ the metric that couples to it. In principle, $\hat{g}_{\mu\nu}$ can be different for different matter species, but we consider the universal $\hat{g}_{\mu\nu}$ here for simplicity. 

The Jordan-frame metric $\hat{g}_{\mu\nu}$ and the Einstein-frame metric $g_{\mu\nu}$ are related to each other by the following conformal scaling
\begin{equation}\label{eq:conformal_metric}
    \hat{g}_{\mu\nu} = A^2(\phi)g_{\mu\nu}.
\end{equation}
Here we work in the \textit{Einstein frame}, in which the effect of the scalar field is in the matter sector, i.e., modified geodesics for matter particles, while the left-hand sides of the Einstein equations keep their standard form. This is in contrast to the \textit{Jordan frame}, where the scalar field manifestly modifies the curvature terms on the left side of the Einstein equation. However, in a classical sense the physics is the same in these two frames. Note that the relation between $g_{\mu\nu}$ and $\hat{g}_{\mu\nu}$ can be more complicated, e.g., including a disformal term, but these possibilities are beyond the scope of the present work.

The scalar field is a dynamical and physical degree of freedom in this model, which is governed by the following equation of motion
\begin{equation}\label{eq:csf_eom_general}
    \nabla^\mu\nabla_\mu\phi = \frac{{\rm d}A(\phi)}{{\rm d}\phi}\left[\rho_{\rm m}-3P_m\right] + \frac{{\rm d}V(\phi)}{{\rm d}\phi},
\end{equation}
where $\rho_{\rm m}$ and $P_m$ are respectively the density and pressure of non-relativistic matter (radiation species do not contribute due to the conformal nature of Eq.~\eqref{eq:conformal_metric}). We also define the coupling strength $\beta(\phi)$ as a dimensionless function of $\phi$:
\begin{equation}\label{eq:csf_beta_general}
    \beta(\phi) \equiv M_{\rm Pl}\frac{{\rm d}\ln A(\phi)}{\phi}.
\end{equation}
Note the $M_{\rm Pl}$ in this definition, which is because $\phi$ has dimensions of mass. For later convenience, we shall define a dimensionless scalar field as
\begin{equation}
    \varphi \equiv \frac{\phi}{M_{\rm Pl}}. 
\end{equation}
We can see from Eq.~\eqref{eq:csf_eom_general} that, in addition to the self-interaction of the scalar field $\phi$, described by its potential energy, $V(\phi)$, the matter coupling means that the dynamics of $\phi$ is also affected by the presence of matter. We can therefore define an \textit{effective total potential} of the scalar field, $V_{\rm eff}(\phi)$, as
\begin{equation}\label{eq:Veff}
    V_{\rm eff}(\phi) \equiv A(\phi)\rho_{\rm m} + V(\phi),
\end{equation}
where we have used $P_m=0$ for matter. With appropriate choices of $V(\phi)$ and $A(\phi)$, the effective potential $V_{\rm eff}(\phi)$ may have one or more minima, i.e., ${\rm d}V_{\rm eff}/{\rm d}\phi=0$ at $\phi=\phi_{\rm min}$. Provided that the shape of $V_{\rm eff}(\phi)$ is sufficiently steep around $\phi_{\rm min}$, as in some classes of models to be studied below, the scalar field can oscillate around it, and we can define a scalar field mass, $m$, as
\begin{equation}
    m^2 \equiv \frac{{\rm d}^2V_{\rm eff}\left(\phi_{\rm min}\right)}{{\rm d}\phi^2}.
\end{equation}

For non-relativistic matter particles, the interaction with the scalar field introduces new terms in their geodesic equations,
\begin{equation}\label{eq:csf_particle_geodesic_general}
    \dot{u}^\mu + \frac{\dot{\phi}}{M_{\rm Pl}}u^\mu = -c\frac{\beta(\phi)}{M_{\rm Pl}}\nabla^\mu\phi,
\end{equation}
where $u^\mu\equiv{\rm d}x^{\mu}/{\rm d}\tau$ is the 4-velocity, and overdot denotes the time derivative.

In the weak-field limit where the metric $g_{\mu\nu}$ can be written through the following line element,
\begin{equation}
    {\rm d}s^2 = -(1+2\Phi)c^2{\rm d}t^2 + (1-2\Phi){\rm d}x^i{\rm d}x_i,
\end{equation}
where $\Phi$ is the Newtonian potential, we can approximately write Eq.~\eqref{eq:csf_particle_geodesic_general} as
\begin{equation}\label{eq:csf_particle_geodesic_qsa}
    \ddot{\boldsymbol{r}} = -\boldsymbol{\nabla}\Phi - c^2\frac{\beta(\phi)}{M_{\rm Pl}}\boldsymbol{\nabla}\phi - \frac{\beta(\phi)}{M_{\rm Pl}}\dot{\phi}\dot{\boldsymbol{r}},
\end{equation}
where $\boldsymbol{r}$ is the physical coordinate of the particle and $\boldsymbol{\nabla}$ is the gradient with respect to the physical coordinate. 

\revision{The gravitational potential $\Phi$ and the \textcolor{black}{perturbation to the} MG scalar field have small values in Newtonian $N$-body simulations.}  
\textcolor{black}{Some relativistic cosmological simulation codes, such as \textsc{gramses} \citep{Barrera-Hinojosa:2020JCAP...01..007B,Barrera-Hinojosa:2020JCAP...04..056B}, go beyond the weak-field approximation by including higher-order terms of the gravitational potentials, but find the effect on small scales is indeed small.}

Eq.~\eqref{eq:csf_particle_geodesic_qsa} summarises three of the key effects that a coupled scalar field can have on cosmic structure formation: (1) a \textit{fifth force}, as given by the gradient of $\phi$, (2) a \textit{frictional force} that is proportional to $\dot{\phi}$ and the particle's velocity $\dot{\boldsymbol{r}}$ -- this is similar to the usual \revision{`frictional'} force caused by the Hubble expansion $H$, but because $H$ can be modified by the coupled scalar field too, we have a third effect through a modified $H$, which is implicit in Eq.~\eqref{eq:csf_particle_geodesic_qsa}.

In the same limit, the scalar field equation of motion, Eq.~\eqref{eq:csf_eom_general}, can be simplified as
\begin{equation}\label{eq:csf_eom_qsa}
    c^2\boldsymbol{\nabla}^2\phi \approx V_\phi(\phi) - V_\phi(\bar{\phi}) + A_\phi(\phi)\rho_{\rm m} - A_\phi(\bar{\phi})\bar{\rho}_{\rm m},
\end{equation}
where an overbar denotes the background value of a quantity, and $V_\phi\equiv{\rm d}V(\phi)/{\rm d}\phi$, $A_\phi\equiv{\rm d}A(\phi)/{\rm d}\phi$. In deriving Eq.~\eqref{eq:csf_eom_qsa} we have used the weak field approximation, as well as the quasi-static approximation which enables use to neglect the time derivative of the scalar field perturbation, $\delta\phi\equiv\phi-\bar{\phi}$, compared with its spatial gradient, i.e., $|\ddot{\delta\phi}|\simeq|H\dot{\delta\phi}|\ll|\boldsymbol{\nabla}^2\delta\phi|=|\boldsymbol{\nabla}^2\phi|$, where $H\equiv\dot{a}/a$ is the Hubble expansion rate. \revision{It is important to note that we do not assume that $\ddot{\bar{\phi}} \ll |\boldsymbol{\nabla}^2 \phi|$, because $\ddot{\bar{\phi}}$ and $H \dot{\bar{\phi}}$ can be significant in certain models such as coupled quintessence, where $\bar{\phi}$ can evolve by a large amount throughout the cosmic history.}

\revision{The quasi-static approximation has been tested for the modified gravity theories considered in this paper, such as $f(R)$ gravity \cite{Oyaizu:2008PhRvD..78l3523O,Bose:2015JCAP...02..034B} and symmetron \cite{Llinares:2014PhRvD..89h4023L}. Ref.~\cite{Oyaizu:2008PhRvD..78l3523O} performed a consistency check of this approximation for Hu-Sawicki $f(R)$ gravity \cite{Hu:2007nk}, where the simulations were run in the quasi-static limit but it was checked that the time derivative of the scalar field perturbation is generally $5$--$6$ orders of magnitude smaller than its spatial derivative in amplitude. Ref.~\cite{Bose:2015JCAP...02..034B} directly examined this approximation by running full simulations including the time derivative terms, and found that the effects of the scalar field time derivative terms can be safely ignored in  Hu-Sawicki $f(R)$ gravity. For the symmetron model, the quasi-static approximation has also been widely used in previous literature, e.g., \citep{Davis:2011pj,2012JCAP...10..002B}. 
Ref.~\cite{Llinares:2014PhRvD..89h4023L} ran simulations with non-static terms and found very little difference in the matter power spectrum with the quasi-static simulations. However, the local power spectrum (defined as the $P(k)$ for the filtered matter field) shows deviations of the order of $1\%$. Therefore, it is expected that the quasi-static approximation is valid for usual cosmological probes such as power spectra which we are interested in, but other properties may be affected.}

{According to these researches, the quasi-static approximation is valid for our cosmological analyses.
The effects of the scalar field time derivatives are small enough that can be safely ignored for the nonlinear evolution of dark matter fields.
}

Finally, the Newtonian potential $\Phi$ is governed by the following Poisson equation, again written under the weak-field and quasi-static approximations,
\begin{equation}\label{eq:csf_poisson_qsa}
    \boldsymbol{\nabla}^2\Phi \approx 4\pi{G}A(\bar{\phi})\left(\rho_{\rm m} - \bar{\rho}_{\rm m}\right),
\end{equation}
where we note the presence of $A(\bar{\phi})$ in front of $\rho_{\rm m}$, which is because the coupling to the scalar field $\phi$ can cause a time evolution of the particle masses of non-relativistic species, therefore affecting the depth of the resulting potential well $\Phi$. This is the fourth key effect a coupled scalar field can have on cosmic structure formation. In the models considered in this paper, either the scalar field perturbation is small such that $A(\phi)\simeq A(\bar{\phi})$, or the scalar field has a small amplitude ($|{\varphi}|\ll1$) in the entire cosmological regime so that $A(\phi)\simeq1$ and $A(\bar{\phi})\simeq1$. 

Eqs.~(\ref{eq:csf_particle_geodesic_qsa}, \ref{eq:csf_eom_qsa}, \ref{eq:csf_poisson_qsa}) are the three key equations to be solved in our $N$-body simulations.

\subsection{Coupled quintessence}
\label{subsect:csf}

The behaviour of the coupled scalar field, as well as its effect on the cosmological evolution, is fully specified with concrete choices of the coupling function $A(\phi)$ and scalar potential $V(\phi)$. Such models are known as \textit{coupled quintessence} \cite{Amendola:1999er}, and have been studied extensively in the literature, including simulation analyses.

With some choices of $A(\phi)$ and $V(\phi)$, the scalar field dynamics can become highly nonlinear, such as in the symmetron and chameleon models described below. These models are often display very little evolution of the background scalar field ($|\Delta\varphi|\ll1$) throughout the cosmic history so that the background expansion rate closely mimics that of $\Lambda$CDM; the spatial perturbations of $\varphi$ can reach $|\delta\varphi|\simeq|\bar{\varphi}|\ll1$. In other, more general, cases, the scalar field can have a substantial dynamical evolution, $|\Delta\varphi|\sim\mathcal{O}(1)$ and $|\delta\varphi|\ll|\bar{\varphi}|$, which allows deviations from the $\Lambda$CDM expansion history, and the fifth force behaves in a less nonlinear way. This latter case is the focus in this subsection. 

We consider an exponential coupling function
\begin{equation}\label{eq:csf_coupling_function}
    A(\phi) = \exp\left(\beta\frac{\phi}{M_{\rm Pl}}\right) = \exp\left(\beta\varphi\right),
\end{equation}
and an inverse power-law potential
\begin{equation}\label{eq:csf_potential}
    V(\phi) = \frac{M^\alpha_{\rm Pl}\Lambda^4}{\phi^\alpha} = \frac{\Lambda^4}{\varphi^\alpha},
\end{equation}
where $\alpha,\beta$ are dimensionless model parameters, and $\Lambda$ is a model parameter with mass dimension 1 which represents a new energy scale related to the cosmic acceleration. For convenience, we define a dimensionless order-unity parameter $\lambda$ as
\begin{equation}\label{eq:csf_param_lambda}
    \frac{\Lambda^4}{M_{\rm Pl}^2} = H_0^2\lambda^2.
\end{equation}
We consider parameters $\alpha>0$, so that $V(\phi)$ is a runaway potential and the scalar field rolls down $V(\phi)$, and $\beta<0$ so that the effective potential $V_{\rm eff}(\phi)$ has no minimum and the scalar field can keep rolling down $V_{\rm eff}(\phi)$ if not stopped by other effects. This means that we can have $|\bar{\varphi}|\sim\mathcal{O}(1)$ at late times (as mentioned in the previous paragraph) and kinetic energy makes up a substantial fraction of the scalar field's total energy (so that its equation of state $w_\phi$ can deviate substantially from $-1$). 

While we specialise to Eqs.~(\ref{eq:csf_coupling_function}, \ref{eq:csf_potential}) for the coupled quintessence models in this paper, the {\sc mg}-{\sc glam} code that we will illustrate below using this model can be applied to other choices of $A(\phi)$ and $V(\phi)$ with minor changes in a few places, to allow fast, inexpensive and accurate simulations for generic coupled quintessence models.

For completeness and convenience of later discussions, we also present here the linear growth equation for matter density contrast $\delta$ (or equivalently the linear growth factor $D_+$ itself) in the above coupled quintessence model:
\begin{equation}\label{eq:csf_lin_growth}
    \delta'' + \left[\frac{a'}{a}+\frac{{\rm d}\ln{A}(\bar{\varphi})}{{\rm d}\varphi}\bar{\varphi}'\right]\delta' - 4\pi{G}\bar{\rho}_{\rm m}(a){a}^2A(\bar{\varphi})\left(1+2\beta^2\right)\delta = 0,
\end{equation}
where $'$ denotes the derivative with respect to the conformal time $\tau$. According to this equation, there are 4 effects that the coupled scalar field has on structure formation: (\textit{i}) a modified expansion history, $a'/a$; (\textit{ii}) a fifth force whose ratio with respect to the strength of the standard Newtonian force is given by $2\beta^2$; (\textit{iii}) a rescaling of the matter density field by $A(\varphi)\neq1$ in the Poisson equation, implying that the matter particle mass is effectively modified; and (\textit{iv}) a velocity-dependent force described by the term involving $\left({\rm d}\ln A/{\rm d}\varphi\right)\bar{\varphi}'\delta'$. The ratio between the fifth and Newtonian forces can be derived as follows: Eq.~\eqref{eq:csf_eom_qsa} can be approximately rewritten as
\begin{equation}\label{eq:csf_eom_qsa_approx}
    \boldsymbol{\nabla}^2\left(c^2\delta\varphi\right) \approx 8\pi G \beta A(\bar{\varphi})\left[\rho_{\rm m} - \bar{\rho}_{\rm m}\right],
\end{equation}
where we have used $A_\phi=\frac{\beta}{M_{\rm Pl}}\exp(\beta\varphi)$, $M_{\rm Pl}^{-2}=8\pi G$, and neglected the contribution the scalar field potetial $V(\phi)$ in the field perturbation $\delta\varphi$. Then, from Eqs.~\eqref{eq:csf_poisson_qsa} and \eqref{eq:csf_particle_geodesic_qsa}, it follows that the ratio of the two forces is $2\beta^2$, which means that the fifth force always boosts the total force experienced by matter particles in this model. In addition, since $\beta$ is a constant, from Eq.~\eqref{eq:csf_lin_growth} we can conclude that the enhancement to linear matter growth, i.e., in the linear growth factor and matter power spectrum, will be scale-independent.

\subsection{Symmetrons}
\label{subsect:sym}

The \textit{symmetron} \cite{Hinterbichler:2010es,Hinterbichler:2011ca} model features the following potential $V(\phi)$ and coupling function $A(\phi)$ for the scalar field:
\begin{eqnarray}\label{eq:sym_potential}
    V(\phi) &=& V_0 - \frac{1}{2}\mu^2\phi^2 + \frac{1}{4}\zeta\phi^4,\\
    \label{eq:sym_coupling_function} A(\phi) &=& 1+ \frac{1}{2}\frac{\phi^2}{M^2},
\end{eqnarray}
where $\mu, M$ are model parameters of mass dimension 1, $\zeta$ is a dimensionless model parameter and $V_0$ is a constant parameter of mass dimension 4, which represents vacuum energy and acts to accelerate the Hubble expansion rate.

We can define 
\begin{equation}
    \phi_\ast \equiv \frac{\mu}{\sqrt{\zeta}},
\end{equation}
which represents the local minimum of the Mexican-hat-shaped symmetron potential $V(\phi)$. The total effective potential of the scalar field, however, is given in Eq.~\eqref{eq:Veff}. Because $A(\phi)$ is a quadratic function of $\phi$, when $\rho_{\rm m}$ is large, the effective potential is dominated by $A(\phi)\rho_{\rm m}$, with single global minimum at $\phi=0$; but when $\rho_{\rm m}$ is small, the effective potential is dominated by $V(\phi)$ and has two minima, $\pm\phi_{\rm min}$.  Explicitly, it can be shown that $\phi_{\rm min}=0$ when $\bar{\rho}_{\rm m}>\mu^2M^2\equiv\rho_\ast$ in background cosmology, while otherwise the symmetry in $V(\phi)$ is broken and the symmetron field solutions are given by
\begin{equation}\label{eq:phi_min_sym}
    \pm\phi_{\rm min} = \sqrt{\frac{1}{\zeta{M}^2}\left(\rho_\ast-\bar{\rho}_{\rm m}\right)},
\end{equation}
from which we can confirm the above statement that as $\bar{\rho}_{\rm m}\rightarrow0$ we have $\phi_{\rm min}\rightarrow\phi_\ast$. Because $\rho_\ast$ has the dimension of density, it is more convenient to express it in terms of a characteristic scale factor $a_\ast$ or redshift $z_\ast$ corresponding to the time of symmetry breaking in $V_{\rm eff}(\phi)$:
\begin{equation}
    \rho_\ast = \bar{\rho}_{m0}a^{-3}_\ast,
\end{equation}
where $\bar{\rho}_{m0}$ is the background matter density today. According to Eq.~\eqref{eq:phi_min_sym}, as $\rho_{\rm m}\rightarrow0$, $\phi_{\rm min}\rightarrow\phi_\ast$, i.e., $\phi_{\rm min}$ approaches the minimum of $V(\phi)$. Therefore, we must have $\phi_{\rm min}\in[0,\phi_\ast]$. For this reason we can define the following dimensionless variable
\begin{equation}
    u \equiv \frac{\phi}{\phi_\ast}\in[0,1),
\end{equation}
Note that this is only true for background $u$, while in the perturbed case it is possible to have $u>1$ in certain regions. Also, $u>0$ is just a choice, because the symmetron field has two physically identical branches of solutions which differ by sign, and we choose the positive branch for simplicity\footnote{Indeed, it is possible that $u$ can have different signs in different regions of the Universe, which are separated by domain walls, but we do not consider this more realistic possibility in this paper, as it does not have a big impact on the observables of interest to us.}. In terms of the dimensionless scalar field $\varphi$, we have \cite{Brax:2012a}
\begin{equation}\label{eq:varphi_min_symm}
    \varphi_{\rm min}(a) = \varphi_\ast\sqrt{1-\left(\frac{a_\ast}{a}\right)^3},
\end{equation}
with
\begin{equation}\label{eq:varphi_ast}
    \varphi_\ast \equiv \frac{\phi_\ast}{M_{\rm Pl}} =  6\Omega_m\beta_\ast\xi^2a_{\ast}^{-3},
\end{equation}
where $\Omega_m$ is the matter density parameter today, $\xi\equiv H_0/m_\ast$ with $m_\ast$ being the `mass' of the scalar field at $\phi_\ast$, given by
\begin{equation}
    m_\ast^2 \equiv \frac{{\rm d}^2V(\phi_\ast)}{{\rm d}^2\phi} = -\mu^2 + 3\zeta\phi^2_\ast = 2\mu^2,
\end{equation}
and $\beta_\ast$ is a dimensionless parameter defined through
\begin{equation}
    M_{\rm Pl}\frac{{\rm d}A}{{\rm d}\phi} = \frac{M_{\rm Pl}\phi}{M^2} \equiv \beta_\ast\frac{\phi}{\phi_\ast},
\end{equation}
which can be further expressed as
\begin{equation}
    \beta_\ast \equiv \frac{M_{\rm Pl}}{M^2}\frac{\mu}{\sqrt{\zeta}} = \frac{M_{\rm Pl}m_\ast^2}{2\rho_\ast}\phi_\ast.
\end{equation}

Therefore, the model can be fully specified by three dimensionless parameters -- $\beta_\ast$, $a_\ast$ (or $z_\ast$) and $\xi$ -- as opposed to the original, dimensional, parameters $\mu, \zeta$, $M$. We are interested in the regime of $\beta_\ast, a_\ast\sim\mathcal{O}(0.1)$ and $\xi\sim\mathcal{O}\left(10^{-3}\right)$. It is then evident from Eqs.~(\ref{eq:varphi_ast}) that $\varphi_\ast\ll1$ and therefore $\varphi_{\rm min}(a)\ll1$, confirming our claim above that in this model the scalar field has little evolution throughout the cosmic history. For simplicity we will assume that in the background the scalar field always follows $\varphi_{\rm min}$, namely $\bar{\varphi}(a)=\varphi_{\rm min}(a)$\footnote{In practice, because $\varphi_{\rm min}(a)$ evolves with time, when trying to track it, $\bar{\varphi}$ can have oscillations around $\varphi_{\rm min}$ because $m_\ast\gg H(a)\simeq H_0$. Following most literature on the symmetron model, we will neglect these oscillations.}. Further, because $\varphi_{\rm min}\simeq\varphi_\ast\ll1$, we have
\begin{equation}
    A(\phi) = 1 + \frac{1}{2}\beta_\ast\frac{\varphi}{\varphi_\ast}\varphi \simeq 1,
\end{equation}
which implies that the time variation of particle mass is negligible in this model, and 
\begin{equation}\label{eq:sym_coupling_strength}
    \beta(\phi) = M_{\rm Pl}\frac{{\rm d\ln A(\phi)}}{{\rm d}\phi} \simeq \frac{{\rm dA(\varphi)}}{{\rm d\varphi}} = \beta_\ast\frac{\varphi}{\varphi_\ast} = \beta_\ast u,
\end{equation}
so that $\beta_\ast$ characterises the coupling strength between the scalar field and matter in this model. 

With all the newly-defined variables, the scalar field equation of motion, Eq.~\eqref{eq:csf_eom_general}, in this model can be simplified as 
\begin{equation}
    c^2\nabla^2\frac{\varphi}{\varphi_\ast} = \frac{1}{2}\xi^{-2}H_0^2a^2\frac{\varphi}{\varphi_\ast}\left(\frac{\varphi^2}{\varphi_\ast^2}-1\right) + \frac{1}{2}\xi^{-2}H_0^2a_\ast^3\frac{\varphi}{\varphi_\ast}\frac{{\rho}_m}{\bar{\rho}_{\rm m}}a^{-1},
\end{equation}
or equivalently
\begin{equation}\label{eq:sym_eom}
    c^2\nabla^2u = \frac{1}{2}\xi^{-2}H_0^2a^2u\left(u^2-1\right) + \frac{1}{2}\xi^{-2}H_0^2a_\ast^3u(1+\delta)a^{-1},
\end{equation}
where the density contrast is defined as
\begin{equation}
    \delta \equiv \frac{\rho_{\rm m}}{\bar{\rho}_{\rm m}}-1.
\end{equation}

The symmetron model and its extensions have been studied with the help of numerical simulations in several works \cite{Davis:2011pj,Brax:2012a}, but the large computational cost has so far made it impossible to run large, high-resolution simulations for a very large number of parameter combinations, which is why we are implementing it in {\sc mg}-{\sc glam}. This model features the \textit{symmetron screening mechanism} \cite{Hinterbichler:2010es}, which helps to suppress the fifth force in dense environments by driving $\varphi\rightarrow0$ so that the coupling strength $\beta(\phi)\rightarrow0$, cf.~Eq.~\eqref{eq:sym_coupling_strength}. This essentially decouples the scalar field from matter and therefore eliminates the fifth force in these environments, such that the model could evade stringent local and Solar System constraints. The \textit{dilaton screening mechanism} \cite{Brax:2010gi} is another class of coupled scalar field models with a screening mechanism that works similarly, so in this paper we shall focus on the symmetron model only.

\subsection{Chameleon $f(R)$ gravity}
\label{subsect:fR}

$f(R)$ gravity \cite{Sotiriou:2008rp,DeFelice:2010aj} is a very popular class of modified gravity models, which can be described by the following gravitational action
\begin{equation}\label{eq:fR_action}
    S = \frac{M^2_{\rm Pl}}{2}\int{\rm d}^4x\sqrt{-g}\left[R+f(R)\right],
\end{equation}
simply replacing the cosmological constant $\Lambda$ with an algebraic function of the Ricci scalar, $f(R)$. It is well known that this theory can be equivalently rewritten as a scalar-tensor theory after a change of variable, and is therefore mathematically and physically equivalent to a coupled scalar field model in which the scalar field has a universal coupling to different matter species. Therefore it belongs to the general models introduced in the beginning of this section. The model is fully specified by fixing the function $f(R)$, with different choices of $f(R)$ equivalent to coupled scalar field models with different forms of the scalar potential $V(\phi)$. Meanwhile, the coupling strength of the scalar field is a constant $\beta=1/\sqrt{6}$ for all $f(R)$ models\footnote{This means that the ratio between the strengths of the fifth and the standard Newtonian forces is at most $1+2\beta^2=1/3$. For more details see below.}, independent of $f(R)$. Despite this limitation, this model still has very rich phenomenology, and in this paper we will study it in the original form given by Eq.~\eqref{eq:fR_action}, instead of studying its equivalent coupled scalar field model.

With certain choices of the function $f(R)$, the model can have the so-called \textit{chameleon screening} mechanism \cite{Khoury:2003aq,Khoury:2003rn,Mota:2006fz,Brax:2008hh}, which can help the fifth force to hide from experimental detections in dense environments where $\rho_{\rm m}$ is high and the scalar field acquires a large mass $m$ and therefore its strength decays exponentially and essentially vanishes beyond a typical distance of order $m^{-1}$. Of course, not all choices of $f(R)$ can lead to a viable chameleon screening mechanism, and in this paper we will focus only on those where the chameleon mechanism works, and we call the latter \textit{chameleon $f(R)$ gravity}. 

In $f(R)$ gravity, the Einstein equation is modified to
\begin{equation}\label{eq:fR_modified_einstein_eqn}
    G_{\mu\nu} - X_{\mu\nu} = 8\pi{G}T_{\mu\nu},
\end{equation}
where $T_{\mu\nu}$ is the energy-momentum tensor, $G_{\mu\nu}\equiv{R}_{\mu\nu}-\frac{1}{2}g_{\mu\nu}R$ is the Einstein tensor with $R_{\mu\nu}$ being the Ricci tensor, and $X_{\mu\nu}$ is defined as
\begin{equation}\label{eq:Xmn}
    X_{\mu\nu} \equiv -f_RR_{\mu\nu} + \frac{1}{2}\left[f(R)-\nabla^\lambda\nabla_\lambda f_R\right]g_{\mu\nu} + \nabla_\mu\nabla_\nu f_R,
\end{equation}
where $f_R\equiv{\rm d}f(R)/{\rm d}R$ is a new dynamical scalar degree of freedom, with the following equation of motion
\begin{equation}\label{eq:fR_eom}
    \nabla^\mu\nabla_\mu{f}_R = \frac{1}{3}\left[R-f_RR+2f(R)-8\pi{G}\rho_{\rm m}\right].
\end{equation}

One of the leading choices of the function $f(R)$ was the one proposed by Hu \& Sawicki \cite{Hu:2007nk}. In this paper, instead of using the original function form provided in \cite{Hu:2007nk}, we present it in an approximate form which will allow us to generalise it. Let's start with the following expression of $f_R(R)$,
\begin{equation}\label{eq:fR_HS}
    f_R(R) = -\left|f_{R0}\right|\left(\frac{\bar{R}_0}{R}\right)^{n+1},
\end{equation}
where $f_{R0}$ is the present-day value of the background $f_R$, $\bar{R}_0$ is the background Ricci scalar today, and $n\geq0$ is an integer. For $n>0$, the functional form $f(R)$ can be written as
\begin{equation}\label{eq:fR_HS_original}
    f(R) \approx -6H_0^2\Omega_\Lambda + \frac{1}{n}\left|f_{R0}\right|\left(\frac{\bar{R}_0}{R}\right)^{n+1}R,
\end{equation}
where $\Omega_\Lambda=1-\Omega_m$ and the first term represents a cosmological constant that is responsible for the cosmic acceleration. For $n=0$, we have
\begin{equation}\label{eq:fR_log}
    f(R) \approx -6H_0^2\Omega_\Lambda + \left|f_{R0}\right|\bar{R}_0\ln\left(\frac{\bar{R}_0}{R}\right).
\end{equation}
Most of the simulation works to date have been performed for the case of $n=1$, while the cases of $n=2$ and $n=0$ are not as well explored. In this paper we implement all three cases into {\sc mg}-{\sc glam}. 

In general, for the model to have a viable chameleon screening, the parameter $f_{R0}$ in Eq.~\eqref{eq:fR_HS} should satisfy $|f_{R0}|\ll1$. At late times, when $\bar{R}(a)\simeq\bar{R}_0$, we can see from Eqs.~(\ref{eq:fR_HS_original}, \ref{eq:fR_log}) that the relation $f(R)\simeq-6H_0^2\Omega_\Lambda$ holds. On the other hand, from Eq.~\eqref{eq:fR_HS} we have $|f_R|\ll1$ throughout the cosmic history, i.e., it has a negligible evolution in time. This implies that all the terms in $X_{\mu\nu}$ in Eq.~\eqref{eq:Xmn} other than $f(R)$ can be neglected compared with the $f(R)$ term, and so the model behaves approximately like $\Lambda$CDM in the background expansion rate, with the background Ricci scalar given by
\begin{equation}\label{eq:R_bar}
    \bar{R}(a) = 3\mathbb{M}^2\left(a^{-3}+4\frac{\Omega_\Lambda}{\Omega_m}\right),
\end{equation}
and $\mathbb{M}^2\equiv{H}_0^2\Omega_m$. This is compatible with what we mentioned above, i.e., in the coupled scalar field model that is equivalent to these $f(R)$ models, the scalar field $\phi$ has little time evolution and therefore has an equation of state which is very close to $-1$. It also implies that the weak-field approximation, where we can neglect the time evolution of the scalar degree of freedom $f_R$, is a good approximation, so that in an inhomogeneous Universe we have
\begin{equation}\label{eq:fR_eom_qsa}
    \boldsymbol{\nabla}^2f_R \approx \frac{1}{3c^2}\left[\delta R-8\pi G\delta\rho_{\rm m}\right]a^2,
\end{equation}
where $\boldsymbol{\nabla}$ is the gradient with respect to the comoving coordinate, as before, $\delta\rho_{\rm m}\equiv\rho_{\rm m}-\bar{\rho}_{\rm m}=\bar{\rho}_{\rm m}\delta$, and 
\begin{equation}
    \delta R = R - \bar{R}.
\end{equation}
By realising that Eq.~\eqref{eq:fR_HS} can be inverted to give
\begin{equation}\label{eq:R_fR}
    R = \bar{R}_0\left(\frac{f_{R0}}{f_R}\right)^{\frac{1}{n+1}}.
\end{equation}
With Eq.~\eqref{eq:R_fR}, Eq.~\eqref{eq:fR_eom_qsa} becomes a nonlinear dynamical equation for $f_R$. 

Also under the quasi-static and weak-field approximations, the Poisson equation takes the following modified form
\begin{equation}\label{eq:fR_Poisson_qsa}
    \boldsymbol{\nabla}^2\Phi \approx \frac{16\pi{G}}{3}\delta\rho_{\rm m}a^2 - \frac{1}{6}\delta{R}a^2 = 
    4\pi{G} \bar{\rho}_{\rm m} a^2 \delta - \frac{1}{2} c^2\boldsymbol{\nabla}^2 f_R \,,
\end{equation}
where in the second step we have used Eq.~\eqref{eq:fR_eom_qsa}.

One can have a quick peek into two opposite regimes of solutions for Eqs.~(\ref{eq:fR_eom_qsa}, \ref{eq:fR_Poisson_qsa}). In the large field limit, when $|f_R|$ is relatively large (e.g., in the case of large $|f_{R0}|$), the perturbation $\delta f_R$ is small compared to the background field $|\bar{f}_R|$, and $|\delta R|\ll8\pi{G}\delta\rho_{\rm m}$, so that the Poisson equation \eqref{eq:fR_Poisson_qsa} can be approximated as 
\begin{equation}
    \boldsymbol{\nabla}^2\Phi \approx \frac{16\pi{G}}{3}\delta\rho_{\rm m}a^2.
\end{equation}
Comparing this with the standard Poisson equation in $\Lambda$CDM,
\begin{equation}\label{eq:GR_Poisson_qsa}
    \boldsymbol{\nabla}^2\Phi \approx {4\pi{G}}\delta\rho_{\rm m}a^2,
\end{equation}
we confirm that the fifth force, i.e., the enhancement of gravity, is $1/3$ of the strength of the standard Newtonian force. In the opposite, small-field, limit where $|f_R|$ takes very small values, the left-hand side of Eq.~\eqref{eq:fR_eom_qsa} is negligible and so we have $\delta{R}\approx8\pi{G}\delta\rho_{\rm m}$, and plugging this into Eq.~\eqref{eq:fR_Poisson_qsa} we recover Eq.~\eqref{eq:GR_Poisson_qsa}: this is the screened regime where the fifth force is strongly suppressed.

\subsection{Summary and comments}
\label{subsect:model_summary}

In this section we have briefly summarised the essentials of the three classes of scalar field modified gravity models to be considered in this work. Among these, coupled quintessence is technically more trivial, because the fifth force is unscreened nearly everywhere, while $f(R)$ gravity and symmetrons are both representative thin-shell screening models \cite{Brax:2012gr} featuring two of the most important screening mechanisms respectively. Compared with previous simulation work, we will consider $f(R)$ models with more values of the parameter $n$: as discussed below, instead of the common choice of $n=1$, we will also look at $n=0,2$ to see how the phenomenology of the model varies.  

We remark that, even with the additional modified gravity models implemented in this paper, as well as the models implemented in the twin paper \cite{Hernandez-Aguayo:2021_twin_paper}, we are still far from covering all possible models. Changing the coupling function $A(\varphi)$ or the scalar field potential $V(\varphi)$, as an example, will lead to new models. However, our objective is to have an efficient simulation code that covers different \textit{types} of models, which serves as a `prototype' that can be very easily modified for any other models belonging to the same type. This differs from the model-independent \cite{Srinivasan:2021gib} or parameterised modified gravity \cite{Hassani:2020rxd} approaches adopted elsewhere, and we perfer this approach since there is a direct link to some fundamental Lagrangian here, and because, any parameterisation of models, one its parameters specified, also corresponds to a fixed model.

\section{Numerical Implementations}
\label{sect:numerics}

This section is the core part of this paper, where will describe in detail how the different theoretical models of \S\ref{sec:theories} can be incorporated in a numerical simulation code, so that the scalar degree of freedom can be solved at any given time with any given matter density field. This way, the various effects of the scalar field on cosmic structure formation can be accurately predicted and implemented. 

\subsection{The {\sc glam} code}
\label{subsect:glam}

The \textsc{glam} code is presented in \cite{Klypin:2017iwu}, and is a promising tool to quickly generate $N$-body simulations with reasonable speed and acceptable resolution, which are suitable for the massive production of galaxy survey mocks.

As a PM code, \textsc{glam} solves the Poisson equation for the gravitational potential in a periodic cube using fast Fourier Transformation (FFT). The code uses a 3D mesh for density and potential estimates, and only one mesh is needed for the calculation: the density mesh is replaced with the potential. The gravity solver uses FFT to solve the discrete analogue of the Poisson equation, by applying it first in $x$- and then to $y$-direction, and finally transposing the matrix to improve data locality before applying FFT in the third ($z$-)direction. After multiplying this data matrix by the Green's function, an inverse FFT is applied, performing one matrix transposition and three FFTs, to compute the Newtonian potential field on the mesh. The potential is then differentiated using a standard three-point finite difference scheme to obtain the $x,y$ and $z$ force components at the centres of the mesh cells. These force components are next interpolated to the locations of simulation particles, which are displaced using a leapfrog scheme. A standard Cloud-in-Cell (CIC) interpolation scheme is used for both the assignment of particles to calculate the density values in the mesh cells and the interpolation of the forces.  

A combination of parameters that define the resolution and speed of the \textsc{glam} code are carefully selected. For example, it uses the \textsc{FFT}5 code (the Fortran 90 version of \textsc{FFTpack}5.1) because it has an option of real-to-real FFT that uses only half of the memory as compared to \textsc{FFTW}. It typically uses $1/2$--$1/3$ of  the number of particles (in 1D) as compared with the mesh size---given that the code is limited by available RAM, this is a better combination than using the same number of particles and mesh points.

\textsc{glam} uses \textsc{openmp} directives to parallelise the solver. 
Overall, the code scales nearly perfectly, as has been demonstrated by tests run with different mesh sizes and on different processors (later in the paper we will present some actual scaling test of \textsc{mg-glam} as well, which again is nearly perfect). \textsc{mpi} parallelisation is used only to run many realisations on different supercomputer nodes with very little inter-node communications. Load balance is excellent since theoretically every realisation requires the same number of CPUs. 

Initial conditions are generated on spot by \textsc{glam}, using the standard Zel'dovich approximation \cite{Zeldovich:1970A&A.....5...84Z,Efstathiou:1985ApJS...57..241E} from a user-provided linear matter power spectrum $P(k)$ at $z=0$. The code backscales this $P(k)$ to the initial redshift $z_{\rm ini}$ using the \textcolor{black}{scale-independent} linear growth factor for $\Lambda$CDM with the specified cosmological parameters. As the Zel'dovich approximation is less accurate at low redshifts \cite{Crocce:2006ve}, the simulation is started at an initial redshift $z_{\rm ini}\geq100$. \textcolor{black}{Starting at a higher redshift such as $z_{\rm ini}=100$ also has the additional advantage that, for the MG models of interest here, the effect of the scalar field is smaller at earlier times, which means that it is an increasingly better approximation to use the same initial conditions in the MG models as in the $\Lambda$CDM model with the same cosmological parameters, as we practice throughout this work. If, as in the general scenarios, there is non-negligible MG effect prior to $z_{\rm ini}$, such effect should be taken into account in the generation of initial conditions, e.g., \citep{Li:2011PhRvD..83b4007L}. We note that using $\Lambda$CDM initial conditions in the MG simulations means that we do not need to backscale the linear $P(k)$ (e.g., at $z=0$) of the corresponding MG models, which are usually scale-dependent --- this latter approach has been checked for clustering dark energy models in \citep{Hassani:2019JCAP...12..011H}, where it is found to be unable to give the correct matter and gravitational potential power spectra at late times simultaneously (see \citep{Hassani:2019wed} for a way to overcome this issue).}


\textsc{glam} uses a fixed number of time steps, but this number is user-specified. The standard choice is about $150$--$200$. Here, we have compared the model difference of the matter power spectra between modified gravity \textsc{mg-glam} and $\Lambda$CDM \textsc{glam} simulations and found that the result is converged with $160$ time steps. Doubling the number of steps from $160$ to $320$ makes negligible difference. 

The code generates the density field, including peculiar velocities, for a particular cosmological model. Nonlinear matter power spectra and halo catalogues at user-specified output redshifts (snapshots) are measured on the fly. For the latter, \textsc{glam} employs the Bound Density Maximum (BDM; \cite{Klypin:1997sk,Riebe:2011arXiv1109.0003R}) algorithm to get around the usual limitations placed on the completeness of low-mass haloes by the lack of force resolution in PM simulations. Here we briefly describe the idea behind the BDM halo finder, and further details can be found in \cite{Riebe:2011arXiv1109.0003R,Knebe:2011MNRAS.415.2293K}. The code starts by calculating a local density at the positions of individual particles, using a spherical tophat filter containing a constant number $N_{\rm filter}$ (typically 20) of particles. It then gathers all the density maxima and, for each maximum, finds a sphere that contains a mass $M_\Delta = \frac{4}{3}\pi\Delta\rho_{\rm crit}(z)R_\Delta^3$, where $\rho_{\rm crit}(z)$ is the critical density at the halo redshift $z$, and $\Delta$ is the overdensity within the halo radius $R_\Delta$. Throughout this work we will use the virial density definition for $\Delta$ given by \cite{Bryan:1997dn}
\begin{equation}
    \Delta_{\rm vir}(z) = 18\pi^2 + 82\left[\Omega_{\rm m}(z)-1\right] - 39\left[\Omega_{\rm m}(z)-1\right]^2,
\end{equation}
where $\Omega_{\rm m}(z)$ is the matter density parameter at $z$. To find distinct haloes, the BDM halo finder still needs to deal with overlapping spheres. To this end, it treats the density maxima as halo centres and finds the one sphere, amongst a group of overlapping ones, with the deepest Newtonian potential. This is treated as a distinct, central, halo. The radii and masses of the haloes which correspond to the other (overlapping) spheres are then found by a procedure that guarantees a smooth transition of the properties of small haloes when they fall into the larger halo to become subhaloes of the latter. The latter is done by defining the radius of the infalling halo as $\max(R_1, R_2)$, where $R_1$ is its distance to the surface of the larger, soon-to-be host, central halo, and $R_2$ is its distance to the nearest density maximum in the spherical shell  $[\min(R_\Delta,R_1),\max(R_\Delta,R_1)]$ centred around it (if no density maximum exists in this shell, $R_2=R_\Delta$). The BDM halo finder was compared against a range of other halo finders in \cite{Knebe:2011MNRAS.415.2293K}, where good agreement was found. 


\textsc{mg-glam} extends \textsc{glam} to a general class of modified gravity theories by adding extra modules for solving MG scalar field equations, which will be introduced in the following subsection.

\subsubsection{The \textsc{glam} code units}
\label{sec:glam_units}

Like most other $N$-body codes, \textsc{glam} uses its own internal unit system. The code units are designed such that the physical equations can be cast in dimensionless form, which is more convenient for numerical solutions.

Let the box size of simulations be $L$ and the number of grid points in one dimension be $N_{\rm g}$. We can introduce dimensionless coordinates $\tilde{\boldsymbol{x}}$, momenta $\tilde{\boldsymbol{p}}$ and potentials $\tilde{\Phi}$ using the following relations \citep{Klypin:2017iwu} 
\begin{equation}\label{eq:code_units}
\tilde{\boldsymbol{x}} = \left( \frac{N_{\rm g}}{L}\right) {\boldsymbol{x}} \,, \qquad
\tilde{\boldsymbol{p}} = \left( \frac{N_{\rm g}}{H_0 L}\right) {\boldsymbol{p}}\,, \qquad
\tilde\Phi = \left( \frac{N_{\rm g}}{H_0 L}\right)^2\Phi\,.
\end{equation}
Having the dimensionless momenta, we can find the peculiar velocity,
\begin{equation}
{\boldsymbol{v}}_{\rm pec} = 100 \left(\frac{L}{N_{\rm g}}\right)\left(\frac{\tilde{\boldsymbol{p}}}{a}\right)\,{\rm km}~{\rm s}^{-1}\,,
\end{equation}
where we assumed that box size $L$ is given in units of $\Mpch$.
Using these notations, we write the particle equations of motion and the Poisson equation as
\begin{align}
\frac{{\rm d}\tilde{\boldsymbol{p}}}{{\rm d} a} &= -\left(\frac{H_0}{\dot{a}}\right)\tilde{\boldsymbol{\nabla}}\tilde{\Phi}\,,\\
\frac{{\rm d}\tilde{\boldsymbol{x}}}{{\rm d} a} &= -\left(\frac{H_0}{\dot{a}}\right)\frac{\tilde{\boldsymbol{p}}}{a^2}\,,\\
\label{eq:GR_poisson_codeunit}\tilde{\nabla}^2\tilde{\Phi} &= \frac{3}{2}\Omega_{\rm m} a^{-1} \tilde{\delta},
\end{align}
where $\tilde{\delta}$ is the code unit expression of the density contrast $\delta$.

From Eqs.~\eqref{eq:code_units} we can derive the following units,
\begin{equation}\label{eq:code_units2}
\tilde{\boldsymbol{\nabla}} = \left(\frac{L}{N_{\rm g}}\right) \boldsymbol{\nabla}\,, \quad {\rm d} \tilde t = H_0 {\rm d} {t}\,, \quad \tilde{\rho}_{\rm m} = \left( \frac{a^3}{\rho_{\rm crit, 0}\Omega_{\rm m}}\right) \rho_{\rm m}\,, \quad \tilde{\delta} = \delta\,.
\end{equation}
In what follows, we will also use the following definition
\begin{equation}
    \tilde{c} = \left(\frac{N_{\rm g}}{H_0 L}\right) c
\end{equation}
for the code-unit expression of the speed of light, $c$.

{\sc glam} uses a regularly spaced three-dimensional mesh of size $N_{\rm g}^3$ that covers the cubic domain $L^3$ of a simulation box. The size of a cell, $\Delta x =L/N_{\rm g}$, and the mass of each particle, $m_{\rm p}$, define the force and mass resolution respectively:
\begin{eqnarray}
m_{\rm p} &=& \Omega_{\rm m} \, \rho_{\rm crit,0}\left[\frac{L}{N_{\rm p}}\right]^3 = 8.517\times 10^{10}\left[\frac{\Omega_{\rm m}}{0.30}\right]
\left[\frac{L/\Gpch}{N_{\rm p}/1000}\right]^3h^{-1}M_\odot, \label{eqn:mass_resolution_def}\\
\Delta x &=& \left[\frac{L/\Gpch}{N_{\rm g}/1000}\right]\Mpch, \label{eqn:force_resolution_def}
\end{eqnarray}
where $N_{\rm p}^3$ is the number of particles and $\rho_{\rm crit,0}$ is the critical density of the universe at present.

\subsection{Solvers for the extra degrees of freedom}
\label{subsect:extradof}

We have seen in \S\ref{sec:theories} that in modified gravity models we usually need to solve a new, dynamical, degree of freedom, which is governed by some nonlinear, elliptical type, partial differential equation (PDE). Being a nonlinear PDE, unlike the linear Poisson equation solved in default {\sc glam}, the equation can not be solved by a one-step fast Fourier transform\footnote{This does not mean that FFT cannot be used under any circumstances. For example, Ref.~\cite{Chan:2009ew} used a FFT-relaxation method to solve nonlinear PDEs iteratively. In each iteration, the equation is treated as if it were linear (by treating the nonlinear terms as a `source') and solved using FFT, but the solution in the previous step is used to update the `source', for the PDE to be solved again to get a more accurate solution, until some convergence is reached.} but requires a \textit{multigrid relaxation} scheme to obtain a solution.

For completeness, we will first give a concise summary of the relaxation method and its multigrid implementation (\S\ref{subsubsect:relaxation}). Next, we will specify the practical side, discussing how to efficiently arrange the memory in the computer, to allow the same memory space to be used for different quantities at different stages of the calculation, therefore minimising the overall memory requirement (\S\ref{subsubsect:code_struc}), and also saving the time for frequently allocating and deallocating operations. 
After that, in \S\ref{subsubsect:csf_imp}--\S\ref{subsubsect:fR_imp}, we will respectively discuss how the nonlinear PDEs in general coupled quintessence, symmetron and $f(R)$ models can be solved most efficiently. Much effort will be devoted to replacing the common Newton-Gauss-Seidel relaxation method by a nonlinear Gauss-Seidel, which has been found to lead to substantial speedup of simulations \cite{Bose:2016wms} (but we will generalise this to more models than focused on in Ref.~\cite{Bose:2016wms}). For the coupled quintessence model, we will also briefly describe how the background evolution of the scalar field is numerically solved as an integral part of {\sc mg}-{\sc glam}, to further increase its flexibility. 

\subsubsection{Multigrid Gauss-Seidel relaxation}
\label{subsubsect:relaxation}

Let the partial differential equation (PDE) to be solved take the following form:
\begin{equation}
    \mathcal{L}(u) = 0,
\end{equation}
where $u$ is the scalar field and $\mathcal{L}$ is the PDE operator. To solve this equation numerically, we use finite difference to get a discrete version of it on a mesh\footnote{In this paper we consider the simplest case of cubic cells.}. Since {\sc mg}-{\sc glam} is a particle-mesh (PM) code, it has a uniform mesh resolution and does not use adaptive mesh refinement (AMR). When discretised on a uniform mesh with cell size $h$, the above equation can be denoted as
\begin{equation}\label{eq:general_pde}
    \mathcal{L}^h({u}^h) = {f}^h,
\end{equation}
where we have added a nonzero right-hand side, $f^h$, for generality (while $f^h=0$ on the mesh with cell size $h$, later when we discrete it on coarser meshes needed for the multigrid implementation, $f$ is no longer necessarily zero). Both $u^h$ and $f^h$ are evaluated at the cell centres of the given mesh.

The solution we obtain numerically, $\hat{u}$, is unlikely to be the true solution $u^h$ to the discrete equation, and applying the PDE operator on the former gives the following, slightly different, equation:
\begin{equation}\label{eq:general_pde_numerical}
    \mathcal{L}^h(\hat{u}^h) = \hat{f}^h.
\end{equation}
Taking the difference between the above two equations, we get
\begin{equation}\label{eq:numerical_error}
    \mathcal{L}^h({u}^h) - \mathcal{L}^h(\hat{u}^h) = f^h-\hat{f}^h = -d^h,
\end{equation}
where 
\begin{equation}\label{eq:residual_d}
    d^h \equiv \hat{f}^h-f^h,
\end{equation}
is the {\it local residual}, which characterises the inaccuracy of the solution $\hat{u}^h$ (this is because if $\hat{u}^h=u^h$, we would expect $\hat{f}^h=f^h$ and hence there is zero `inaccuracy'). $d^h$ is also evaluated at cell centres. Later, to check if a given set of numerical solution $\hat{u}^h$ is acceptable, we will use a \textit{global residual}, $\epsilon^h$, which is a single number for the given mesh of cell size $h$. In this work we choose to define $\epsilon^h$ as the root-mean-squared of $d^h$ in all mesh cells (although this is by no means the only possible definition). We will call both $d^h$ and $\epsilon^h$ `residual' as the context will make it clear which one is referred to.

Relaxation solves Eq.~\eqref{eq:general_pde} by starting from some approximate trial solution to $u^h$, $\hat{u}^h_{\rm old}$, and check if it satisfies the PDE. If not, this trial solution can be updated using a method that is similar to the Newton-Ralphson iterative method to solve nonlinear algebraic equations
\begin{equation}\label{eq:relaxation_iteration}
    \hat{u}^h_{\rm new} = \hat{u}^h_{\rm old} - \frac{\mathcal{L}^h\left(\hat{u}^h_{\rm old}\right)-\hat{f}^h}{\partial\mathcal{L}^h\left(\hat{u}^h_{\rm old}\right)/\partial\hat{u}^h}.
\end{equation}
This process can be repeated iteratively, until the updated solution satisfies the PDE to an acceptable level, i.e., $\epsilon^h$ becomes small enough. In practice, because we are solving the PDE on a mesh, Eq.~\eqref{eq:relaxation_iteration} should be performed for all mesh cells, which raises the question of how to order this operation for the many cells. We will adopt the Gauss-Seidel `black-red chessboard' approach, where the cells are split into two classes, `black' and `red', such that all the six direct neighbours\footnote{The direct neighbours of a given cell are the six neighbouring cells which share a common face with that cell.} of a `red' cell are black and vice versa. The relaxation operation, Eq.~\eqref{eq:relaxation_iteration}, is performed in two sweeps, the first for `black' cells (i.e., only updating $\hat{u}^h$ in `black' cells while keeping their values in `red' cells untouched), while the second for all the `red' cells. This is a standard method to solve nonlinear elliptical PDEs by using relaxation, known as the \textit{Newton-Gauss-Seidel method}. However, although this method is generic, it is not always efficient, and later we will describe a less generic alternative which is nevertheless more efficient. 

Relaxation iterations are useful at reducing the Fourier modes of the error in the trial solution $\hat{u}^h$, whose wavelengths are comparable to that of the size of the mesh cell $h$. If we do relaxation on a fine mesh, this means that the short-wave modes of the error are quickly reduced, but the long-wave modes are generally much slower to decrease, which can lead to a slow convergence of the relaxation iterations. A useful approach to solve this problem is by using {\it multigrid}: after a few iterations on the fine level, we `move' the equation to a coarser level where the cell size is larger and the longer-wave modes of the error in $\hat{u}^h$ can be more quickly decreased. The discretised PDE on the coarser level is given by
\begin{equation}\label{eq:general_pde_numerical_coarse}
    \mathcal{L}^H(u^H) = \mathcal{L}\left(\mathcal{R}\hat{u}^h\right) - \mathcal{R}d^h \equiv S^H,
\end{equation}
where the superscript $^H$ denotes the coarse level where the cell size is $H$ (in our case $H=2h$), and $\mathcal{R}$ denotes the \textit{restriction} operator which interpolates quantities from the fine level to the coarse level. In our numerical implementation, a coarse (cubic) cell contains 8 fine (cubic) cells of equal volume, and the restriction operation can be conveniently taken as the arithmetic average of the values of the quantity to be interpolated in the 8 fine cells. 

Eq.~\eqref{eq:general_pde_numerical_coarse} can be solved using relaxation similarly to Eq.~\eqref{eq:general_pde_numerical}, for which the numerical solution is denoted as $\hat{u}^H$. This can be used to `correct' and `improve' the approximate solution $\hat{u}^h$ on the fine level, as
\begin{equation}\label{eq:solution_correction}
    \hat{u}^{h,{\rm new}} = \hat{u}^{h,{\rm old}} + \mathcal{P}\left(\hat{u}^H - \mathcal{R}\hat{u}^h\right),
\end{equation}
where $\mathcal{P}$ is the \textit{prolongation} operation which does the interpolation from the coarse to the fine levels. In this work we shall use the following definition of the prolongation operation: for a given fine cell,
\begin{enumerate}
    \item find its parent cell, i.e., the coarser cell that contains the fine cell;
    \item find the seven neighbours of the parent cell, i.e., the coarser cells which share a face (there are 3 of these), an edge (there are 3 of these) or a vertex (just 1) with the above parent coarser cell;
    \item calculate the fine-cell value of the quantity to be interpolated from the coarse to the fine levels, as a weighted average of the corresponding values in the 8 coarse cells mentioned above: $27/64$ for the parent coarse cell, and $9/64$, $3/64$ and $1/64$ respectively for the coarse cells sharing a face, an edge and a vertex with the parent cell. 
\end{enumerate}

The above is a simple illustration of how multigrid works for two levels of mesh resolution, $h$ and $H$. In principle, multigrid can be and is usually implemented using more than two levels. In this paper we will use a hierarchy of increasingly coarser meshes with the coarsest one having $4^3$ cells. 

\begin{figure}
    \centering
    \includegraphics[width=\textwidth]{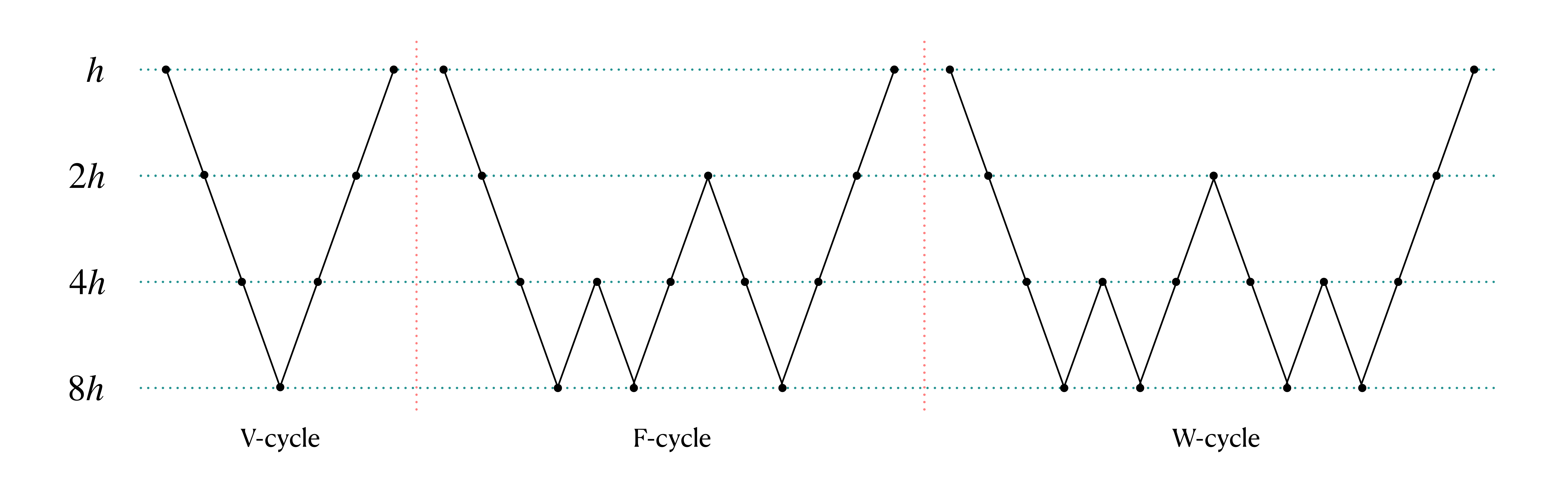}
\caption{An illustration of the three different arrangements of multigrid relaxation method used in this paper: from left to right, V-cycle, F-cycle and W-cycle. The horizontal dotted lines depict 4 multigrid levels of mesh, with the finest mesh (denoted by its cell size $h$) on top, and the coarsest mesh (with cell size $8h$) at the bottom. The relaxation always starts on the finest level, and the solid lines show how the multigrid solver walks through the different levels, performing Gauss-Seidel relaxation iterations at each level (denoted by the circles), called smoothing. Only one single full cycle is shown for each case. The solver walks over the multigrid levels more times in W-cycle than in F-cycle and V-cycle, and thus it requires fewer cycles in the former case to arrive at a converged solution. However, it is also computationally more expensive. We will compare the performances of the three different arrangements in real cosmological simulations in Sect.~\ref{subsect:convergence_tests}.}
    \label{fig:multigrid_cycles}
\end{figure}

There are flexibilities in how to arrange the relaxations at different levels. The most-commonly used arrangement is the so-called V-cycle, where one starts from the finest level, moves to the coarsest one performing relaxation iterations on each of the intermediate levels (cf.~Eq.~\eqref{eq:general_pde_numerical_coarse}), and then moves straight back to the finest performing corrections using Eq.~\eqref{eq:solution_correction} on each of the intermediate levels. Other arrangements, such as F-cycle and W-cycle (cf.~Fig.~\ref{fig:multigrid_cycles}), are sometimes more efficient in improving the convergence rate of $\hat{u}^h$ to $u^u$, and we have implemented them in {\sc mg}-{\sc glam} as well. 

\subsubsection{Memory usage}
\label{subsubsect:code_struc}

{\sc glam} uses a single array to store mesh quantities, such as the matter density field and the Newtonian potential, because at any given time only one of these is needed. The Newtonian force at cell centres is calculated by finite-differencing the potential and then interpolated to the particle positions. To be memory efficient, \textsc{glam} also opts not to create a separate array to store the forces at the cell centres, but instead directly calculates them at the particle positions immediately before updating the particle velocities.

With the new scalar field to be solved in modified gravity models, we need two additional arrays of size $N_{\rm g}^3$, where $N_{\rm g}^3$ is the number of cells of the PM grid (i.e., there are $N_{\rm g}$ cells in each direction of the cubic simulation box). This leads to three arrays. \texttt{Array 1} is the default array in {\sc glam}, which is used to store the density field $\rho$ and the Newtonian potential $\Phi$ (at different stages of the simulation). Note that the density field is also needed when solving the scalar field equation of motion during the relaxation iterations, and so we cannot use this array to also store the scalar field. On the other hand, we will solve the Newtonian potential after the scalar field, by when it is safe to overwrite this array with $\Phi$. \texttt{Array2} is exclusively used to store the scalar field solution $\hat{u}^h$ on the PM grid, which will be used to calculate the fifth force. \texttt{Array3} is used to store the various intermediate quantities which are created for the implementation of the multigrid relaxation, such as $d^h$, $\hat{u}^H$, $\mathcal{R}\hat{u}^h$, $\mathcal{R}d^h$, $S^H$ and $\rho^H$, the last of which is the density field on the coarser level $^H$, which appears in the coarse-level discrete PDE operator $\mathcal{L}^H$.

To be concrete, we imagine the 3D array (\texttt{Array3}) as a cubic box with $N_{\rm g}^3$ cubic cells of equal size. An array element, denoted by $(i,j,k)$, represents the $i$th cell in the $x$ direction, $j$th cell in the $y$ direction and $k$th cell in the $z$ direction, with $i,j,k=1,\cdots,N_{\rm g}$. We divide this array into 8 sections, each of which can be considered to correspond to one of the 8 octants that equally divide the volume of the cubic box. The range of $(i,j,k)$ of each section and the quantity stored in that section of \texttt{Array3} are summarised in the table below: 
\begin{center}
    \begin{tabular}{|c|p{3cm}|p{3cm}|p{3cm}|p{3cm}|}
    \hline
    Section & $i$ range & $j$ range & $k$ range & Quantity\\ 
    \hline
    \hline
    1 & $1,\cdots,N_{\rm g}/2$ & $1,\cdots,N_{\rm g}/2$ & $1,\cdots,N_{\rm g}/2$ & $d^\ell$, $\mathcal{R}d^\ell$ \\ 
    \hline
    2 & $N_{\rm g}/2+1,\cdots,N_{\rm g}$ & $1,\cdots,N_{\rm g}/2$ & $1,\cdots,N_{\rm g}/2$ & $d^\ell$, $\rho^{\ell-1} = \mathcal{R}\rho^\ell$ \\ 
    \hline
    3 & $1,\cdots,N_{\rm g}/2$ & $N_{\rm g}/2+1,\cdots,N_{\rm g}$ & $1,\cdots,N_{\rm g}/2$ & $d^\ell$, $\mathcal{R}\hat{u}^\ell$ \\ 
    \hline
    4 & $N_{\rm g}/2+1,\cdots,N_{\rm g}$ & $N_{\rm g}/2+1,\cdots,N_{\rm g}$ & $1,\cdots,N_{\rm g}/2$ & $d^\ell$, $\hat{u}^{\ell-1}$ \\ \hline
    5 & $1,\cdots,N_{\rm g}/2$ & $1,\cdots,N_{\rm g}/2$ & $N_{\rm g}/2+1,\cdots,N_{\rm g}$ & $d^\ell$, recursion \\ 
    \hline
    6 & $N_{\rm g}/2+1,\cdots,N_{\rm g}$ & $1,\cdots,N_{\rm g}/2$ & $N_{\rm g}/2+1,\cdots,N_{\rm g}$ & $d^\ell$, $d^{\ell-1}$ \\ \hline
    7 & $1,\cdots,N_{\rm g}/2$ & $N_{\rm g}/2+1,\cdots,N_{\rm g}$ & $N_{\rm g}/2+1,\cdots,N_{\rm g}$ & $d^\ell$, $S^{\ell-1}$ \\ \hline
    8 & $N_{\rm g}/2+1,\cdots,N_{\rm g}$ & $N_{\rm g}/2+1,\cdots,N_{\rm g}$ & $N_{\rm g}/2+1,\cdots,N_{\rm g}$ & $d^\ell$ \\
    \hline
    \end{tabular}
\end{center}

Let us explain this more explicitly. First of all, the whole \texttt{Array3}, of size $N_{\rm g}^3$, will be used to store the residual value $d^h$ on the PM grid (which has $N_{\rm g}^3$ cells). From now on, we label this grid by `level-$\ell$', and use `level-($\ell-m$)' to denote the grid that are $m$ times coarser, i.e., if the cell size of the PM grid is $h$, then the cells in this coarse grid have a size of $2^mh$. 
In the table above we have used $d^\ell$ to denote the $d^h$ on level-$\ell$, and so on. 
Note that we always use $N_{\rm g}=2^\ell$.

The local residual $d^h$ on a fine grid is only needed for two purposes: (1) to calculate the global residual on that grid, $\epsilon^h$, which is needed to decide convergence of the relaxation, and (2) to calculate the coarse-level PDE operator $\mathcal{L}^H$ that is needed for the multigrid acceleration, as per Eq.~\eqref{eq:general_pde_numerical_coarse}. This suggests that $d^h$ does not have to occupy \texttt{Array3} all the time, and so this array can be reused to store other intermediate quantities (see the last column of the above table) \textit{after} we have obtained $\epsilon^h$. 

In our arrangement, Section 1 stores the residual $\mathcal{R}d^\ell$, Section 2 stores the restricted density field $\rho^{\ell-1}=\mathcal{R}\rho^\ell$, Sections 3 and 4 store, respectively, the restricted scalar field solution $\mathcal{R}\hat{u}^\ell$ and the coarse-grid scalar field solution $\hat{u}^{\ell-1}$ --- the former is needed to calculate $S^{\ell-1}$ in Eq.~\eqref{eq:general_pde_numerical_coarse} and to correct the fine-grid solution using Eq.~\eqref{eq:solution_correction}, which is fixed after calculation, while the latter is updated during the coarse-grid relaxation sweeps\footnote{We use $\mathcal{R}\hat{u}^\ell$ as the initial guess for $\hat{u}^{\ell-1}$ for the Gauss-Seidel relaxations on the coarse level.}. Section 7 stores the coarse-grid source $S^{\ell-1}$ for the PDE operator $\mathcal{L}^{\ell-1}$ as defined in Eq.~\eqref{eq:general_pde_numerical_coarse}, and finally Section 6 stores the residual on the coarse level, $d^{\ell-1}$. Note that all these quantities are for level-$(\ell-1)$, so that they can be stored in section of \texttt{Array3} of size $\left(N_{\rm g}/2\right)^3$. Section 8 is not used to store anything other than $d^\ell$. 

We have not touched Section 5 so far --- this section is reserved to store the same quantities as above, but for level-$(\ell-2)$, which are needed if we want to use more than two levels of multigrid. It is further divided into 8 section, each of which will play the same roles as detailed in the table above\footnote{The exception is that, as $d^{\ell-1}$ is already stored in Section 6, it does not have to be stored in Section 5 again.}. In particular, the (sub)Section 5 of Section 5 is reserved for quantities on level-$(\ell-3)$, and so on. In this way, there is no need to create separate arrays of various sizes to store the intermediate quantities on different multigrid levels which therefore saves memory. 

There is a small tricky issue here: as we mentioned above, the local residual $d^\ell$ on the PM grid is needed to calculate the coarse-grid source $S^{\ell-1}$ using Eq.~\eqref{eq:general_pde_numerical_coarse}, thus we will be using the quantity $d^\ell$ stored in \texttt{Array3} to calculate $\mathcal{R}d^{\ell}$ and then write it to (part of) the same array, running the risk of overwriting some of the data while it is still needed. To avoid this problem, we refrain from using the $d^\ell$ data already stored in \texttt{Array3}, but instead recalculate it in the subroutine to calculate $\mathcal{R}d^\ell$ (this only needs to be done for level-$\ell$). With a bit of extra computation, this enables use to avoid creating another array of similar size to \texttt{Array3}.

Since \texttt{Array3} stores different quantities in different parts, care must be excised when assessing these data. There is a simple rule for this: suppose that we need to read or write the quantities on the coarse grid of level-$(\ell-m)$ with $m\geq1$. These are 3-dimensional quantities with the three directions labelled by $I,J,K$, which run over $1,\cdots,2^{\ell-m}$, and we have
\begin{align}
    \mathcal{R}(d^{\ell-m+1}) \left[I,J,K\right] &\leftrightarrow \texttt{Array3}[i=I,&j=J\phantom{\ +2^{\ell-m}},k=K + \left(2^{m}-2\right)\cdot2^{\ell-m}],\nonumber\\
    \mathcal{R}(\rho^{\ell-m+1})\left[I,J,K\right] &\leftrightarrow \texttt{Array3}[i=I+2^{\ell-m},&j=J\phantom{\ +2^{\ell-m}},k=K + \left(2^{m}-2\right)\cdot2^{\ell-m}],\nonumber\\
    \mathcal{R}(u^{\ell-m+1})\left[I,J,K\right] &\leftrightarrow\texttt{Array3}[i=I,&j=J+2^{\ell-m},k=K + \left(2^{m}-2\right)\cdot2^{\ell-m}],\nonumber\\
    \hat{u}^{\ell-m}\left[I,J,K\right] &\leftrightarrow \texttt{Array3}[i=I+2^{\ell-m},&j=J+2^{\ell-m},k=K + \left(2^{m}-2\right)\cdot2^{\ell-m}],\nonumber\\
    d^{\ell-m}\left[I,J,K\right] &\leftrightarrow \texttt{Array3}[i=I+2^{\ell-m},&j=J\phantom{\ +2^{\ell-m}},k=K + \left(2^{m}-1\right)\cdot2^{\ell-m}],\nonumber\\
    S^{\ell-m}[I,J,K] &\leftrightarrow \texttt{Array3}[i=I,&j=J+2^{\ell-m},k=K + \left(2^{m}-1\right)\cdot2^{\ell-m}],
\end{align}
where $i,j,k=1,\cdots,N_{\rm g}$ run over the entire \texttt{Array3}.

We can estimate the required memory for \textsc{mg-glam} simulations as follows. As mentioned above, the code uses a 3D array of single precision to store both the density field and the Newtonian potential, and one set of arrays for particle positions and velocities. In addition, two arrays are added 
to store the scalar field solution (\texttt{Array2}) and various intermediate quantities in the multigrid relaxation solver (\texttt{Array3}). In the cosmological simulations described in this paper, we have used double precision for the two new arrays, and we have checked that using single precision slightly speeds up the simulation, while agreeing with the double-precision results within $0.001\%$ and $0.5\%$ respectively for the matter power spectrum and halo mass function. Given its fast speed and its shared-memory nature, memory is expected to be the main limiting factor for large \textsc{mg-glam} jobs. For this reason, we assume that all arrays are set to be single precision for future runs, and this leads to the following estimate of the total required memory:
\begin{align}
M_{\rm tot} &= 12N^3_{\rm g} + 24N^3_{\rm p} \, {\rm bytes}\,,\nonumber\\
&= 89.41 \qty(\frac{N_{\rm g}}{2000})^3 + 22.35\qty(\frac{N_{\rm p}}{1000})^3 \, {\rm GB}\,,\nonumber\\
&\approx 112 \, \qty(\frac{N_{\rm p}}{1000})^3 \, {\rm GB}\,,\quad {\rm for} \  N_{\rm g}=2N_{\rm p}\,,
\end{align}
where we have used $1~\mathrm{GB} = 1024^3~\mathrm{bytes}$. This is slightly more than twice the memory requirement of the default \textsc{glam} code, which is $52\left(N_{\rm p}/1000\right)^3$ GB \citep{Klypin:2017iwu}.

\subsubsection{Implementation of coupled quintessence}
\label{subsubsect:csf_imp}

Defining the code unit of the dimensionless scalar field perturbation, $\delta\varphi=\varphi-\bar{\varphi}$, as\footnote{Note that, for brevity, we have slightly abused the notations, by using the same symbol $\varphi$ with a tilde for the code-unit expression of $\delta\varphi$. Given that the code-unit quantity always comes with a tilde, this should not cause any confusion with, e.g., the background scalar field $\bar{\varphi}$, or the total dimensionless scalar field $\varphi$ in physical units.}
\begin{equation}
    \tilde{\varphi} \equiv \frac{c^2N_{\rm g}}{H_0^2L^2}\delta\varphi = \tilde{c}^2\delta\varphi,
\end{equation}
with $\delta\varphi$ being the perturbation to $\varphi$, we can rewrite its equation of motion as
\begin{equation}\label{eq:csf_eom_code}
    \tilde{\boldsymbol{\nabla}}^2\tilde{\varphi} = 3\beta{\Omega_{\rm m}}{a^{-1}}e^{\beta\bar{\varphi}}\left[\exp\left(\beta\frac{\tilde{\varphi}}{\tilde{c}^2}\right)(1+\tilde{\delta})-1\right] -\alpha\lambda^2a^2\left[\frac{1}{\left(\bar{\varphi}+\tilde{c}^{-2}\tilde{\varphi}\right)^{1+\alpha}}-\frac{1}{\bar{\varphi}^{1+\alpha}}\right],
\end{equation}
where $\bar{\varphi}$ is the background value of $\varphi$, and $\lambda$ is defined in Eq.~\eqref{eq:csf_param_lambda}. The Poisson equation becomes
\begin{equation}\label{eq:csf_poisson_code}
    \Tilde{\boldsymbol{\nabla}}^2\tilde{\Phi}_{\rm N} = \frac{3}{2}\Omega_{\rm m}a^{-1}e^{\beta\bar{\varphi}}\left[\exp\left(\beta\frac{\tilde{\varphi}}{\tilde{c}^2}\right)(1+\tilde{\delta})-1\right] + \lambda^2a^2\left[\frac{1}{\left(\bar{\varphi}+\tilde{c}^{-2}\tilde{\varphi}\right)^{\alpha}}-\frac{1}{\bar{\varphi}^{\alpha}}\right].
\end{equation}
In practice, as we know that the scalar field density perturbation is small in the models of interest, the second term on the right-hand side of the Poisson equation can be dropped approximately. We have also chosen to neglect the term $\exp\left(\beta\tilde{c}^{-2}\tilde{\varphi}\right)$ in front of $(1+\tilde{\delta})$, to simplify the simulation --- this is again justified because $|\delta\varphi|\ll|\bar{\varphi}|\simeq\mathcal{O}(1)$ at late times, although including this in the simulation is trivial.

The modified particle coordinate and velocity updates can be rewritten as
\begin{eqnarray}
    \label{eq:csf_particle_geodesic_code1}\frac{{\rm d}\tilde{\boldsymbol{x}}}{{\rm d}a} &=& \frac{H_0}{a^2\dot{a}}\tilde{\boldsymbol{p}},\\
    \label{eq:csf_particle_geodesic_code2}\frac{{\rm d}\tilde{\boldsymbol{p}}}{{\rm d}a} &=& -\frac{H_0}{\dot{a}}\left[\tilde{\boldsymbol{\nabla}}\tilde{\Phi}_{\rm N}+\beta\tilde{\boldsymbol{\nabla}}\tilde{\varphi}\right] - \beta\frac{{\rm d}\bar{\varphi}}{{\rm d}a}\tilde{\boldsymbol{p}}.
\end{eqnarray}
Here we can observe more explicitly the effect of a modified background expansion history in coupled quintessence models, encoded in the $H_0\dot{a}^{-1}$ terms.

In \textsc{mg}-\textsc{glam}, Eq.~\eqref{eq:csf_eom_code} is solved using the Newton-Gauss-Seidel method described in \S\ref{subsubsect:relaxation}. Eq.~\eqref{eq:csf_poisson_code} is not directly solved, but instead we solve the (standard) Poisson equation not having $e^{\beta\bar{\varphi}}$: since this is a background quantity, we instead multiply it when calculating the Newtonian force from $\tilde{\Phi}_{\rm N}$. Eqs.~(\ref{eq:csf_particle_geodesic_code1}, \ref{eq:csf_particle_geodesic_code2}) are then solved --- the fifth force $\beta\tilde{\boldsymbol{\nabla}}\tilde{\varphi}$ is incorporated by first summing up the two potentials, $\tilde{\Phi}_{\rm N}+\beta\tilde{\varphi}$, and then doing the finite difference.

\subsubsection*{\textsc{mg-glam} background cosmology solver}

As Eqs.~(\ref{eq:csf_particle_geodesic_code1}, \ref{eq:csf_particle_geodesic_code2}) contain background quantities $\dot{a}$ and ${\rm d}\tilde{\varphi}/{\rm d}a$, for every given coupled quintessence model we need to solve its background evolution. This is governed by the following system of equations --- the equation of motion for the background scalar field $\bar{\varphi}$:  
\begin{equation}\label{eq:csf_background_eom}
    \ddot{\bar{\varphi}} + 3\frac{\dot{a}}{a}\dot{\bar{\varphi}} + \frac{{\rm d}V(\bar{\varphi})}{{\rm d}\varphi} + \frac{{\rm d}A\left(\bar{\varphi}\right)}{{\rm d}\varphi}8\pi G\bar{\rho}_{\rm m} = 0,
\end{equation}
the Friedmann equation (with a flat Universe, $k=0$, being assumed)
\begin{equation}\label{eq:csf_background_Friedmann}
    H^2 = \left(\frac{\dot{a}}{a}\right)^2 = \frac{8\pi{G}}{3}\left[\bar{\rho}_{\rm r}(a) + A(\bar{\varphi})\bar{\rho}_{\rm m}(a)\right] + \frac{1}{6}\dot{\bar{\varphi}}^2+H_0^2\frac{\lambda^2}{\bar{\varphi}^\alpha},
\end{equation}
and the Raychaudhuri equation 
\begin{equation}\label{eq:csf_background_Raychaudhuri}
    3\left(\dot{H}+H^2\right) = -4\pi{G}\left[2\bar{\rho}_{\rm r}(a) + A(\bar{\varphi})\bar{\rho}_{\rm m}(a)\right] - \dot{\bar{\varphi}}^2+H_0^2\frac{\lambda^2}{\bar{\varphi}^\alpha},
\end{equation}
where $\bar{\rho}_{\rm r}$ denotes the background density of radiations (we assume that all three species of neutrinos are massless and thus counted as radiation). Note that in Eqs.~(\ref{eq:csf_background_Friedmann}, \ref{eq:csf_background_Raychaudhuri}), to see the dimensions of the different terms clearly, we have already explicitly substituted the inverse-powerlaw potential and used the definition of $\lambda$ in Eq.~\eqref{eq:csf_param_lambda}. In \textsc{mg-glam} the scalar field equation is solved by a fifth-sixth order continuous Runge-Kutta method\footnote{For this numerical integrator we have used \texttt{subroutine dverk} from the \href{https://camb.info/}{\textsc{camb} code}, originally developed in Fortran 66 by K.~R.~Jackson.}. 

For numerical solutions in background cosmology, instead of directly working with Eqs.~(\ref{eq:csf_background_eom}, \ref{eq:csf_background_Friedmann}, \ref{eq:csf_background_Raychaudhuri}), it is convenient to use $N\equiv\ln(a)$ as the time variable, for which we have
\begin{equation}
    \bar{\varphi}' = \mathcal{H}\frac{{\rm d}\bar{\varphi}}{{\rm d}N}, \quad \bar{\varphi}'' = \mathcal{H}^2\frac{{\rm d}^2\bar{\varphi}}{{\rm d}N^2} + \mathcal{H}'\frac{{\rm d}\bar{\varphi}}{{\rm d}N},
\end{equation}
where, as mentioned in the Introduction, $'$ is the derivative with respect to the conformal time $\tau$ and $\mathcal{H}\equiv a'/a$. In this convention, the background quintessence field equation of motion, Eq.~\eqref{eq:csf_background_eom}, can be written as
\begin{equation}\label{eq:csf_background_eom_codeunit}
    \frac{\mathcal{H}^2}{H_0^2}\frac{{\rm d}^2\bar{\varphi}}{{\rm d}N^2} + \left[2\frac{\mathcal{H}^2}{H^2_0}+\frac{\mathcal{H}'}{H_0^2}\right]\frac{{\rm d}\bar{\varphi}}{{\rm d}N} - 3\alpha\lambda{\rm e}^{2N}{\bar{\varphi}^{-(1+\alpha)}} + 3\beta{\rm e}^{-N}\Omega_{\rm m}\exp(\beta\bar{\varphi}) = 0,
\end{equation}
where the quantities $\mathcal{H}^2/H_0^2$ and $\mathcal{H}'/H_0^2$ can be obtained from Eqs.~(\ref{eq:csf_background_Friedmann}, \ref{eq:csf_background_Raychaudhuri}) as
\begin{eqnarray}
    \frac{\mathcal{H}^2}{H^2_0} &=& \left[1-\frac{1}{6}\left(\frac{{\rm d}\bar{\varphi}}{{\rm d}N}\right)^2\right]^{-1}\left[\Omega_{\rm r}{\rm e}^{-2N}+\exp(\beta\bar{\varphi})\Omega_{\rm m}{\rm e}^{-N}+\lambda{\rm e}^{2N}\bar{\varphi}^{-\alpha}\right],\\
    \frac{\mathcal{H}'}{H^2_0} &=& -\frac{1}{3}\left(\frac{{\rm d}\bar{\varphi}}{{\rm d}N}\right)^2\frac{\mathcal{H}^2}{H_0^2} + \lambda{\rm e}^{2N}\bar{\varphi}^{-\alpha} - \Omega_{\rm r}{\rm e}^{-2N} - \frac{1}{2}\Omega_{\rm m}{\rm e}^{-N}\exp(\beta\bar{\varphi}).
\end{eqnarray}
Here $\Omega_{\rm r}$ denotes the present-day radiation density parameter, with `radiation' including CMB photons with a present-day temperature of $2.7255$ K and $3.046$ flavours of massless neutrinos; we defer the implementation of massive neutrinos, both as a non-interacting particle species and in the context of coupling to scalar fields, to future works.

We note that $\lambda$ is \textit{not} a free parameter of the model. Rather, once the density parameters $\Omega_{\rm m}$, $\Omega_{\rm r}$ and $H_0$ are specified (or equally once the present-day densities of matter and radiation are specified), $\lambda$, which quantifies the size of the potential energy of the scalar field, must take some certain value in order to ensure consistency. If $\lambda$ is too large, the predicted $H(a=1)$ will be larger than the desired (input) value of $H_0$, and vice versa. In practice, \textsc{mg}-\textsc{glam} starts from a trial value of $\lambda=1$, evolves Eqs.~(\ref{eq:csf_background_eom}, \ref{eq:csf_background_Friedmann}, \ref{eq:csf_background_Raychaudhuri}) from some initial redshift ($z_{\rm i}=10^5$) to $z=0$, and checks if the calculated value of $H(a=1)$ is equal to the desired value $H_0$ (within a small relative error of order $\mathcal{O}\left(10^{-6}\right)$); if the predicted $H(a=1)$ is larger than the desired $H_0$, $\lambda$ is decreased, and vice versa. This process is repeated iteratively to obtain a good approximation to $\lambda$ with a relative error smaller than $10^{-6}$. The initial conditions of $\bar{\varphi}$ and $\dot{\bar{\varphi}}$ at $z_{\rm i}=10^5$ are not important, as long as their values are small enough. Once the value of $\lambda$ has been determined in this way, it is stored to be used in other parts of the code; also stored are a large array of the various background quantities such as $H, \dot{H}, \bar{\varphi}$ and $\dot{\bar{\varphi}}$ --- if needed at any time by the scalar field solver of \textsc{mg}-\textsc{glam}, these quantities can be linearly interpolated in the scale factor $a$.  

\subsubsection{Implementation of symmetrons}
\label{subsubsect:sym_imp}

The scalar field equation of motion in the symmetron model, Eq.~\eqref{eq:sym_eom}, can be written in code unit as
\begin{equation}\label{eq:sym_eom_code_unit}
    \tilde{c}^2\tilde{\nabla}^2u = \frac{a^2}{2\xi^2}\left[\tilde{\rho}\frac{a_\ast^3}{a^3}-1\right]u + \frac{a^2}{2\xi^2}u^3.
\end{equation}
While this equation can be solved similarly to the case of coupled quintessence by using the standard Newton-Gauss-Seidel relaxation method we described in \S\ref{subsubsect:csf_imp}, the `Newton' approximation of this method, Eq.~\eqref{eq:relaxation_iteration}, is indeed unnecessary, as can be seen from the following derivation. Defining
\begin{equation}\label{eq:Lijk}
    L_{i,j,k}(u) \equiv u_{i+1,j,k}+u_{i-1,j,k}+u_{i,j+1,k}+u_{i,j-1,k}+u_{i,j,k+1}+u_{i,j,k-1},
\end{equation}
where a subscript $_{i,j,k}$ denotes the value of a quantity in a cell that is the $i$th ($j$th, $k$th) in the $x$ ($y$, $z$) direction, the discretised version of Eq.~\eqref{eq:sym_eom_code_unit}, after some rearrangement, can be written as
\begin{equation}
    u^3_{i,j,k}+\left[\tilde{\rho}_{i,j,k}\frac{a^3_\ast}{a^3}-1\right]u_{i,j,k} + \frac{12}{h^2}\frac{\tilde{c}^2\xi^2}{a^2}u_{i,j,k} - \frac{2}{h^2}\frac{\tilde{c}^2\xi^2}{a^2}L_{i,j,k} = 0.
\end{equation}
We can define
\begin{eqnarray}
    p &\equiv& \tilde{\rho}_{i,j,k}\frac{a^3_\ast}{a^3}-1 + \frac{12}{h^2}\frac{\tilde{c}^2\xi^2}{a^2},\\
    q &\equiv& - \frac{2}{h^2}\frac{\tilde{c}^2\xi^2}{a^2}L_{i,j,k},
\end{eqnarray}
so that the above equation can be simplified as
\begin{equation}\label{eq:sym_cubic_eqn}
    u^3_{i,j,k} + pu_{i,j,k} + q = 0.
\end{equation}
This is similar to the discrete equation of motion in the Hu-Sawicki $f(R)$ gravity model with $n=1$, as discussed in Ref.~\cite{Bose:2016wms}, which can be treated as a cubic equation of $u_{i,j,k}$ that can be solved exactly (analytically). Therefore, given the (approximate) values of the field $u$ in the six direct neighbouring cells of $(i,j,k)$, we can calculate $u_{i,j,k}$ analytically, and there is no need to solve it using the Newton approximation as in Eq.~\eqref{eq:relaxation_iteration}. The relaxation iterations are still needed, since the values of $u$ in the six direct neighbours are \textit{approximate} and therefore need to be updated iteratively, but the replacement of the Newton solver with an exact solution of $u_{i,j,k}$ (therefore the name \textit{nonlinear Gauss-Seidel} as opposed to \textit{Newton Gauss-Seidel}) has been found to significantly improve the convergence speed of the relaxation \cite{Bose:2016wms}. This method for the symmetron model was briefly mentioned in an Appendix of Ref.~\cite{Bose:2016wms} but no numerical implementation was shown there.

The solution to Eq.~\eqref{eq:sym_cubic_eqn} can be found as
\begin{align}\label{eq:sym_cubic_solns}
    u_{i,j,k} = \begin{cases}
        \displaystyle -\frac{1}{3} \left( C + \frac{\Delta_0}{C} \right) \ , & \Delta > 0  \ , \\
        \displaystyle \sqrt[3]{-q} \ , & \Delta = 0 \ , \\
        \displaystyle -\frac{2}{3} \sqrt{\Delta_0} \cos \left( \frac{\Theta}{3} + \frac{2\pi}{3} \right) \ , & \Delta < 0 \ ,
    \end{cases} 
\end{align}
where we have defined $\Delta_0 \equiv -3p$, $\Delta_1 \equiv 27 q$, $\Delta \equiv \Delta_1^2 - 4 \Delta_0^3$ and 
\begin{eqnarray}
    C &\equiv& \sqrt[3]{\frac{1}{2} \left[ \Delta_1 + \sqrt{\Delta_1^2 - 4 \Delta_0^3} \right]},\\ 
    \Theta &\equiv& \arccos \left( \frac{\Delta_1}{2\sqrt{\Delta_0^3}}\right).
\end{eqnarray}
It can be shown that all the 3 branches of solutions in Eq.~\eqref{eq:sym_cubic_solns} can be the physical solution in certain regimes, depending on model parameters, density values, mesh size, and so on. In our implementation in \textsc{mg}-\textsc{glam}, we have used Eq.~\eqref{eq:sym_cubic_solns} instead of Eq.~\eqref{eq:relaxation_iteration} for the symmetron model. 

The acceleration on particles, Eq.~\eqref{eq:csf_particle_geodesic_qsa}, can be written as following in the symmetron model:
\begin{eqnarray}
    \frac{{\rm d}\tilde{\boldsymbol{x}}}{{\rm d}a} &=& \frac{H_0}{a^2\dot{a}}{\tilde{\boldsymbol{p}}},\\
    \frac{{\rm d}\tilde{\boldsymbol{p}}}{{\rm d}a} &=& \tilde{\boldsymbol{F}}_{\rm N}+\tilde{\boldsymbol{F}}_5+\tilde{\boldsymbol{F}}_\times,
\end{eqnarray}
where $\tilde{\boldsymbol{F}}_{\rm N}$, $\tilde{\boldsymbol{F}}_5$ and $\tilde{\boldsymbol{F}}_\times$ denote, respectively, the standard Newtonian acceleration, the fifth force acceleration and the frictional force acceleration, in code units, given by
\begin{eqnarray}
    \label{eq:F_N_codeunit}\tilde{\boldsymbol{F}}_{\rm N} &=& -\frac{H_0}{\dot{a}}\tilde{\boldsymbol{\nabla}}\tilde{\Phi}_{\rm N},\\
    \label{eq:F_5_codeunit}\tilde{\boldsymbol{F}}_5 &=& -6\frac{H_0}{\dot{a}}\xi^2\Omega_{\rm m}\beta_\ast^2\tilde{c}^2a_{\ast}^{-3}u\tilde{\boldsymbol{\nabla}}u = -3\frac{H_0}{\dot{a}}\xi^2\Omega_{\rm m}\beta_\ast^2\tilde{c}^2a_{\ast}^{-3}\tilde{\boldsymbol{\nabla}}\left(u^2\right),\\
    \label{eq:F_x_codeunit}\tilde{\boldsymbol{F}}_\times &=& -9\Omega_{\rm m}\beta_\ast^2\xi^2\sqrt{1-\left(\frac{a_\ast}{a}\right)^3}\frac{1}{a^4}u\tilde{\boldsymbol{p}}.
\end{eqnarray}
In practice, as mentioned earlier, the frictional force is much weaker than the other two force components because of the very slow time evolution of the symmetron field. Likewise, any time variation of the matter particle mass due to the coupling with the symmetron field must be tiny and negligible. Therefore, for the Poisson equation, which governs $\Phi_{\rm N}$ and thus $\boldsymbol{F}_{\rm N}$, we simply approximate it to be the same as in $\Lambda$CDM.

\subsubsection{Implementation of $f(R)$ gravity}
\label{subsubsect:fR_imp}

In \S\ref{subsect:fR} we have introduced a class of $f(R)$ models with an (inverse) power-law function $f_R$, Eq.~\eqref{eq:fR_HS}, and mentioned that we will focus on the cases of $n=0$, $1$, $2$. In this subsection, we shall first derive equations that apply to general values of $n$, and then specialise to these three cases, for which we will develop case-specific algorithms of nonlinear Gauss-Seidel relaxation. 

In code unit, the $f_R$ equation of motion of this model, Eq.~\eqref{eq:fR_eom_qsa}, can be written as
\begin{equation}
\label{eq:fR_eom_codeunit1}\tilde{c}^2\tilde{\boldsymbol{\nabla}}^2\tilde{f}_R = -\Omega_{\rm m}{a}^{-1}\left(1+\tilde{\delta}\right) + \frac{1}{3} \tilde{\bar{R}}(a) a^2 \left(\frac{\tilde{\bar{f}}_{R}}{\tilde{f}_R}\right)^{\frac{1}{n+1}} - 4\Omega_\Lambda a^2,
\end{equation}
where $\tilde{f}_R \equiv f_R$ and $\tilde{\bar{f}}_R\equiv\bar{f}_R(a)$ is the background value of $f_R$. 
The Newtonian force is still given by Eq.~\eqref{eq:F_N_codeunit} with $\tilde{\Phi}_{\rm N}$ governed by Eq.~\eqref{eq:GR_poisson_codeunit}. On the other hand, the fifth force in code unit can be written as
\begin{equation}\label{eq:F5_fR_codeunit}
\tilde{\boldsymbol{F}}_5 = \frac{1}{2}\tilde{c}^2\tilde{\boldsymbol{\nabla}}\tilde{f}_R,
\end{equation}
It is more convenient to define the following new, positive-definite, scalar field variable \cite{Bose:2016wms}
\begin{equation}\label{eq:fR_u}
    u \equiv (-{f}_R)^{1/(n+1)},
\end{equation}
where the minus sign is because $f_R<0$. Eq.~\eqref{eq:fR_eom_codeunit1} then becomes
\begin{equation}\label{eq:fR_eom_codeunit2}
    -\tilde{c}^2\tilde{\boldsymbol{\nabla}}^2\left(u^{n+1}\right) + \frac{\Omega_{\rm m}}{a}\delta + \frac{1}{3}\tilde{\bar{R}}(a)a^2  - \frac{1}{3} \tilde{\bar{R}}(a)a^2\left[-\bar{f}_R (a)\right]^{1/(n+1)}\frac{1}{u} = 0,
\end{equation}
where we have defined the following dimensionless background quantity:
\begin{equation}
    \tilde{\bar{R}}(a) \equiv \frac{\bar{R}(a)}{H_0^2} = 3 \qty(\Omega_{m} a^{-3} + 4\Omega_\Lambda),
\end{equation}
with $\bar{R}(a)$ being the background value of the Ricci scalar at scale factor $a$. Eq.~\eqref{eq:fR_eom_codeunit2} can be further simplified to
\begin{equation}\label{eq:fR_eom_codeunit}
    u^{n+2}_{i,j,k} + p u_{i,j,k} + q = 0,
\end{equation}
where 
\begin{eqnarray}
    p &\equiv& \frac{h^2}{6 \tilde{c}^2}\left[\frac{\Omega_{\rm m}}{a}\delta_{i,j,k} + \frac{1}{3} \tilde{\bar{R}}(a)a^2\right] - \frac{1}{6}L_{i,j,k},\\
    q &\equiv& -\frac{h^2}{6\tilde{c}^2}\frac{1}{3} \tilde{\bar{R}}(a)a^2\left[-\bar{f}_R(a)\right]^{1/(n+1)}
\end{eqnarray}
where $L_{i,j,k}$ was defined in Eq.~\eqref{eq:Lijk}, and we have neglected the tilde in $\tilde{\bar{f}}_R(a)$ because $\tilde{\bar{f}}_R=\bar{f}_R$ anyway.

Eq.~\eqref{eq:fR_eom_codeunit} is a polynomial for $u_{i,j,k}$, which can be analytically solved for the cases of $n=0$, $1$ and $2$. The case of $n=1$ has been discussed in Ref.~\cite{Bose:2016wms}, while cases of $n=0,2$ have not been studied before using nonlinear Gauss-Seidel schemes\footnote{The case of $n=2$ has been studied using simulations based on Newton-Gauss-Seidel relaxation \cite[e.g.,][]{Li:2011uw}.}. Here we discuss all three cases with equal details.

\begin{itemize}
    \item {\underline{\textbf{The case of} $n=2$}} 

    In this case, Eq.~\eqref{eq:fR_eom_codeunit} is a quartic equation of $u_{i,j,k}$. Define 
    \begin{eqnarray}
    \Delta_0 &\equiv& 12q,\nonumber\\ 
    \Delta_1 &\equiv& 27p^2.
    \end{eqnarray} 
    We see that $q<0$ and so $\Delta_0<0$ and $\Delta_1>0$. Eq.~\eqref{eq:fR_eom_codeunit} has 4 branches of analytical solutions:
    \begin{eqnarray}
    \label{eq:branch1}u_{i,j,k} &=& -S \pm \frac{1}{2}\sqrt{-4S^2+\frac{p}{S}},\\
    \label{eq:branch2}u_{i,j,k} &=&  S \pm \frac{1}{2}\sqrt{-4S^2-\frac{p}{S}},
    \end{eqnarray}
    where we have defined
    \begin{eqnarray}
    S &\equiv& \frac{1}{2}\sqrt{\frac{1}{3}\left(Q+\frac{\Delta_0}{Q}\right)},\nonumber\\
    Q &\equiv& \sqrt[3]{\frac{1}{2}\left[\Delta_1+\sqrt{\Delta_1^2-4\Delta_0^3}\right]}.
    \end{eqnarray}
    We need to find the correct branch of solution. First, note that $S$ is a square root, and so we can show that if the quantity under the square root is a positive number, then $S>0$. This is straightforward, as.
    \begin{equation}
    12S^2 = Q + \frac{\Delta_0}{Q} = \sqrt[3]{\frac{1}{2}\left[\sqrt{\Delta_1^2-4\Delta_0^3}+\Delta_1\right]} - \sqrt[3]{\frac{1}{2}\left[\sqrt{\Delta_1^2-4\Delta_0^3}-\Delta_1\right]} > 0,
    \end{equation}
    
    Consider first the limit $p\rightarrow0$. From the above equation we have
    \begin{equation}
    12S^2 \approx \sqrt[3]{\left(-\Delta_0\right)^{3/2}+\frac{1}{2}\Delta_1} - \sqrt[3]{\left(-\Delta_0\right)^{3/2}-\frac{1}{2}\Delta_1} \approx -\frac{1}{3}\frac{\Delta_1}{\Delta_0} = -\frac{3}{4}\frac{p^2}{q},
    \end{equation}
    which means that $S\simeq |p|\rightarrow0$ but $p/S \rightarrow \pm4\sqrt{-q}$ depending on the sign of $p$. This leads to the solution $u_{i,j,k}=\sqrt[4]{-q}$.
    
    Given that $S>0$, if $p>0$, Eq.~\eqref{eq:branch2} cannot be the physical branch because $u_{i,j,k}$ in this branch is complex. The `$-$' branch of Eq.~\eqref{eq:branch1} cannot be chosen either, because $u_{i,j,k}<0$, inconsistent with the requirement that $u_{i,j,k}>0$.
    
    If $p<0$, Eq.~\eqref{eq:branch1} cannot be the physical branch because $u_{i,j,k}$ in this branch is complex. Out of the two branches of Eq.~\eqref{eq:branch2}, we should choose `$+$', because this guarantees that when $p\rightarrow0^-$ we still have $u_{i,j,k}>0$.
    
    Therefore, the analytical solution can be summarised as
    \begin{align}\label{eq:fR_quartic_solns}
        u_{i,j,k} = \begin{cases}
            \displaystyle -S+\frac{1}{2}\sqrt{-4S^2+\frac{p}{S}}, & p>0, \\
            \displaystyle \sqrt[4]{-q}, & p=0, \\
            \displaystyle S+\frac{1}{2}\sqrt{-4S^2-\frac{p}{S}}, & p<0.
        \end{cases} 
    \end{align}
    Note that it can be shown that $8S^3<|p|$ because $\Delta_1=27p^2$ and $\Delta_0=12q<0$. This fact guarantees that in Eqs.~\eqref{eq:fR_quartic_solns} the square roots are real; it also guarantees that in the $p>0$ branch the condition $u_{i,j,k}>0$ is satisfied (in the $p<0$ branch it is satisfied automatically). 
    
    The existence of analytical solutions Eq.~\eqref{eq:fR_quartic_solns} indicates that, like in the symmetron model, in the $n=2$ case of $f(R)$ gravity here, it is not necessary to use the Newton approximation within the Gauss-Seidel relaxation, but the solution $u_{i,j,k}$ of cell $(i,j,k)$ can be solved given the density field in this cell and the approximate solutions of $u$ in the neighbouring cells. 
    
    \item {\underline{\textbf{The case of} $n=1$}} 
    
    In this case, Eq.~\eqref{eq:fR_eom_codeunit} is a cubic equation of $u_{i,j,k}$ \cite{Bose:2016wms}. Define $\Delta_0 \equiv -3p$, $\Delta_1 \equiv 27 q$ and the discriminant 
    \begin{equation}
        \Delta \equiv \Delta_1^2-4\Delta_0^3.
    \end{equation}
    We see that $q<0$ and so $\Delta_1<0$. The solution is given by
    \begin{align}\label{eq:fR_cubic_solns}
        u_{i,j,k} = \begin{cases}
            \displaystyle -\frac{1}{3}\left(C+\frac{\Delta_0}{C}\right), & \Delta>0, \\
            \displaystyle \sqrt[3]{-q}, & \Delta=0, \\
            \displaystyle -\frac{2}{3}\sqrt{\Delta_0}\cos\left(\frac{\Theta}{3} + \frac{2\pi}{3}\right), & \Delta<0,
        \end{cases} 
    \end{align}
    where 
    \begin{eqnarray}
        C &\equiv& \sqrt[3]{\frac{1}{2}\left[\Delta_1+\sqrt{\Delta_1^2-4 \Delta_0^3}\right]},\\
        \Theta &\equiv& \arccos \left(\frac{\Delta_1}{2\sqrt{\Delta_0^3}}\right).
    \end{eqnarray}
    Again, the exact analytical solutions given in Eq.~\eqref{eq:fR_cubic_solns} eliminates the need for Newton-Gauss-Seidel relaxations, and this has led to a significant improvement in the speed and convergence properties of simulations of this model compared with previous simulations \cite{Bose:2016wms}.
    
    \item {\underline{\textbf{The case of} $n=0$}} 
    
    In this case, Eq.~\eqref{eq:fR_eom_codeunit} is a quadratic equation of $u_{i,j,k}$. The solution in this case is simple and the physical branch is given by
    \begin{equation}\label{eq:fR_quadratic_solns}
        u_{i,j,k} = \frac{1}{2}\left[-p+\sqrt{p^2-4q}\right],
    \end{equation}
    which satisfies $u_{i,j,k}>0$.
    
\end{itemize}

\section{Code tests}
\label{sect:code_tests}

In this section, we present various code test results to demonstrate the reliability of the equations, algorithms and implementations described in the previous sections. We follow the code test framework of the \textsc{ecosmog} code papers \citep{Li:2011_ECOSMOG_code_paper,Li:2013_ECOSMOGV_code_paper}. Apart from the background cosmology test, all the tests shown in this section were performed on a cubic box with size $256 \, h^{-1} \mathrm{Mpc}$ and $512$ grid cells in each direction, and all background quantities are calculated at $a = 1$. 

\subsection{Background cosmology tests}
\label{subsect:bg_tests}

Of the models considered in this work, only the coupled quintessence model can substantially affect the background expansion history, while for (viable) $f(R)$ gravity and symmetron models the expansion rate is practically indistinguishable from that of $\Lambda$CDM. In \textsc{mg}-\textsc{glam}, the background cosmology in the coupled quintessence model is solved numerically, as described in Sect.~\ref{subsubsect:csf_imp}. 

\begin{figure}
    \centering
    \includegraphics[width=\textwidth]{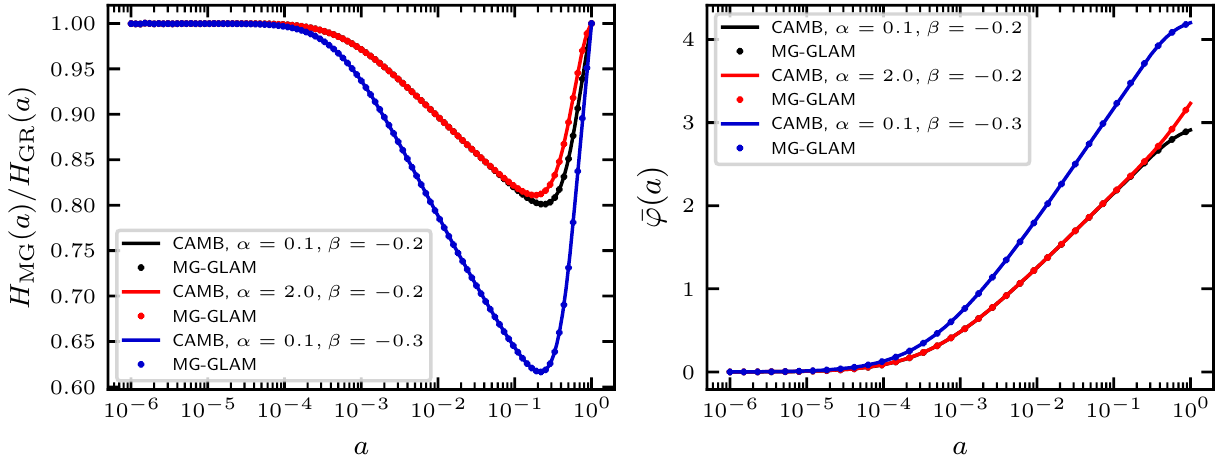}
\caption{
    Cosmological background evolution tests.
    \textit{Left panel:} The ratio of the Hubble parameters between the coupled quintessence and the \ac{GR} models, from the modified \href{https://camb.info/}{\textsc{camb}} (lines) and \textsc{mg-glam} (dots) codes for three kinds of model parameter values as labeled.  
    \textit{Right panel:} The evolution of the background scalar field in the coupled quintessence from \href{https://camb.info/}{\textsc{camb}} (lines) and \textsc{mg-glam} codes.
}
    \label{fig:code_test_background_csf}
\end{figure}

To check the numerical implementation, we have compared the predictions of certain background quantities by \textsc{mg}-\textsc{glam} with the results produced by the modified \href{https://camb.info/}{\textsc{camb}} code, for the same coupled quintessence model, described in \cite{Li:2010re}. The results are presented in Fig~\ref{fig:code_test_background_csf}, where the left panel shows the ratio between the background expansion rates of three coupled quintessence models and that of a $\Lambda$CDM model with the same (non-MG) cosmological parameters, while the right panel shows the background evolution of the scalar field, $\bar{\varphi}(a)$, for the same three models. Lines are from the modified \href{https://camb.info/}{\textsc{camb}} code and symbols are for \textsc{mg-glam}. We see that the background cosmology solver of \textsc{mg-glam} agrees with the \href{https://camb.info/}{\textsc{camb}} code very well in all cases.

There are two additional interesting features displayed in Fig.~\ref{fig:code_test_background_csf}. First, the results are much more sensitive to $\beta$ than to $\alpha$, as can be observed by comparing the closeness between the black vs red lines, and the large difference between the black vs blue lines. This shows that the coupling to matter has a stronger impact on the scalar field background evolution than the potetial itself.

Second, as discussed in Sect.~\ref{subsect:csf}, the scalar field affects structure formation through a combination of the following four effects:
\begin{itemize}
    \item \textit{modified expansion rate}: in the models studied here, the expansion rate is slowed down, which can lead to enhancement of structure formation.
    \item \textit{fifth force}: the fifth-force-to-Newtonian-gravity ratio is a constant $2\beta^2$, and this boosts structure formation.
    \item \textit{velocity-dependent force}: from the right panel of Fig.~\ref{fig:code_test_background_csf}, we see that the scalar field is positive and grows over time such that, with $\beta<0$, the term $\left({\rm d}\ln A(\bar{\varphi})/{\rm d}\varphi\right)\bar{\varphi}'<0$, which means that the velocity-dependent force is in the same direction as the particle velocity, i.e., it is essentially an `anti-friction' force which tends to strengthen structure formation.
    \item \textit{time variation of effective particle mass}: since the particle mass effectively depends on $\exp(\beta\bar{\varphi})$, with $\beta<0$ and $\bar{\varphi}>0$, at late times the effective mass decreases, which tends to weaken structure formation.
\end{itemize}
Therefore, the 4 effects work in different directions, and the net effect on structure formation---whether it is boosted or weakened---will need to be calculated numerically for specific models.

\subsection{Density tests}

This subsection is devoted to the tests of the multigrid solvers for the $f(R)$, symmetron and coupled quintessence models, using different density configurations for which the scalar field solution can be solved analytically or using a different numerical code.


\subsubsection{Homogeneous matter density field}
\label{subsubsec:homogeneous_rhom_test}

In a homogeneous density field the \ac{MG} scalar field should also be homogeneous and exactly equal to its background value if the matter field is homogeneous, i.e.,
\begin{align}\label{eq:field_uniform_dens_test}
    \tilde{\delta}(\tilde{x}) \equiv 0 \longrightarrow \begin{cases}
        f_R(\tilde{x}) / \bar{f}_R \equiv 1, & \text{$f(R)$ gravity;} \\
        \varphi(\tilde{x}) / \varphi_\ast = \bar{\varphi}/\varphi_\ast \leq 1, & \text{symmetron;} \\
        \tilde{c}^2\delta\varphi(\tilde{x}) \equiv 0, & \text{coupled quintessence.}
    \end{cases}
\end{align}
This offers a very simple test for the relaxation solvers described above, that is particularly useful for checking the implementation of multigrid.

\begin{figure}
    \centering
    \includegraphics[width=0.7\textwidth]{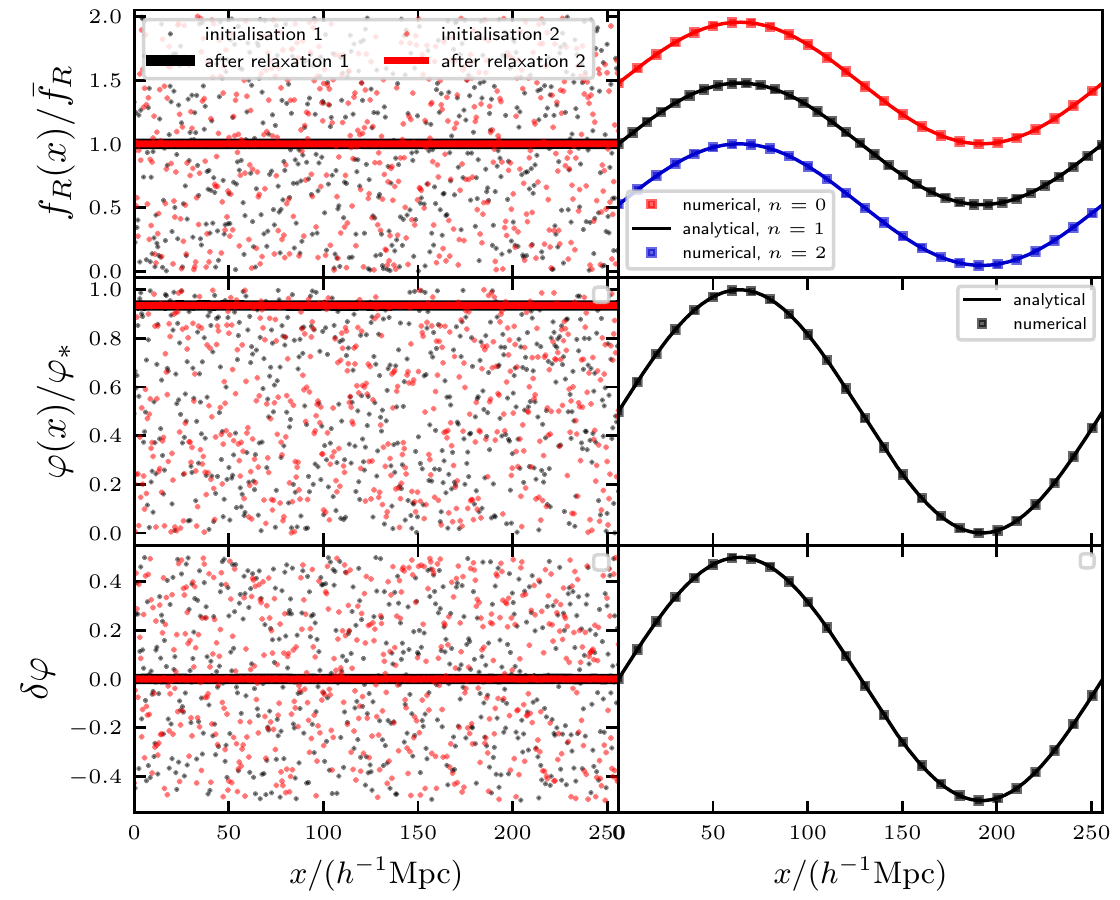}
    \caption{The uniform and one-dimensional code test results. The two columns show the cases with homogeneous (left) and sine (right) scalar fields, whilst the different rows represent the $f(R)$ gravity (upper), symmetron (middle) and coupled quintessence (bottom) models. \textit{Left Panels:} Uniform density test, where the symbols represent the random initial guesses of the \ac{MG} scalar field in the ranges of $[0, 2]$ ($f(R)$ gravity), $[0, 1]$ (symmetron) and $[-0.5, 0.5]$ (coupled quintessence), respectively. The solid lines show the numerical solutions after multigrid relaxation. Two random initialisations have been displayed in red and black. \textit{Right Panels:} Sine field tests. The squares show the numerical results and the lines show the analytical solutions. The upper right panel shows the $f(R)$ gravity test results with $n = 0, 1$ and $2$ as labeled.
    }
    \label{fig:code_test_1D_3x2}
\end{figure}

We show the test results for a homogeneous density field in the left-hand panels of Fig.~\ref{fig:code_test_1D_3x2}, where we display the scalar field values along the $\tilde{x}$ direction for fixed $\tilde{y}, \tilde{z}$ coordinates before (symbols) and after (lines) the multigrid relaxation, for two initial guesses (black and red). The three rows, from top to bottom, are respectively for the $f(R)$, symmetron and coupld quintessence models. For $f(R)$ gravity, the initial guesses are randomly generated from a uniform distribution within $\xi=f_R(\tilde{x}) / \bar{f}_R\in[0,2]$, and the model parameters used are $n=1$ and $f_{R0}=-10^{-5}$; for the symmetron model, the random initial guesses are generated from a uniform distribution $\varphi(\tilde{x})/\varphi_\ast\in[0,1]$ and the model parameters adopted are $a_\ast = 0.5, \xi = 10^{-3}$, $\beta_\ast = 0.1$; for coupled quintessence we consider the model parameters $\alpha = 0.1, \beta = -0.2$, and the initial guesses are from a uniformation distribution $\delta\varphi(\tilde{x})\in[-0.5,0.5]$.

In all cases, we find that the solutions after relaxation agree very well to the analytical predictions given in Eq.~\eqref{eq:field_uniform_dens_test}.

\subsubsection{One-dimensional code tests}
\label{subsubsect:1D_tests}

In the case of one spatial dimension, the scalar field satisfies ordinary differential equations. Therefore, we can construct a density field that has a known analytical solution of the scalar field, to check if the code returns the correct numerical solution, according to the scalar field equations of $f(R)$ gravity (Eq.~\eqref{eq:fR_eom_codeunit2}), symmetron (Eq.~\eqref{eq:sym_eom_code_unit}) and coupled quintessence (Eq.~\eqref{eq:csf_eom_code}) in code units. In practice this can be achieved by choosing a functional form of the scalar field in 1D, and applying the above equations to derive analytical expressions for $\delta(\tilde{x})$. For example, we can design density configurations in $f(R)$ gravity by manipulating Eq.~\eqref{eq:fR_eom_codeunit2} in the 1D case as 
\begin{align}
    \tilde{\delta}(\tilde{x}) = -\frac{a}{\Omega_{\rm m}} \qty{ -\tilde{c}^2 \tilde{\nabla}^2 \qty[u^{n+1}] - \frac{1}{3} \tilde{\bar{R}}(a) a^2 \qty[-\bar{f}_R (a)]^{1/(n+1)} \frac{1}{u} + \frac{1}{3} \tilde{\bar{R}}(a) a^2 } \ . \label{eq:fR_eom_codeunit2_CodeTestForm}
\end{align}
We have designed such tests where the scalar field solution is a sine function. 

For $f(R)$ gravity, the scalar field takes the following sine-function form, 
\begin{align}\label{eq:sine_field_fR}
    \frac{f_R(\tilde{x})}{\bar{f}_R} = 1 + A \sin\frac{2\pi \tilde{x}}{N_{\rm g}}\,,
\end{align}
if the density field is given by 
\begin{equation}
    \tilde{\delta}(\tilde{x}) = \frac{a}{\Omega_{\rm m}} \qty[\tilde{c}^2 \bar{f}_R \qty(\frac{2\pi}{N_{\rm g}})^2A \sin\frac{2\pi \tilde{x}}{N_{\rm g}} + \frac{1}{3} \tilde{\bar{R}}(a) a^2 \qty(1 + A \sin\frac{2 \pi \tilde{x}}{N_{\rm g}})^{-\frac{1}{n+1}} - \frac{1}{3} \tilde{\bar{R}}(a) a^2],
\end{equation}
where $A$ is a constant and $|A| < 1$ as $f_R/\bar{f}_R$ should be positive. We have again adopted $f_{R0}=-10^{-5}$ and considered the three cases of $n=0,1,2$ respectively. 

For the symmetron model, we have taken the following form of the scalar field
\begin{equation}\label{eq:sine_field_sym}
    u(\tilde{x}) = \frac{\varphi(\tilde{x})}{\varphi_\ast} = \frac{1}{2}+A\sin\frac{2\pi\tilde{x}}{N_{\rm g}},
\end{equation}
which corresponds to the following overdensity field,
\begin{equation}
    \tilde{\delta}\left(\tilde{x}\right) = \frac{a^3}{a^3_\ast}\left[1 - \left(\frac{1}{2}+A\sin\frac{2\pi\tilde{x}}{N_{\rm g}}\right)^2 - \frac{2}{a^2}\xi^2\tilde{c}^2A\left(\frac{2\pi}{N_{\rm g}}\right)^2\frac{\sin\frac{2\pi\tilde{x}}{N_{\rm g}}}{\frac{1}{2}+A\sin\frac{2\pi\tilde{x}}{N_{\rm g}}}\right]-1.
\end{equation}
The model parameter used here are the same as in the uniform density test above.

For the coupled quintessence model, we have taken the following form of the scalar field
\begin{equation}\label{eq:sine_field_cs}
    \tilde{\varphi}(\tilde{x}) = \tilde{c}^2\delta\varphi(\tilde{x}) = A\sin\frac{2\pi\tilde{x}}{N_{\rm g}},
\end{equation}
which corresponds to the following overdensity field,
\begin{equation}
    \tilde{\delta}\left(\tilde{x}\right) = -\frac{a}{3\beta\Omega_{\rm m}}\exp(\beta\bar{\varphi})\left[A\left(\frac{2\pi}{N_{\rm g}}\right)^2\sin\frac{2\pi\tilde{x}}{N_{\rm g}}-\frac{\lambda^2a^2}{\left(\bar{\varphi}+\tilde{c}^{-2}A\sin\frac{2\pi\tilde{x}}{N_{\rm g}}\right)^\alpha} + \frac{\lambda^2a^2}{\bar{\varphi}^\alpha}\right].
\end{equation}
The model parameter used here are the same as in the uniform density test above.

The panels in the right column of Fig.~\ref{fig:code_test_1D_3x2} present the sine field test results for the three classes of models, in the same order as in the left column. The numerical solutions from \textsc{mg-glam} (squares) agree well with the analytical solutions of Eqs.~(\ref{eq:sine_field_fR}, \ref{eq:sine_field_sym}, \ref{eq:sine_field_cs}), shown by lines, indicating that the code works accurately to solve the scalar field equations. In all the tests shown here we have taken $A=0.5$, but we have checked other values of $A$, as well as sine functions with more than one oscillation period, and found similar agreements in all cases.



\subsubsection{Three-dimensional density tests}
\label{subsubsect:3D_tests}

As the final part of our tests of the multigrid relaxation solver, we consider slightly more complicated density configurations than the uniform and 1D density fields used previously. In order to get analytical and numerical solutions that can be compared with the predictions by \textsc{mg-glam}, we still would like to use density fields that have certain symmetries. To this end, we have done tests using a point mass (for $f(R)$ gravity) and spherical tophat overdensity (for the symmetron and coupled quintessence models). These tests will see the scalar field values vary in $x, y$ and $z$ directions, and they are therefore proper 3D tests.

\subsubsection*{Point mass}

For the first test in 3D space, we consider the solution of the scalar field around a point mass placed at the origin, for which we have approximated analytical solution that is valid in the regions far from the mass. This test has been widey performed in previous MG code papers such as \citep{2008PhRvD..78l3523O,2011PhRvD..83j4026B,Li:2011_ECOSMOG_code_paper,Arnold:2019_MGAREPO_code_paper}. The matter overdensity array is constructed as
\begin{align}
    \tilde{\delta}_{i,j,k} = \begin{cases}
        10^{-4} (N_{\rm g}^3 - 1) , & i=j=k=1; \\
        -10^{-4} , & \text{otherwise.}
    \end{cases} \label{eqn:deltaijk_point_mass}
\end{align}
where $i,j,k = 1, ..., N_{\rm g}$ are the cell indices in $x,y,z$ directions, respectively.

In $f(R)$ gravity, with this density configuration, the scalaron equation Eq.~\eqref{eq:fR_eom_qsa} in regions far from the point mass simplifies to 
\begin{align}
    \boldsymbol{\nabla}^2 \delta f_R \approx m_{\rm eff}^2 \delta f_R \ , \label{eqn:fR_eom_pointmass_far}
\end{align}
where $\delta f_R (\bm{x}) \equiv f_R(\bm{x}) - \bar{f}_R$, and the effective mass of the scalar field, $m_{\rm eff}$, is given by
\begin{equation}
    m^2_{\rm eff} \equiv -\frac{1}{3(n+1)}\frac{\bar{R}_0}{c^2\bar{f}_{R0}}\left(\frac{\bar{R}}{\bar{R}_0}\right)^{n+2} = \frac{H_0^2\Omega_{\rm m}}{c^2(n+1) (-\bar{f}_{R0})} \frac{\qty(a^{-3} + 4\frac{\Omega_{\Lambda}}{\Omega_{\rm m}})^{n+2}}{\qty(1 + 4\frac{\Omega_{\Lambda}}{\Omega_{\rm m}})^{n+1}}\,.
\end{equation}
At $a=1$, this only depends on the combination $(n+1)\bar{f}_{R0}$. For a sphericially symmetric case such as the one considered here, the equation can be recast in the following form,
\begin{equation}
    \frac{1}{r^2}\frac{{\rm d}}{{\rm d}r}\left[r^2\frac{{\rm d}\delta{f}_R}{{\rm d}r}\right] = m^2_{\rm eff}\delta{f}_R,
\end{equation}
or equivalently 
\begin{equation}
    \frac{{\rm d}^2}{{\rm d}r^2}\left[r\delta{f}_R(r)\right] = m^2_{\rm eff}\cdot r\delta{f}_R(r),
\end{equation}
where $r$ is the distance from the central point mass. This equation has the solution 
\begin{equation}
    r\delta{f}_R(r) = \alpha_1\exp\left(-m_{\rm eff}r\right) + \alpha_2\exp\left(m_{\rm eff}r\right),
\end{equation}
where $\alpha_{1}, \alpha_2$ are constants of integral, and we must have $\alpha_2=0$ to prevent the solution from diverging at $r\rightarrow\infty$. This leads to the following solution 
\begin{equation}\label{eq:fR_point_mass_soln}
    \delta{f}_R(r) \propto \frac{1}{r}\exp\left(-m_{\rm eff}r\right),
\end{equation}
which in code unit can be rewritten as
\begin{equation}\label{eq:fR_point_mass_soln_code_unit}
    \delta{f}_R\left(\tilde{r}\right) \propto \frac{1}{\tilde{r}}\exp\left(-\tilde{m}_{\rm eff}\tilde{r}\right),
\end{equation}
where the $\tilde{m}_{\rm eff}$ is the scalar field mass $m_{\rm eff}$ in code unit, given by
\begin{equation}
    \tilde{m}_{\rm eff}^2 \equiv \frac{\Omega_{\rm m}}{\tilde{c}^2(n+1) (-\bar{f}_{R0})} \frac{\qty(a^{-3} + 4\frac{\Omega_{\Lambda}}{\Omega_{\rm m}})^{n+2}}{\qty(1 + 4\frac{\Omega_{\Lambda}}{\Omega_{\rm m}})^{n+1}}\,.
\end{equation}
Note that we have neglected the tilde for $\delta{f}_R$ since in our code units $\tilde{f}_R = f_R$ and $\tilde{\bar{f}}_{R0} = \bar{f}_{R0} \equiv f_{R0}$.

\begin{figure}
    \centering 
    \includegraphics[width=\textwidth]{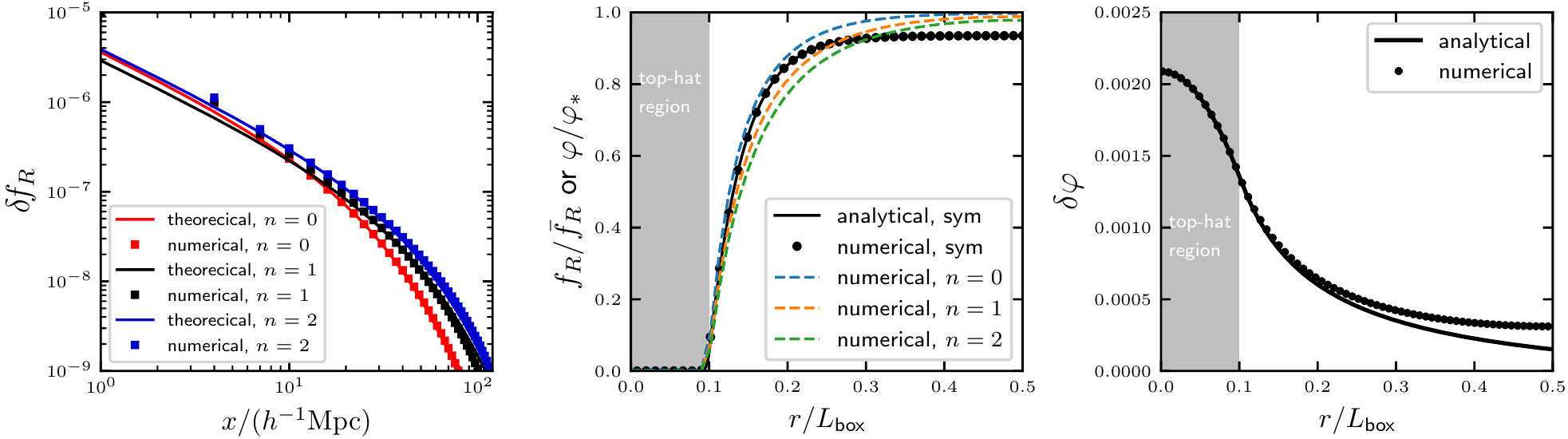}
    \caption{The three-dimensional code test results. 
    \textit{Left Panel:} The numerical (squares) and analytical (lines) solutions to $\delta f_R \equiv f_R - \bar{f}_R$ around a point mass located at $(x,y,z)=(0,0,0)$, for three $f(R)$ gravity models with $f_{R0} = -10^{-5}$ and $n = 0$ (red), $1$ (black) and $2$ (blue), respectively. The analytical approximations are only valid far from the point mass. Only the solutions along the $x$-axis are shown. 
    \textit{Middle Panel:} Top-hat overdensity test for the symmetron (black) and $f(R)$ gravity models (dashed lines). 
    The lines correspond to the analytical solutions and the dots represent the numerical results. 
    The quantities shown on the $y$-axis are $\varphi (r) / \varphi_*$ for the symmetron model, and $f_R (r) / \bar{f}_R$ for the $f(R)$ model.
    \textit{Right Panel:} The same as the middle panel but for the coupled quintessence model.
    }
    \label{fig:CodeTests3D}
\end{figure}

In the left panel of Fig.~\ref{fig:CodeTests3D}, we show the numerical solutions from \textsc{mg-glam} and the analytical results given in Eq.~\eqref{eq:fR_point_mass_soln_code_unit}. Notice that the latter has an unknown coefficient, which we have tuned to match the amplitude of the \textsc{mg-glam} solution. Once that is done, the two agree very well for all three $f(R)$ gravity models with $f_{R0}=-10^{-5}$ for $n = 0, 1$ and $2$ resepctively, except on scales smaller than $\simeq 5 \, h^{-1} \mathrm{Mpc}$ since Eq.~\eqref{eqn:fR_eom_pointmass_far} is not valid near the point mass, and far from the point mass where the \textsc{mg-glam} solution starts to see the effect of periodic boundary condition, which is absent in Eq.~\eqref{eq:fR_point_mass_soln_code_unit}.

\subsubsection*{Spherical tophat overdensity}

For the symmetron and coupled quintessence models, instead of a point mass test, we have considered a spherical tophat overdensity with radius $\tilde{R}_{\rm TH}$ located at the centre of the simulation box $(\tilde{x}, \tilde{y}, \tilde{z}) = (N_{\rm g}/2, N_{\rm g}/2, N_{\rm g}/2)$. Note that code units are used here. The overdensity field is given by 
\begin{align}
    \tilde{\delta}_{\rm TH} \left(\tilde{r}\right) = \begin{cases}
        \tilde{\delta}_{\rm in}, & \tilde{r} \le \tilde{R}_{\rm TH} \\
        \tilde{\delta}_{\rm out}, & \tilde{r} > \tilde{R}_{\rm TH}
    \end{cases},
\end{align}
where $\tilde{r} \equiv \sqrt{(\tilde{x}-N_{\rm g}/2)^2 + (\tilde{y}-N_{\rm g}/2)^2 + (\tilde{z}-N_{\rm g}/2)^2}$ is the distance from the tophat centre, and we have adopted $\tilde{R}_{\rm TH} = 0.1 N_{\rm g}, \tilde{\delta}_{\rm in} = 5000$ and $\tilde{\delta}_{\rm out} = 0$ in our tests. In spherical symmetry, the scalar field equations for the symmetron  (Eq.~\eqref{eq:sym_eom_code_unit}) and coupled quintessence (Eq.~\eqref{eq:csf_eom_code}) models reduce to the following 1D ordinary differential equations,
\begin{align}
    \tilde{c}^2 \frac{1}{\tilde{r}^2} \frac{\dd}{\dd{\tilde{r}}} \qty[\tilde{r}^2 \frac{\dd{u}}{\dd{\tilde{r}}}] &= \frac{a^2}{2\xi^2}\left[ \qty(1 + \tilde{\delta}_{\rm TH}(\tilde{r}))\frac{a_\ast^3}{a^3}-1\right]u + \frac{a^2}{2\xi^2}u^3 \label{eq:sym_eom_code_unit_spherical}
\intertext{and}
    \frac{1}{\tilde{r}^2} \frac{\dd}{\dd{\tilde{r}}} \qty[\tilde{r}^2 \frac{\dd{\tilde{\varphi}}}{\dd{\tilde{r}}}] &= \frac{3\beta\Omega_{\rm m}}{a} e^{\beta\bar{\varphi}}\left[\exp\left(\beta\frac{\tilde{\varphi}}{\tilde{c}^2}\right) \qty(1+\tilde{\delta}_{\rm TH}(r))-1\right] \notag \\
    &\phantom{=} -\alpha\lambda^2a^2\left[\frac{1}{\left(\bar{\varphi}+\tilde{c}^{-2}\tilde{\varphi}\right)^{1+\alpha}}-\frac{1}{\bar{\varphi}^{1+\alpha}}\right] , \label{eq:csf_eom_code_spherical}
\end{align}
respectively.

These two 1D equations are numerically solved using the \textsc{maple} software, with the following boundary conditions on the interval $r \in [0, N_{\rm g}/2]$,
\begin{align}
    u(\tilde{r} = N_{\rm g}/2) = \sqrt{1 - \qty(\frac{a_*}{a})^3}, \quad \frac{\dd{u}}{\dd{\tilde{r}}}(\tilde{r}=0) = 0\,, \\
\intertext{and}
    \tilde{\varphi}(\tilde{r} = N_{\rm g}/2) = 0, \quad \frac{\dd{\tilde{\varphi}}}{\dd{\tilde{r}}}(\tilde{r}=0) = 0\,,
\end{align}
for the symmetron and coupled quintessence models respectively. Note that rigorously speaking the first boundary condition should really have been set at $\tilde{r}\rightarrow\infty$, but for numerical implementation this is impractical and we instead use $N_{\rm g}/2$ as an approximation to $\infty$. 

We have obtained the numerical solutions of these ODEs for $u(\tilde{r})$ and $\tilde{\varphi}(\tilde{r})$, but still call them `analytical' to distinguish from the numerical solutions directly solved from the original PDEs solved by \textsc{mg-glam}. The model parameters are the same as in the uniform and 1D density tests: for the symmetron model we have used $a_\ast = 0.5, \xi = 10^{-3}$, $\beta_\ast = 0.1$, while for coupled quintessence we have used $\alpha = 0.1, \beta = -0.2$.

The analytical and numerical solutions for the symmetron and coupled quintessence models are displayed in the middle and right panels of Fig.~\ref{fig:CodeTests3D}, respectively as the black solid line and black symbols. They agree very well.

As a comparison, in the middle panel of Fig.~\ref{fig:CodeTests3D} we have also shown, with coloured symbols, the \textsc{mg-glam} solutions for the $f(R)$ model with $f_{R0}=-10^{-5}$ and $n=0$ (blue), $1$ (orange) and $2$ (green). This can serve as a quick comparison of the screening efficiencies in these four models. First, we note that in all three $f(R)$ models the solution, $f_R(\tilde{r})/\bar{f}_{R0}$, tends to $1$ far from the spherical tophat, which is expected because the scalar field approaches its background value far from the matter perturbation at the centre. Second, for all four models, the scalar field is strongly suppressed inside the tophat (grey shaded region), but increases sharply immediately outside $\tilde{R}_{\rm TH}$ such that within some small distance from the edge of the tophat it already reaches $\gtrsim50\%$ of the background value: this is what one would expect from the $f(R)$ and symmetron models---both of which are examples of the so-called \textit{thin-shell screened} models \cite{Brax:2012gr}. Third, comparing the solutions of the three $f(R)$ models with the same $f_{R0}$, it seems that increasing the value of $n$ increases the screening efficency, implying that the $n=2$ case has the strongest screening amongst them; we shall see the consequence of this in the cosmological simulations in the next section. Finally, comparing the tested symmetron model with the $f(R)$ ones, it seems that the solution of the former lies somewhere in between the $n=0$ and $n=1$ cases (at least near the tophat); however, we caution that the fifth forces in the two models are obtained in different ways: in $f(R)$ gravity it is directly proportional to $\boldsymbol{\nabla}f_{R}$, while for symmetrons it is proportional to $\boldsymbol{\nabla}\left(u^2\right)$, cf.~Eq.~\eqref{eq:F_5_codeunit}, rather than $\boldsymbol{\nabla}u$.

\subsection{Convergence tests}
\label{subsect:convergence_tests}

As mentioned in Sect.~\ref{subsubsect:relaxation}, we have implemented three different arrangements of the multigrid solver --- V-, F- and W-cycles. To compare them we have run a series of small cosmological simulations for the $f(R)$ gravity model with $f_{R0} = -10^{-5}$ and $n=1$, the symmetron model with $a_\ast = 0.3, \xi = 10^{-3}$ and $\beta_\ast = 0.1$ and the coupled quintessence model with $\alpha = 0.1$ and $\beta = -0.2$. These runs all use $L = 256 \, h^{-1}\mathrm{Mpc}$, $N_{\rm p}^3 = 512^3$ and $N_{\rm g}^3 = 1024^3$ for the smaller simulations. We consider $10$ and $2$ V-cycles (V10 and V2), $1$ F-cycle (F1) and 1 W-cycle (W1) to test the convergence of the \ac{MG} scalar field solutions. The V10 simulation results are used as the benchmark of our test. 
For F- and W-cycles we only conisder one cycle because, as will be shown below, this already gives excellently converged results.

\begin{figure}
    \centering 
    \includegraphics[width=\textwidth]{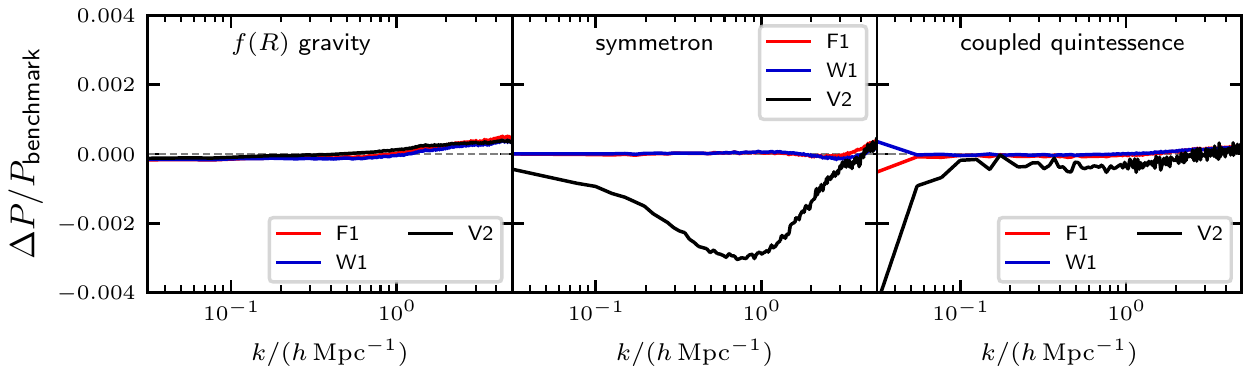}
    \caption{A comparison of the convergence with different multigrid arrangements and numbers of cycles of the Gauss-Seidel relaxation. The fractional differences in the matter power spectra are plotted at $z=0$, obtained for different multigrid schemes (V2, V10, F1, W1) using V10 as the reference. 
    The cases shown are for the $f(R)$ model with $f_{R0} = -10^{-5}$ and $n=1$ (left panel), the symmetron model with $a_* = 0.3, \xi=10^{-3}$ and $\beta_* = 0.1$ (middle panel), and the coupled quintessence model with $\alpha = 0.1, \beta = -0.2$ (right panel).}
    \label{fig:ConvergenceTest}
\end{figure}

Figure \ref{fig:ConvergenceTest} shows the relative differences
of the matter power spectra at $z=0$ between the test simulations described above and the benchmark case (V10), for $f(R)$ gravity (left), and the symmetron (middle) and coupled quintessence (right) models. We find that all the different schemes and different numbers of cycles used to solve the partial differential equations have good agreement on almost all scales probed by the simulations ($\lesssim0.4\%$). However, when more cycles are used, the run time gets longer, and the slowest simulations are those using V10. 
F-cycles and W-cycles are more effective in reducing residuals, both agreeing with V10 by $\lesssim0.05\%$ after only one cycle, which is not surprising since they walk more times across the fine and coarse multigrid levels. As a result, Both F1 and W1 are slower than V2. Therefore, as a compromise between accuracy and cost, we decide to always use V2 in our cosmological runs. 

It is a great achievement for the multigrid solver to reach convergence after just 2 V-cycles (and 2 Gauss-Seidel passings of the entire mesh in each cycle), for nonlinear equations in
the $f(R)$ gravity and symmetron models.

\subsection{Comparisons with previous simulations}
\label{subsect:comparions}

\begin{figure}
    \centering 
    \includegraphics[width=\textwidth]{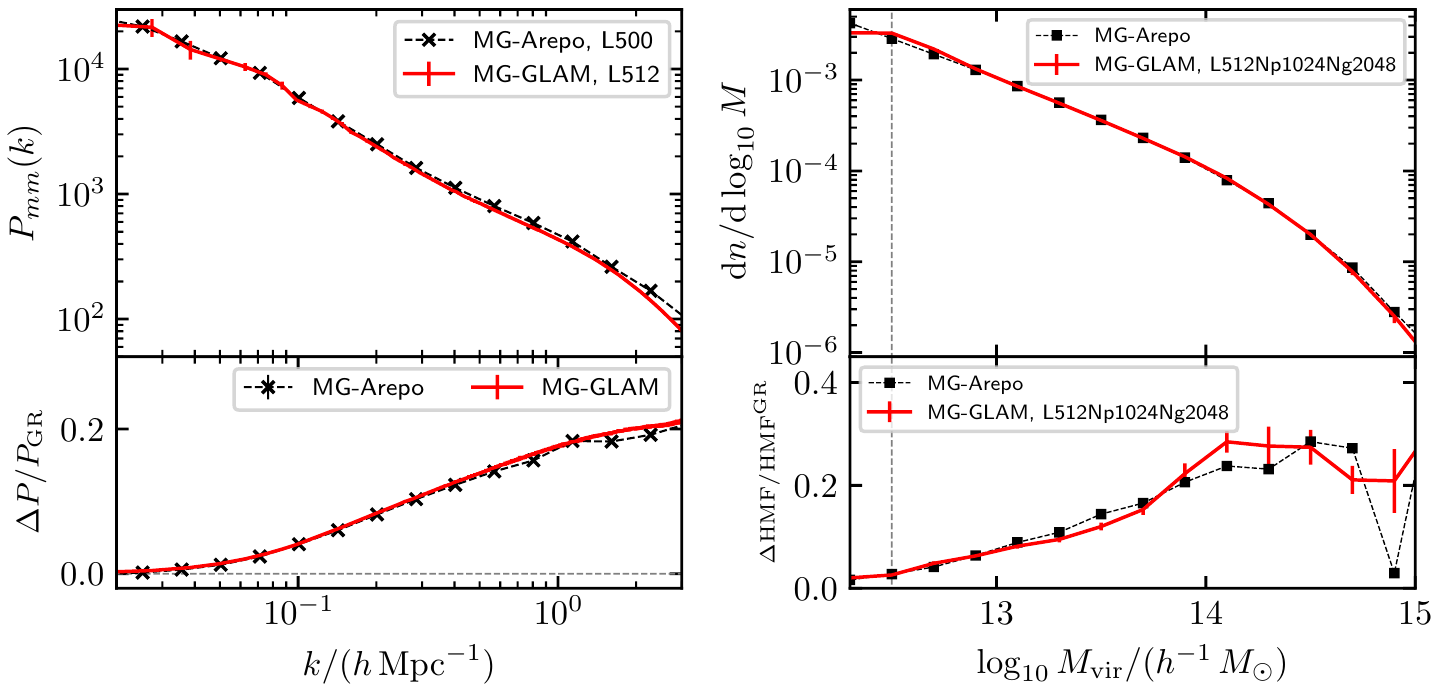}
    \caption{Comparison of matter power spectra (left panel) and halo mass functions (right panel) predicted by simulations with the same box size and particle number, using the \textsc{mg-glam} (black dashed lines with symbols) and \textsc{mg-arepo} (red solid lines) codes for the same $f(R)$ model, $n=1$ and $-f_{R0}=10^{-5}$. The upper subpanels show the absolute measurements from the simulations, while the lower subpanels show the relative differences from the counterpart $\Lambda$CDM runs. The vertical dashed line in the right panel denotes $10^{12.5}h^{-1}M_\odot$. The two codes agree very well above this mass.
    }
    \label{fig:Pk_hmf_MGGLAMvsArepo}
\end{figure}

As a final test of the \textsc{mg-glam} code, we compare its predictions from cosmological simulations with those by other modified gravity codes in the literature. We do this for the $f(R)$ and symmetron models only, since the coupled quintessence model is more trivial: the fifth force in this model is unscreened, and has a nearly constant ratio with the strength of Newtonian gravity in space \cite{Li:2010re}. 

For $f(R)$ gravity, we have run two \textsc{mg-glam} simulations for the model $f_{R0}=-10^{-5}, n=1$, using a box size $L=512 \, h^{-1}\mathrm{Mpc}$ with $N_{\rm p}^3 = 1024^3$ particles and  $N_{\rm g}^3=2048^3$ mesh cells. These are compared with the predictions from a simulation using \textsc{mg-arepo}, with $L = 500 \, h^{-1}\mathrm{Mpc}$ and $N_{\rm p}^3 = 1024^3$. All these simulations have the same cosmological parameters, but they started from different realisations of initial conditions (ICs). Since \textsc{mg-arepo} uses adaptive mesh refinement for the modified gravity force and trees for the Newtonian force with a softening length of $\approx 15 \, h^{-1}\mathrm{kpc}$, it achieves better force resolution as compared with the \textsc{mg-glam} simulations that use a regular mesh with $N_{\rm g}=2048$ giving a force resolution of $0.25 \, h^{-1}\mathrm{Mpc}$. Despite this, we will see that \textsc{mg-glam} can reproduce the \textsc{mg-arepo} results on scales of interest.

In the left panel of Fig.~\ref{fig:Pk_hmf_MGGLAMvsArepo} we compare the matter power spectra, $P_{mm}(k)$, from the \textsc{mg-glam} (lines) and \textsc{mg-arepo} (symbols) simulations. The upper subpanel shows the absolute $P(k)$, where the two codes agree down to $k \approx 1 \, h\,\mathrm{Mpc}^{-1}$. As shown in \cite{Klypin:2017iwu}, with a mesh resolution of $0.25 \, h^{-1} \mathrm{Mpc}$ the \textsc{glam} code is capable of predicting $P_{mm}(k)$ with percent-level accuracy down to $k \approx 1 \, h\,\mathrm{Mpc}^{-1}$. The lower subpanel shows the enhancements of the matter power spectrum due to $f(R)$ gravity. To obtain this, we have also run a counterpart $\Lambda$CDM simulation for each of the $f(R)$ simulations, using the same box size, grid number, particle number, cosmological parameters and initial conditions; we then take the relative difference between an $f(R)$ run and its counterpart $\Lambda$CDM run. We can see an excellent agreement between the two codes, down to $k \approx 3 \, h\,\mathrm{Mpc}^{-1}$ (even though the power spectra themselves agree only down to $k \approx 1 \, h\,\mathrm{Mpc}^{-1}$).  

The right panel of Fig.~\ref{fig:Pk_hmf_MGGLAMvsArepo} extends the comparison to the differential halo mass function (dHMF). The dHMF is a description of the halo abundance; more accurately, it quantifies the number density of haloes, in a spatial volume, that falls into a given halo mass bin. In the upper subpanel we present the dHMFs measured from the \textsc{mg-glam} and \textsc{mg-arepo} simulations, while in the lower subpanel we show the enhancements with respect to their counterpart $\Lambda$CDM runs. As we mentioned above, \textsc{mg-glam} uses the spherical overdensity halo mass definition with the virial halo overdensity, $M_{\rm vir}$. On the other hand, \textsc{mg-arepo} by default uses the $M_{200c}$ halo mass difinition, which is defined by requiring the mean overdensity within the halo radius $R_\Delta$, to be $\Delta=200\rho_{\rm crit}(z)$. To be self-consistent, we have rerun \textsc{mg-arepo}'s halo finder, \textsc{subfind} \cite{Springel2001}, using the $M_{\rm vir}$ definition. 
The upper subpanel shows that, at this specific mesh resolution, the dHMF predicted by \textsc{mg-glam} is complete down to $10^{12.5}h^{-1}M_\odot$, and agrees with \textsc{mg-arepo} for $M>10^{12.5}h^{-1}M_\odot$. In addition, the dHMF enhancements due to $f(R)$ gravity predicted by these two codes also agree very well, despite being noisy at the high-mass end due to the small box sizes used here.

Overall, for the $f(R)$ version, we find very good agreement between \textsc{mg-glam} and \textsc{mg-arepo}. We have also compared the \textsc{mg-glam} simulation results with predictions by \textsc{ecosmog} (although the results are not presented here), and obtained as good agreements as shown in Fig.~\ref{fig:Pk_hmf_MGGLAMvsArepo} for both $P_{mm}(k)$ and the HMF.

\begin{figure}
    \centering 
    \includegraphics[width=0.7\textwidth]{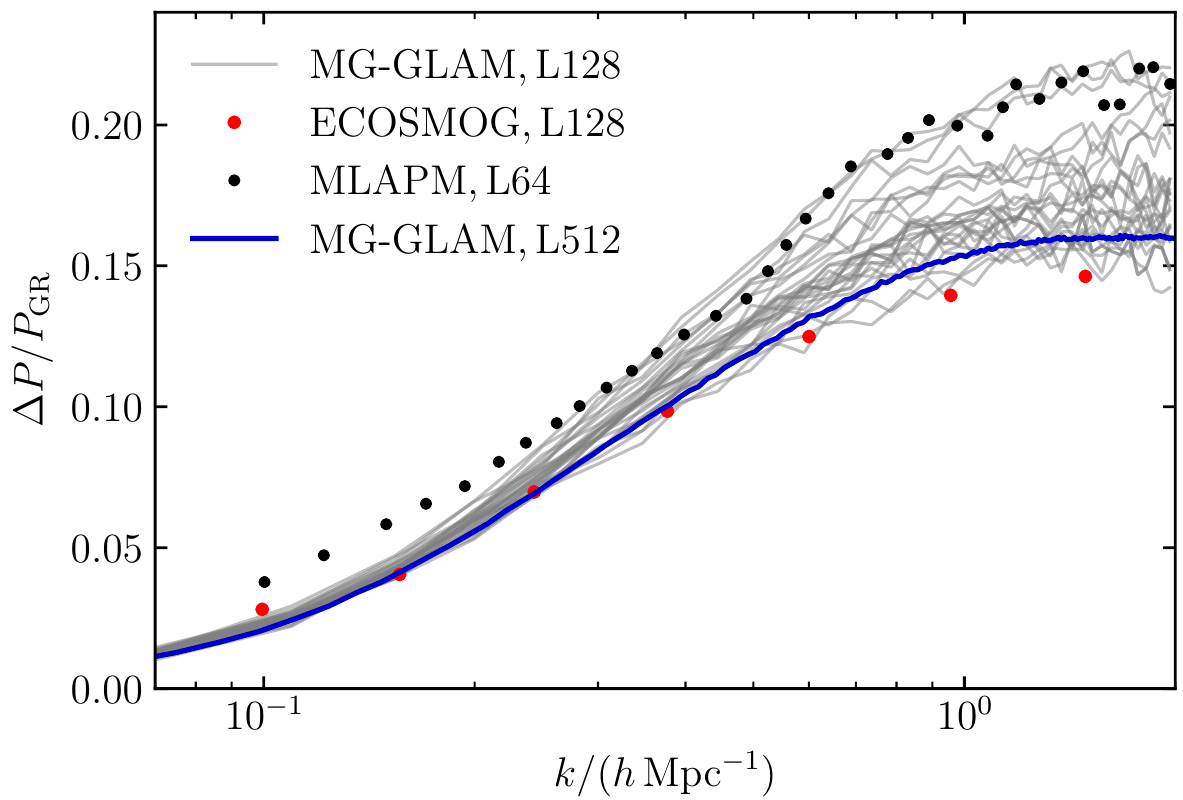}
    \caption{\revision{Comparison of the matter power spectrum enhancements with respect to $\Lambda$CDM, for the symmetron model with $a_\ast=0.33, \beta_\ast=1$ and $\xi=3.34 \times 10^{-4}$ at $z=0$, from two previous simulations run with \textsc{ecosmog} \citep{2012JCAP...10..002B} (red symbols) and \textsc{mlapm} \citep{2012ApJ...756..166W} (black symbols) respectively, and two groups of \textsc{mg-glam} runs with box sizes of $128$ (thin grey lines) and $512 \, h^{-1}\mathrm{Mpc}$ (thick blue line).  The box sizes used in the \textsc{mlapm} and \textsc{ecosmog} runs are respectively $64\,h^{-1}\mathrm{Mpc}$ and $128\,h^{-1}\mathrm{Mpc}$. The $25$ small-box \textsc{mg-glam} realisations have substantial sample variance (as shown by the strong scatters) on small scales due to the box size, and these curves are also much noisier compared with the large-box result. The \textsc{ecosmog} and \textsc{mlapm} results are close to the two limits of the scatters in the 25 \textsc{mg-glam} runs. The large scale behaviour of the \textsc{mlapm} result is likely due to its very small box size.}}
    \label{fig:Pk_sym_MGGLAMvsECOSMOG}
\end{figure}

For the symmetron model, we have three \textsc{mg-glam} runs for the parameter values $a_\ast=0.33, \beta_\ast=1$ and $\xi = 3.34 \times 10^{-4}$, and we compare the measured matter power spectra with those presented in \citep{2012JCAP...10..002B} and \citep{2012ApJ...756..166W} using the adaptive mesh refinements codes \textsc{ecosmog} and \textsc{mlapm}, respectively.
The \textsc{ecosmog} symmetron run followed the evolution of  $N_{\rm p}^3 = 256^3$ particles in a box of size $L = 128 \, h^{-1}\mathrm{Mpc}$, and the domain grid (defined as the finest uniform grid which covers the whole simulation box) has $N_{\rm g}^3 = 256^3$ cells. For the \textsc{mlapm} simulation, the box size is $64 \, h^{-1} \mathrm{Mpc}$, the particle number is $256^3$ and the domain grid cell number is $128^3$. 

\revision{Fig.~\ref{fig:Pk_sym_MGGLAMvsECOSMOG} shows the matter power spectrum enhancement, $\Delta P / P_{\mathrm{GR}}$, between a pair of $\Lambda$CDM and MG simulations starting from the same initial conditions, for the \textsc{mg-glam} runs (solid lines) and the data taken from Ref.~\citep{2012JCAP...10..002B} (red symbols) and \citep{2012ApJ...756..166W} (black symbols). 
For the power spectrum itself, the sample variance should be smaller on smaller scales, which have more $k$~modes than large scales. For $\Delta P / P_{\mathrm{GR}}$, however, we see the opposite behaviour: the sample variance is substantially suppressed on large scales ($k < 0.1 \, h^{-1}\mathrm{Mpc}$) where the evolution is largely linear and different $k$~modes uncoupled to each other. On small scales, different Fourier modes are coupled together, and two different gravity models that have different strengths of gravity would lead to different levels of such coupling.
Hence, the difference between the power spectra in these two models at the same high $k$ can be substantial, especially when the box size is small and therefore the result is more susceptible to rare large objects present in the box.
An example to illustrate this point is the bottom panel of Fig.~5 of Ref.~\cite{2011PhRvD..83d4007Z}, which compares the $\Delta P / P_{\mathrm{GR}}$ from an $f(R)$ simulation and a `linearised' (no-chameleon) $f(R)$ simulation which has its chameleon screening effects removed --- the latter case corresponds to a much stronger gravity and a much larger power spectrum enhancement, and therefore much more significant scatters.  }

\revision{In Fig.~\ref{fig:Pk_sym_MGGLAMvsECOSMOG}, we find that the \textsc{mg-glam} simulations for the $25$ independent realisations with a small box ($L = 128 \, h^{-1}\mathrm{Mpc}$) have strong scatters in $\Delta P / P_{\mathrm{GR}}$ on small scales while not on large scales.
The previous simulation results from \textsc{ecosmog} and \textsc{mlapm} follow into this range of scatters.
Note that the \textsc{mlapm} simulation has a box size of $64 \, h^{-1} \mathrm{Mpc}$ which can likely explain its behaviour on large scales.
We also show the result from a single $512 \, h^{-1} \mathrm{Mpc}$ box.
Within the uncertainties allowed by sample variance, all three codes seem to agree with each other.}

\subsection{Summary}
\label{subsect:code_test_summary}

To quickly sum up this section: we have done a number of tests of different aspects of the \textsc{mg-glam} code. These include the test of the background cosmology solver for the coupled quintessence model (cf.~Sect.~\ref{subsect:bg_tests}), tests of the multigrid relaxation solver of the scalar field equations for different density configurations (cf.~Sects.~\ref{subsubsec:homogeneous_rhom_test}, \ref{subsubsect:1D_tests}, \ref{subsubsect:3D_tests}), convergence property tests of the relaxation solvers with three different multigrid arrangements (V-cycle, F-cycle and W-cycle), and additionally comparisons of \textsc{mg-glam} cosmological simulations with runs using other codes. We see that \textsc{mg-glam} satisfactorily passes all these tests, and gives reasonable results.

\section{Cosmological runs}
\label{sect:cos_runs}

The objective of \textsc{mg-glam} is the very fast generation of $N$-body simulations for a wide range of modified gravity models. In this section, we will present some examples of cosmological runs using this code. In particular, we will run a very large suite of $f(R)$ simulations with different parameter values of $n$ and $f_{R0}$. These simulations only take a small fraction of time of a single high-resolution run of \textsc{mg-arepo} or \textsc{ecosmog} for the box size and mass resolution.

The inventory of the cosmological runs we have performed is 
\begin{itemize}
    \item $f(R)$ gravity runs with $n = 0, 1$ and $2$ and $\log_{10}(|f_{R0}|)$ in $10$ bins linearly spaced in the range $[-6.00, -4.50]$. One realisation for each model.
    \item Ten realisations of $f(R)$ gravity runs with $n=0$ and 1 and $-\log_{10} (|f_{R0}|) = 5.00$.
    \item five symmetron models with fixed $a_\ast=0.33$ and $\beta_\ast=1$, with different values of $\xi$ given by $c\xi/H_0=0.5, 1, 2, 2.5, 3$.
    \item three variants of the coupled quintessence model described in Sect.~\ref{subsect:csf}, with $(\alpha,\beta)$ equal to $(0.1,-0.1)$, $(0.1,-0.2)$ and $(0.5,-0.2)$ respectively. 
    \item For each \ac{MG} simulation, we have a  counterpart $\Lambda$CDM run with the same simulation specifications of cosmological parameters. We will label these runs as `GR' runs, to contrast with `\ac{MG}' runs, even though none of our simulations is really general relativisic.
\end{itemize}
For all simulations, we followed the evolution of $1024^3$ particles in a cubic box with size $512 \, h^{-1}\mathrm{Mpc}$ using a grid with $2048^3$ cells. The non-MG cosmological parameters are from the Planck 2015 \citep{Ade:2015xua} best-fitting  $\Lambda$CDM parameters: $$\{\Omega_{\rm b}, \Omega_{\rm m}, h, n_s, \sigma_8\} = \{0.0486,0.3089,0.6774,0.9667,0.8159\}.$$ The ICs of both the \ac{GR} and \ac{MG} runs are generated on the fly from the same $\Lambda$CDM linear perturbation theory power spectrum at $z_{\rm init} = 100$, which itself is generated using the \href{https://camb.info/}{\textsc{camb} code}. We have used the same ICs for \ac{GR} and \ac{MG} simulations (for the same realisation),  since the \ac{MG} effect is very weak at $z > 100$, so that the linear matter power spectrum at $z_{\rm ini}=100$ is nearly identical to that of $\Lambda$CDM.

\subsection{$f(R)$ gravity}
\label{subsect:fR_runs}

\begin{figure}
    \centering 
    \includegraphics[width=\textwidth]{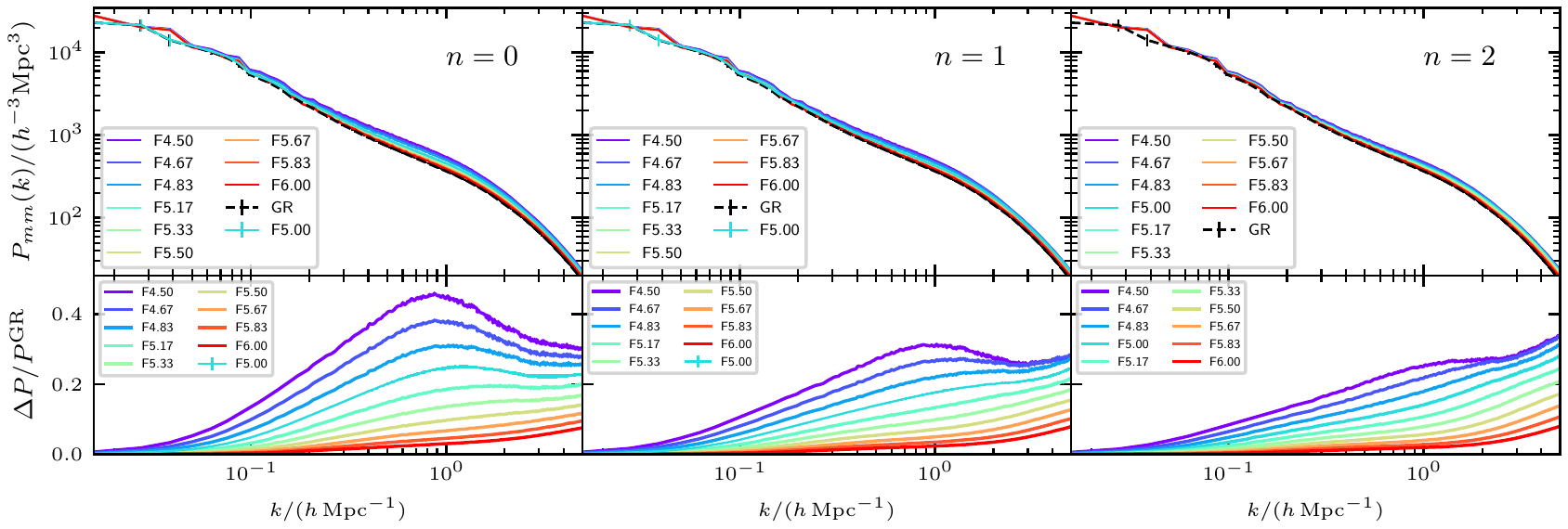}
    \caption{\textit{Upper Panels:} The non-linear matter power spectra at redshift $z=0$, from \textsc{mg-glam} simulations of the $f(R)$ model for $n=0$ (left panel), $n=1$ (middle) and $n=2$ (right), each with $10$ values of $|f_{R0}|$ logarithmically spaced between $10^{-6}$ and $10^{-4.5}$, i.e., $-\log_{10}|f_{R0}| = 4.50, 4.67, \dots, 6.00$. These are indicated with different colours given in the legends. 
    \textit{Lower Panels:} The fractional difference, $\Delta P / P_{\mathrm{GR}}$, between the $f(R)$ and $\Lambda$CDM results, where $\Delta P \equiv P_{\rm MG} - P_{\rm GR}$. The $n=0,1$, $-\log_{10} |f_{R0}| = 5.00$ and $\Lambda$CDM results are the mean of ten independent realisations, while other models only have one realisation.}
    \label{fig:Pk_mm_z0_fRn012}
\end{figure}

\begin{figure}
    \centering 
    \includegraphics[width=\textwidth]{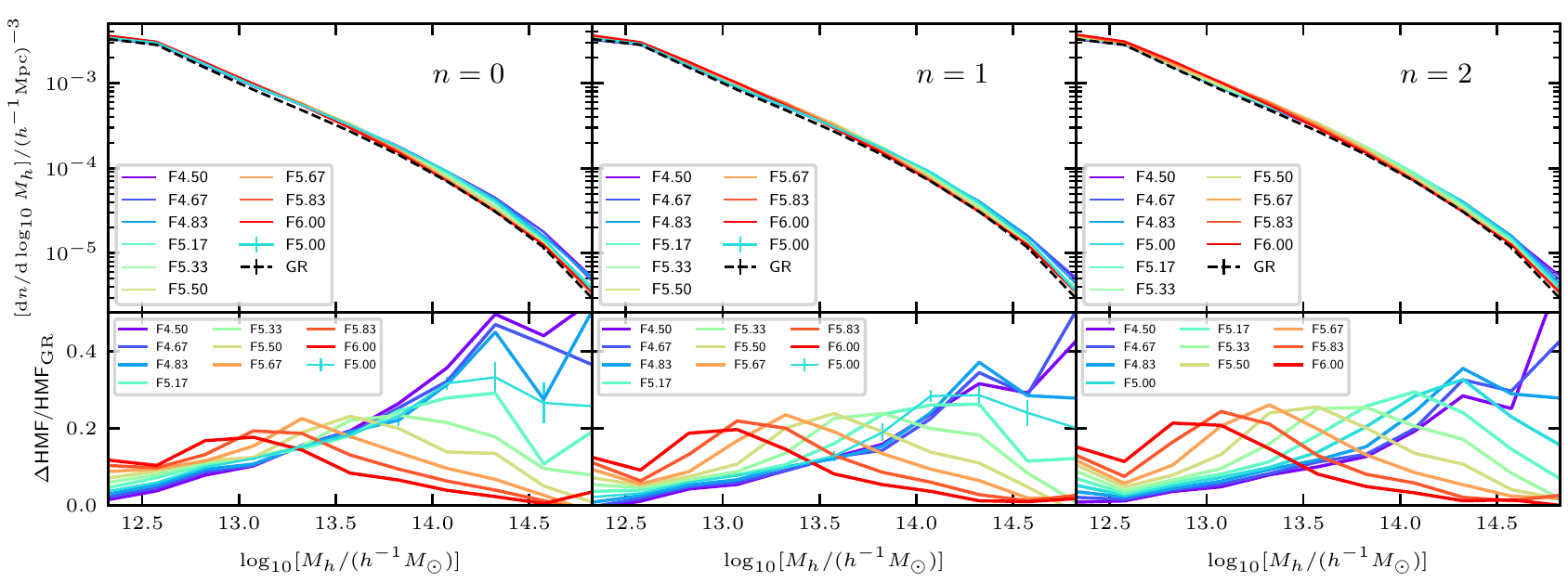}
    \caption{\textit{Upper Panels:} Differential halo mass functions (HMFs) of $f(R)$ gravity for $n=0$ (left panel), $n=1$ (middle) and $n=2$ (right), each with $10$ values of $|f_{R0}|$ logarithmically spaced between $10^{-6}$ and $10^{-4.5}$, i.e., $-\log_{10}|f_{R0}| = 4.50, 4.67, \dots, 6.00$, at redshift $z=0$, from \textsc{mg-glam} cosmological runs.
    \textit{Lower Panels:} The fractional difference $\Delta \text{HMF} / \text{HMF}_{\mathrm{GR}}$ between $f(R)$ and $\Lambda$CDM results, where $\Delta \text{HMF}  \equiv \text{HMF}_{\rm MG} - \text{HMF}_{\rm GR}$.
    The $n=0,1$, $-\log_{10} |f_{R0}| = 5.00$ and $\Lambda$CDM results come from ten realisations (the standard deviation of which is shown as the error bars in the bottom left/central panels), while other models only have one realisation.}
    \label{fig:hmf_z0_fRn012}
\end{figure}

We have measured the matter power spectra $P_{\rm mm} (k)$ and halo mass functions (HMF) at $z = 0$. The results are shown in Fig.~\ref{fig:Pk_mm_z0_fRn012} for the matter power spectra and Fig.~\ref{fig:hmf_z0_fRn012} for the halo mass functions. The relative differences for $P_{\rm mm}(k)$ and HMF between $f(R)$ and \ac{GR} are displayed in the lower subpanels.

Fig.~\ref{fig:Pk_mm_z0_fRn012} shows that the matter clustering is boosted by $3$-$40\%$ due to the fifth force, but the boost is scale-dependent and is weak on very large scales ($k \lesssim 0.03 \, h \, \mathrm{Mpc}^{-1}$).
The $P_{\rm mm}(k)$ enhancement, $\Delta P / P_{\mathrm{GR}}$, depends qualitatively on the value of $|f_{R0}|$.
When $|f_{R0}|$ is small so that the \ac{MG} effect is weak, $\Delta P / P_{\mathrm{GR}}$ increases monotonically with $k$. On the other hand, when the \ac{MG} effect is strong, the fractional difference of matter power spectra no longer monotonically increases with $k$, but goes down at small scales after reaching some peak value at $k \sim 1 \, h \,\mathrm{Mpc}^{-1}$ (although on even smaller scales the $P_{\rm mm}(k)$ enhancement increases again for some models, we only focus on the scales $k \lesssim 3 \, h\,\mathrm{Mpc}^{-1}$ given the fixed simulation resolution, cf.~Sec.~\ref{subsect:comparions}).
This behaviour can be explained in the context of the halo model \citep{Cooray:2002PhR...372....1C}, which assumes that on small scales the matter power spectrum is determined mainly by the matter distribution inside dark matter haloes (the one-halo term).
\begin{itemize}
    \item In the regime of weak \ac{MG} effect, haloes are well screened inside so that particles do not feel the fifth force during most of their evolution. 
    When the haloes become unscreened at late times, the total gravitational potential rapidly becomes $1/3$ deeper, but the particle kinetic energy requires more time to respond, so that these particles tend to fall towards the halo centre, increasing the halo density profile and therefore the one-halo contribution to $P_{\rm mm}(k)$.
    \item When the \ac{MG} effect is strong, particles have been accelerated for a long time (both well before and after they fall into haloes, as the latter are unscreened or less screened) due to the relatively strong fifth force. This means that the accelerations and velocities of particles can be boosted by a similar fraction as the enhancement in the depth of the gravitational potential, and hence the partice kinetic energy can be increased by a larger factor than the deepening of the potential, so that the particles are less likely to be trapped towards the centre of the potential. The small-scale structure can thus be erased out to a certain degree.
    This behaviour of $f(R)$ gravity has been discussed in previous works such as \citep[][]{2011PhRvD..83d4007Z,Li:2013MNRAS.428..743L,Mitchell:2019qke}.
    The explanation also works for other models in which screening has always been weak or absent, such as the coupled quintessence model (the left panel of Fig.~\ref{fig:Pk_HMF_csf}) and the K-mouflage model \citep{Hernandez-Aguayo:2021_twin_paper}; in both cases we see a decay of $\Delta P/P_{\rm GR}$ at $k\gtrsim 1 \, h \, \mathrm{Mpc}^{-1}$. 
\end{itemize}

We note that the parameter $n$ of the $f(R)$ model also has a considerable influence on structure formation. For fixed $f_{R0}$, the larger the value of $n$, the more efficiently the fifth force is screened, as can be seen from Fig.~\ref{fig:Pk_mm_z0_fRn012}, which shows that the matter clustering enhancement is strongest in the $n=0$ while weakest in the $n=2$ case. We have found similar behaviour when we checked the solution of scalar field around a top-hat overdensity in Sec.~\ref{subsubsect:3D_tests}, see the middle panel Fig.~\ref{fig:CodeTests3D}: the $n=2$ case has the strongest screening efficiency.

In \ac{MG} theories, the dark matter halo populations are also affected. One of the elementary halo properties is their abundance, which we quantify using the differential halo mass function (dHMF), $\dd{n}(M)/\dd{\log_{10}M}$, which is defined as the halo number density per unit logarithmic halo mass.
The dHMF result for the $f(R)$ gravity runs at $z=0$ is shown in Fig.~\ref{fig:hmf_z0_fRn012}, where the lower subpanels show the enhancements with respect to $\Lambda$CDM.

Firstly, we note that the abundance of haloes is enhanced due to the enhancement of total gravity. Secondly, for the weaker $f(R)$ models, the relative difference from $\Lambda$CDM is suppressed for massive haloes, where the fifth force is efficiently screened; going to smaller haloes the enhancement increases first, which is due to the less efficient screening and stronger \ac{MG} effect for these objects; but for even smaller haloes the HMF enhancement decreases after reaching a peak, which is due to smaller haloes experiencing more mergers to form larger haloes. Apparently, this trend is not seen for the strong \ac{MG} cases, such as F4.50 (purple) and F4.67 (dark blue), where the HMF enhancement seems to increase monotonoically with halo mass. However, we speculate that the qualitative behaviour for the weaker $f(R)$ models should also hold even in these cases: note that our halo catalogues have been cut off for $M_{\rm vir} \gtrsim 10^{14.7} \, h^{-1} M_{\odot}$ due to the relatively small box size; should the simulations be run with larger box sizes (while keeping the same resolution), we expect the HMF enhancement to dacay to zero for large enough haloes even in the strong \ac{MG} cases.
Finally, we note that the dHMFs are less sensitive to the model parameter $n$ than to $f_{R0}$, compared to the matter power spectra. The shapes and amplitude of dHMFs are similar for $n = 0, 1$ and $2$, though we can still see that they are enhanced slightly more in the case of $n=0$ than the cases of $n=1,2$, for F4.50 and F4.67.

\subsection{Symmetrons and coupled quintessence}
\label{subsect:sym_csf_runs}

\begin{figure}
    \centering 
    \includegraphics[width=\textwidth]{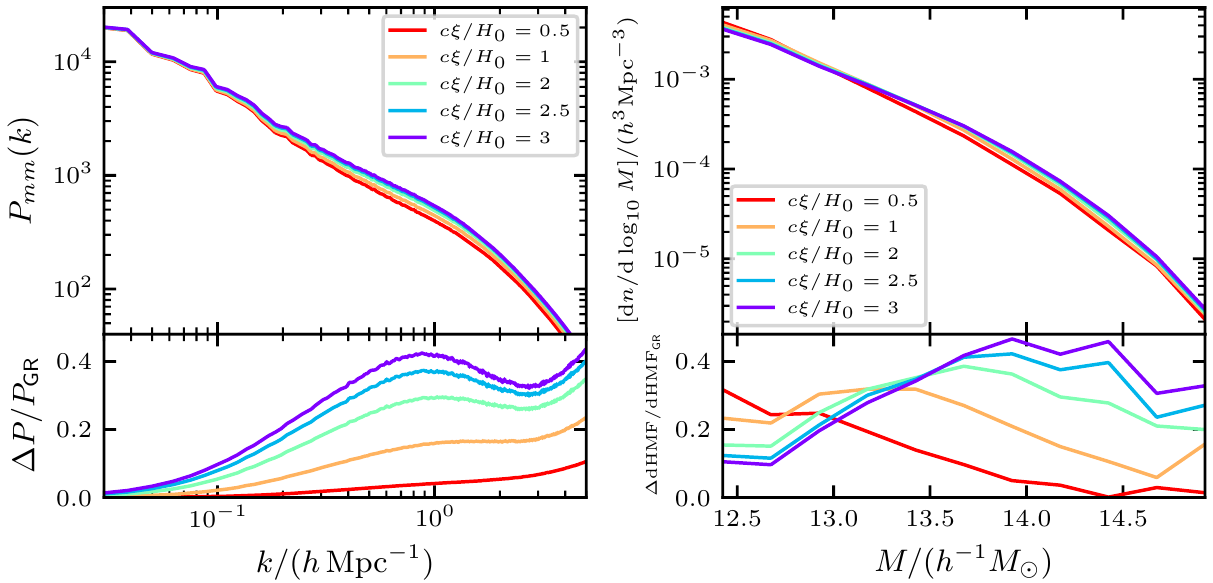}
    \caption{The matter power spectra (left panel) and differential halo mass functions (right) of the symmetron models at $z=0$ for 5 different values of $c\xi/H_0$ as labelled. In all cases the remaining symmetron parameters are fixed as $a_\ast=0.33$ and $\beta_\ast=1.0$. As in Figs.~\ref{fig:Pk_mm_z0_fRn012} and \ref{fig:hmf_z0_fRn012}, the upper subpanels present the absolute measurements from simulations, while the lower subpanels show the relative differences from $\Lambda$CDM.
    }
    \label{fig:Pk_hmf_z0_sym}
\end{figure}

\begin{figure}
    \centering 
    \includegraphics[width=\textwidth]{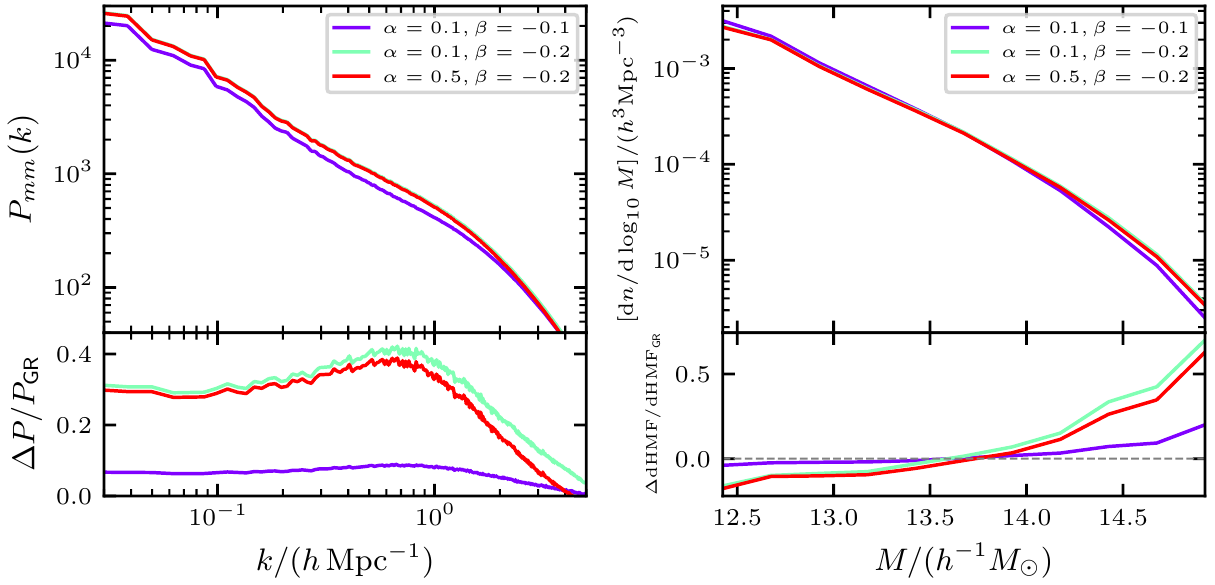}
    \caption{The matter power spectra (left panel) and the differential halo mass functions (right) of the coupled quintessence models at $z=0$, for three different $\alpha$ and $\beta$ values as labelled. As in Figs.~\ref{fig:Pk_mm_z0_fRn012}, \ref{fig:hmf_z0_fRn012} and \ref{fig:Pk_hmf_z0_sym}, the upper subpanels present the absolute measurements from \textsc{mg-glam} simuations, while the lower subpanels show the relative differences from $\Lambda$CDM.
    }
    \label{fig:Pk_HMF_csf}
\end{figure}

We now present the measured matter power spectra and halo mass functions from our symmetron and coupled quintessence runs in Figs.~\ref{fig:Pk_hmf_z0_sym} and \ref{fig:Pk_HMF_csf}, respectively.

Fig.~\ref{fig:Pk_hmf_z0_sym} presents the symmetron model results with $a_\ast = 0.33$, $\beta_\ast = 1.0$ and five $c\xi/H_0$ values of $0.5, 1.0, 2.0, 2.5$ and $3.0$.
The behaviour of the symmetron model is qualitatively similar to that of the $f(R)$ model since both of them are thin-shell screened models \citep{Brax:2012gr}. This agrees with the middle panel of Fig.~\ref{fig:CodeTests3D}, which shows that these two models have qualitatively very similar scalar field profiles for a given spherical tophat overdensity. A smaller value of $c\xi/H_0$ means $m_\ast$, the `mass' of the symmetron field, is larger, which subsequently implies that the field can more easily settle to the potential minimum (which corresponds to $\varphi=0$) in dense regions, and therefore be screened.

In Fig.~\ref{fig:Pk_HMF_csf} we show the $P_{\rm mm}(k)$ and dHMF from our three coupled quintessence models with $(\alpha, \beta) = (0.1, -0.1), (0.1, -0.2)$ and $(0.5, -0.2)$. 
The power spectrum enhancement remains approximately constant at $k\lesssim0.1h\mathrm{Mpc}^{-1}$, which is the linear perturbation regime. This is different from the behaviour seen in the $f(R)$ and symmetron models above, where $\Delta P/P_{\rm GR}$ increases with $k$ in this range, and the difference is because in coupled quintessence there is no screening, so that the fifth force is long ranged, with a ratio to the strength of Newtonian gravity that is almost constant in space. At small scales, $k \gtrsim 1 \, h \, \mathrm{Mpc}^{-1}$, $\Delta P/P_{\rm GR}$ decays with $k$, as we found in the stronger $f(R)$ models in Fig.~\ref{fig:Pk_mm_z0_fRn012}, and the physical reason behind this is the same as there: different from the weaker $f(R)$ models, even inside dark matter haloes the particles still feel a strong fifth force, which is almost in constant proportion to the strength of Newtonian force, and on top of this the direction-dependent force can also speed up the particles; the result of the two forces is that the particles gain higher kinetic energy and tend to move to and stay in the outer region of haloes, thereby reducing matter clustering on small scales compared to $\Lambda$CDM.

The right panel of Fig.~\ref{fig:Pk_HMF_csf} presents the dHMF results. 
We find that the coupled quintessence models studied here produce more high-mass haloes and fewer low-mass haloes than \ac{GR}, which is the consequence of the competition between the four effects discussed in Sect.~\ref{subsect:bg_tests}. Because these effects strongly entangle with each other through the complicated structure formation process, it is difficult to know quantitatively how they lead to the observed behaviour above, except by running simulations with different combinations of them switched on or off. 
Although this is obviously an interesting and important thing to do, it is beyond the scope of this paper, so we will leave it to future works.

\subsection{Summary}
\label{subsect:cosmo_runs_summary}

In this section we have had an initial taste of the \textsc{mg-glam} code, by running a large suite of simulations covering all three classes of models studied in this paper.

One particularly relevant aspect of the \textsc{mg-glam} code is its fast speed. The 40 $f(R)$ simulations described in this section have been run using 56 threads with \textsc{openmp} parallelisation, and we find that the run times vary randomly between $\simeq17,000$ and $\simeq33,000$ seconds, apparently depending on the real-time performance of the computer nodes used. The majority of them took $\sim24,000$ seconds, or equivalently $\simeq375$ core hours. This is roughly 100 times faster than \textsc{mg-arepo}, and $300$ times faster than \textsc{ecosmog}, for the same simulation specifications. With such a high efficiency, we can easily ramp up the simulation programme to include many more models and parameter choices, and increase the size and/or resolution of the runs, e.g., using box size of at least $1 \, h^{-1} \, \mathrm{Gpc}$. For the symmetron and coupled quintessence runs we have found similar speeds, though the run time for coupled quintessence models can perhaps be dramatically reduced if we do not explicitly solve the scalar field and the fifth force, by instead assume that the latter is proportional to the Newtonian force. We have also run a few even larger simulations for $\Lambda$CDM, F5n0 and F5n1 with $L = 512 \, h^{-1}\mathrm{Mpc}$, $N_{\rm p}=2048$ and $N_{\rm g}=4096$ (for the same cosmology as above), and some of the results are presented in Appendix \ref{appendix:resolution} --- these runs took around $42,000$ seconds for $\Lambda$CDM, $80,000$ seconds for F5n0 and $125,000$ seconds (wallclock time) for F5n1 with 128 threads using the SKUN8@IAA supercomputer at the IAA-CSIC in Spain, suggesting that a single run of specification L1000Np2048Ng4096, which would be useful for cosmological (e.g., galaxy clustering and galaxy clusters) analyses should take at most $1$--$1.5$ days to complete and is therefore easily affordable with existing computing resources.

On the other hand, efficiency should not be achieved at the cost of a significant loss of accuracy. For the runs used here, we have used a mesh resolution of $0.25 \, h^{-1} \, \mathrm{Mpc}$, which is sufficient to achieve percent-level accuracy of the matter power spectrum at $k\lesssim 1 \, h \, \mathrm{Mpc}^{-1}$ \cite{Klypin:2017iwu}, matter power spectrum enhancement at $k \lesssim 3 \, h \, \mathrm{Mpc}^{-1}$, and (main) halo mass function down to $\sim \, 10^{12.5} \, h^{-1}M_\odot$. The particle number, $N_{\rm p}^3$, in \textsc{glam} simulations is normally set according to $N_{\rm p}=N_{\rm g}/2$, so that in the simulations here we have used $1024^3$ particles. However, we have checked that increasing the particle number to $2048^3$ has little impact on the halo mass function. We notice that the completeness level of the HMFs here is similar to \textsc{ecosmog} runs with the same simulation specifications, suggesting that \textsc{mg-glam} is capable of striking an optimal balance between cost and accuracy. In Appendix \ref{appendix:resolution}, we present more detailed tests of \textsc{mg-glam}'s power spectrum and HMF predictions at different force and mesh resolutions, including our highest-resolution runs for $\Lambda$CDM and F5n1 with $L = 512 \, h^{-1}\mathrm{Mpc}$, $N_{\rm p}=2048$ and $N_{\rm g}=4096$ (for the same cosmology as above). There we demonstrate that the increase of force resolution can lead to further improvement of the small-scale and low-mass predictions.

\begin{figure}
    \centering 
    \includegraphics[width=\textwidth]{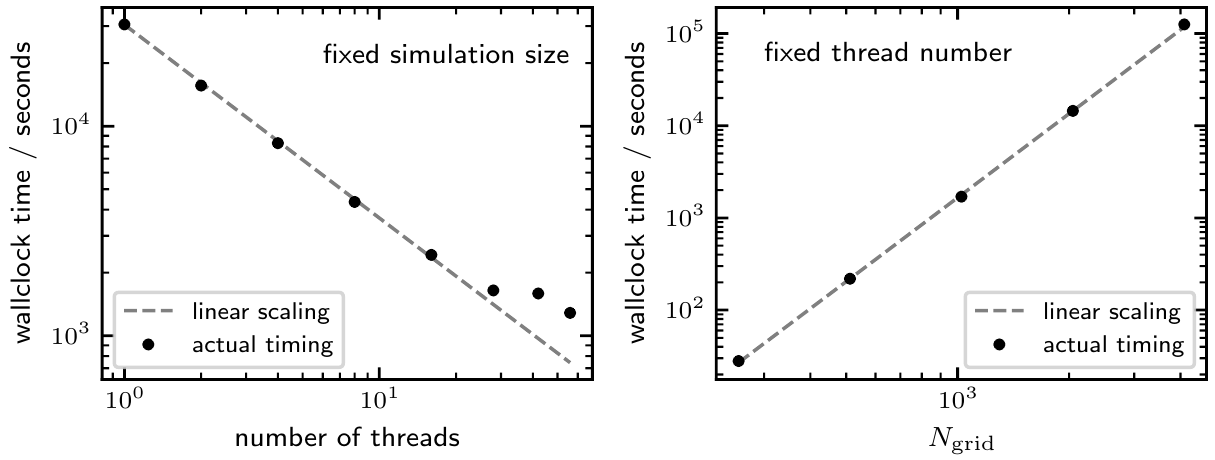}
    \caption{
        \textit{Left panel:} The wallclock time of the \textsc{mg-glam} test runs for the F5n1 model, with fixed simulation size/resolution ($L=512h^{-1}\mathrm{Mpc}$, $N_{\rm p} = 256$ and $N_{\rm g} = 512$), as a function of the number of threads used in \textsc{openmp} parallelisation.
        The scaling between run time and thread number is very close to be perfectly linear for number of threads up to $\sim 30$.
        \textit{Right panel:} The wallclock times of the \textsc{mg-glam} runs for F5n1 with varying simulation sizes and resolutions (from left to right: $N_{\rm g} = 256, 512, 1024, 2048, 4096$, and $N_{\rm p} = N_{\rm g}/2$), while the number of threads is chosen as $128$. Again, the scaling is nearly perfectly linear. In both cases, the symbols denote the times taken by the test runs, and the lines denote the expected results with a `perfect linear scaling'. We find similarly good scaling properties for $\Lambda$CDM and F5n0, but those are not shown here.
    }
    \label{fig:scaling}
\end{figure}

Before concluding this paper, let us briefly describe some tests we have performed to understand how well the parallelisation of \textsc{mg-glam} works. This consists of a series of runs (taking $f(R)$ gravity F5n1 as a representative) to demonstrate the scaling of \textsc{mg-glam}, and these runs were all done on the SKUN6/SKUN8 facility managed by the IAA-CSIC in Spain.The strong scaling tests are presented in the left panel of Fig.~\ref{fig:scaling}. The test simulations employed a fixed resolution of $256^3$ particles and $512^3$ grids, with the same Planck 2015 cosmology as used in the main \textsc{mg-glam} runs of this paper. This plot shows that, when the number of \textsc{openmp} threads ranges between 1 and $\sim 30$, the wallclock time scales linearly with the thread number. The deviation from a perfect linear scaling (black dashed line) when the number of threads exceeds $30$ is possibly due to the small size of the test run. In addition, we have also run a set of simulations of different sizes by varying the resolutions and keeping the number of threads fixed. The wallclock time used is shown in the right panel of Fig.~\ref{fig:scaling}. We see that the time consumption again scales nearly perfectly linearly with the considered resolutions (up to $N_{\rm g} = 4096$ and $N_{\rm p} = N_{\rm g}/2$). These tests demonstrate that \textsc{mg-glam} is well scalable.

\section{Discussions and conclusions}
\label{sect:discuz}

In this work, together with a twin paper \citep{Hernandez-Aguayo:2021_twin_paper}, we have introduced a new, fast and accurate modified gravity simulation code, \textsc{mg-glam}, which is based on the highly-optimised parallel particle-mesh $N$-body code \textsc{glam} \citep{Klypin:2017iwu}.
We have focused on the numerical implementation of three representative classes of conformally coupled scalar field models, including two thin-shell screening models, $f(R)$ gravity and symmetrons, and a coupled quintessence model with no screening. In the case of $f(R)$ gravity, we have extended earlier simulation studies to include more general parameter choices, e.g., $n=0,2$, by generalising an efficient algorithm developed for the $n=1$ case in \cite{Bose:2016wms} to these new cases. The twin paper \citep{Hernandez-Aguayo:2021_twin_paper} explores \ac{MG} models with derivative coupling terms, including the {DGP} and K-mouflage models. Altogether, the \textsc{mg-glam} code not only covers several of the most popular \ac{MG} models in the literature, but can also serve as prototypes that can be 
easily extended to work for other leading classes of \ac{MG} models, such as chameleons, Galileon gravity and coupled quintessence models with other user-specified potentials and coupling functions.

We have performed a series of tests to check that our implementation of the multigrid solvers works correctly, using different density configurations for which we can obtain analytical or independent numerical solutions of the scalar field, and found that the numerical solutions given by \textsc{mg-glam} agree very well with them in all cases. We have shown that, using only two V-cycles, we can achieve convergence for the nonlinear equations in the \ac{MG} models considered. 
Also, we have compared the solutions of the background scalar field and the modified expansion rate in the coupled quintessence model obtained using \textsc{mg-glam} and \href{https://camb.info/}{\textsc{camb}}, finding excellent agreement between both codes. 
Finally, we have compared the power spectrum enhancement and the abundance of dark matter haloes for the $f(R)$ model predicted by \textsc{mg-glam} and the \textsc{mg-arepo} code. 
In general, \textsc{mg-glam} is able to reproduce the predictions of these quantities by \textsc{mg-arepo} and \textsc{ecosmog} simulations with sufficiently high accuracy for the cosmological applications of interest to us, in spite of taking only a tiny fraction of the time needed for the latter codes. For example, with $1024^3$ particles in a box of size $512h^{-1}\mathrm{Mpc}$, \textsc{mg-glam} simulations can accurately predict $\Delta P/P_{\rm GR}$ at $k\lesssim3h\mathrm{Mpc}^{-1}$ and the HMF down to $10^{12.5}h^{-1}M_\odot$, with about $1\%$ of the computational costs for \textsc{mg-arepo} and \textsc{ecosmog}. 

We have run a large suite of $f(R)$ cosmological simulations for 10 models with $|f_{R0}|$ logarithmically spaced in $[-6.00, -4.50]$ and $n = 0, 1, 2$, and carried out the simulations for five symmetron models and three coupled quintessence models. 
With this large suite of \ac{MG} simulations we are able to study in great detail the modified gravity effects, including that of the screening mechanisms, on structure formation, as we have shown in the nonlinear matter power spectra and halo mass functions.
In particular, the large number of $f(R)$ gravity runs demonstrate, with fine details, how the effect of the chameleon screening mechanism depends on not only the present-day scalar field value, $f_{R0}$, but also the parameter $n$ which has been less explored to date.

The development of \textsc{mg-glam} will help in the construction of a large number of galaxy mock catalogues in MG theories for Stage-IV galaxy surveys, such as DESI and Euclid. 
Owing to its high efficiency and accuracy, this code can be used to perform $\mathcal{O} (100)$ large ($L > 1.0 \, h^{-1} \, \mathrm{Gpc}$ at least) and high-resolution ($m_{\rm p} < 10^{10} \, h^{-1} \, M_{\odot}$) simulations for each modified gravity model, with minimal computational cost. 
These will allow for variations of not only the gravitational but also cosmological parameters, and subsequently the construction of accurate emulators for various physical quantities in different gravity models. 
This will open up a wide range of possibilities for future works to test gravity using cosmological observations. 

One of the main potential applications of \textsc{mg-glam} simulations is the study of various galaxy clustering statistics \cite[e.g.][]{Alam:2020jdv} based on the mock galaxy catalogues mentioned above. \textsc{mg-glam} will have the flexibility to be run at different resolutions, tailored to the different observables and/or galaxy types. It will also have the advantage of allowing different classes of \ac{MG}, as well as dynamical dark energy \cite{Klypin:2020tud} models, to be studied with equal depths and fine details. In a series of upcoming papers, we will visit this topic, starting with the prescriptions to populate dark matter haloes with galaxies, as well as a more detailed study of halo properties, including halo clustering.

\acknowledgments


C-ZR, CA and BL are supported by the European Research Council (ERC) through a starting Grant (ERC-StG-716532 PUNCA). BL and CMB are further supported by the UK Science and Technology Funding Council (STFC) Consolidated Grant No.~ST/I00162X/1 and ST/P000541/1. CH-A acknowledges support from the Excellence Cluster ORIGINS which is funded by the Deutsche Forschungsgemeinschaft (DFG, German Research Foundation) under Germany's Excellence Strategy - EXC-2094-390783311. AK and FP thank the support of the Spanish Ministry of Science and Innovation funding grant PGC2018- 101931-B-I00.

This work used the DiRAC@Durham facility managed by the Institute for Computational Cosmology on behalf of the STFC DiRAC HPC Facility (\url{www.dirac.ac.uk}). The equipment was funded by BEIS via STFC capital grants ST/K00042X/1, ST/P002293/1, ST/R002371/1 and ST/S002502/1, Durham University and STFC operation grant ST/R000832/1. DiRAC is part of the UK National e-Infrastructure.

This work used the skun6@IAA facility (\url{www.skiesanduniverses.org}) managed by the Instituto de Astrof\'{i}sica de Andaluc\'{i}a (CSIC). The equipment was funded by the Spanish Ministry of Science EU-FEDER infrastructure grants EQC2018-004366-P and EQC2019-006089-P.

\appendix

\section{Effects of mass and force resolution}
\label{appendix:resolution}

The original \textsc{glam} code has been well tested (cf.~\citep{Klypin:2017iwu}) by examining the effects of time-stepping and force resolution and comparing with the high-resolution MultiDark simulations \citep{Klypin:2016MNRAS.457.4340K} which were performed using the \textsc{gadget} code \citep{Springel:2005_Gadget_code_paper}. In addition, Ref.~\citep{Klypin:2020tud} compared the \textsc{glam} results of halo mass functions and matter power spectra with those of the \textsc{quijote} simulations \citep{Villaescusa-Navarro:2020ApJS..250....2V}, and found good agreement. Denote $k_{1\%}$ as the wavenumber above which the \textsc{glam} matter power spectrum begins to deviate by more than $1\%$ from those of the high-resolution simulations. Based on the comparison with the MultiDark simulations, the authors found that $k_{1\%}$ is related to the force resolution $\Delta x = L_{\rm box}/N_{\rm g}$ (cf. Eq.~\eqref{eqn:force_resolution_def}) as 
\begin{align}
    k_{1\%} = \frac{0.25 \pm 0.05}{(\Delta x) / (h^{-1} \mathrm{Mpc})} \, h \, \mathrm{Mpc}^{-1} \ .
\end{align}
If this relation also works for the \textsc{mg-glam} code, the power spectra of the modified gravity cosmological runs presented in the main text are reliable down to $k_{1\%} \sim 1 \, h \, \mathrm{Mpc}^{-1}$. However, as mentioned in the main text above, the results of the power spectrum enhancement with respect to $\Lambda$CDM are empirically reliable down to larger $k$.

To test the effect of mass and force resolutions, we have performed four $f(R)$ gravity simulations for $f_{R0} = -10^{-5}, n = 1$ with fixed box size $512 \, h^{-1}\mathrm{Mpc}$ and varying particle and grid numbers
\begin{align*}
    (N_{\rm p}, N_{\rm g}) = \begin{Bmatrix}
    (1024, 2048),& (2048, 2048) \\ 
    (1024, 4096),& (2048, 4096)
    \end{Bmatrix}\ ,
\end{align*} 
which correspond to mass and force resolutions of 
\begin{align*}
    \qty( \displaystyle \frac{m_{\rm p}}{10^9 \, h^{-1} M_{\odot}}, \frac{\Delta x}{h^{-1} \mathrm{Mpc}} ) = \begin{Bmatrix}
    (11.0, 0.25\phantom{1}), &(1.37, 0.25\phantom{1}), \\
    (11.0, 0.125), &(1.37, 0.125)
    \end{Bmatrix} \ .
\end{align*}
Here the two runs in the same row (column) have the same force/mesh (mass) resolution. The adopted cosmological parameters are the same as the simulations used in the main text.

We compare the \textsc{mg-glam} matter power spectrum and halo mass function enhancement $\Delta P / P_{\rm GR}$ and $\Delta \mathrm{HMF} / \mathrm{HMF}_{\rm GR}$ with those of the \textsc{mg-arepo} simulations.
We focus on these quantities instead of comparing the absolute $P(k)$ and HMF from (\textsc{mg}-)\textsc{glam} and other codes, for the following reasons: (1) as mentioned above, the reliability of the $\Lambda$CDM results from \textsc{glam} has been carefully tested and established; (2) comparisons between different codes usually suffer from cosmic variance and different implementation details (such as the IC set up, force calculation, time stepping and halo finding), and as a result a large number of runs are needed to make reliable comparisons, after carefully calibrating simulation specifications of the different codes --- such an effort is unnecessary and beyond the scope of this work given (1); (3) in MG simulations, people are often more interested in the enhancement with respect to $\Lambda$CDM, and this is indeed what has been tested in the code papers of the previous MG simulation codes. Also, we note that the \textsc{mg-glam} and \textsc{mg-arepo} simulations presented in this work use slightly different cosmological parameters: we have checked explicitly (by running test simulations with \textsc{mg-glam} using identical cosmological parameters as the \textsc{mg-arepo} runs) that the effect is small (few percent level), but this nevertheless still makes it difficult to justify directly comparing the absolute $P(k)$ or HMF from them; the enhancement, on the other hand, is known empirically to be less sensitive to cosmological parameter values and differences between simulation codes.


\begin{figure}
    \centering 
    \includegraphics[width=\textwidth]{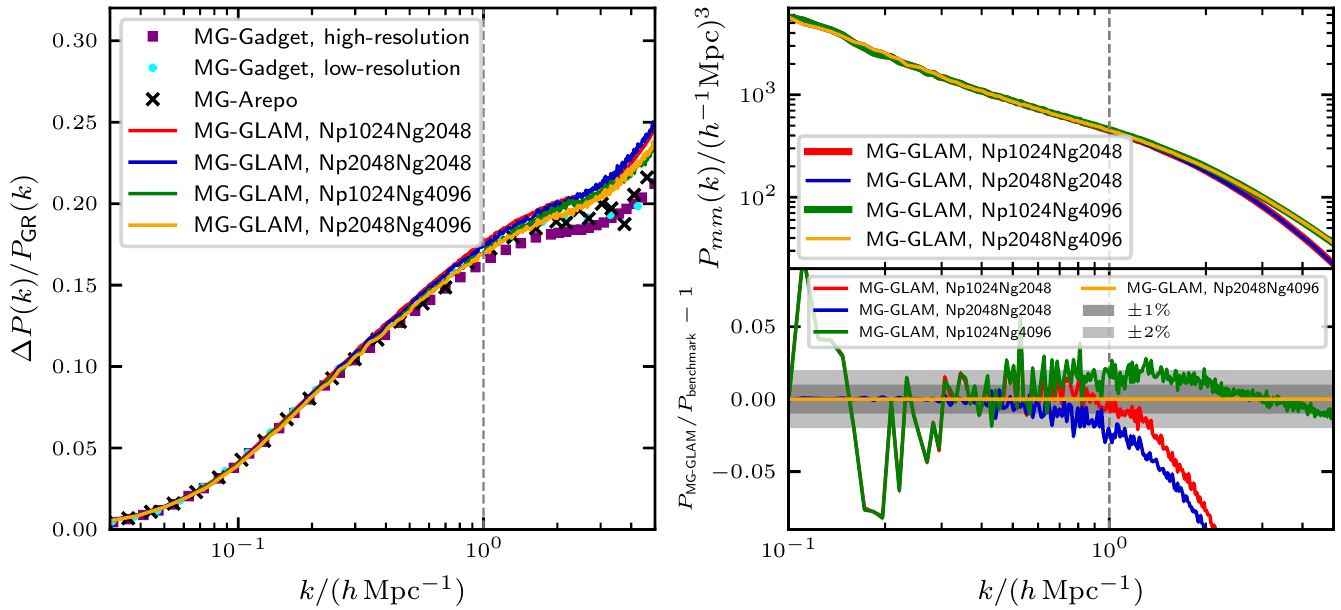}
    \caption{
        Comparison of matter power spectra from \textsc{mg-glam} (lines), \textsc{mg-gadget} (squares and circles) and \textsc{mg-arepo} (crosses) simulations at $z=0$. The left panel shows the matter spectrum enhancement, $\Delta P/P_{\rm GR}$, from the different codes and resolutions, as the legend labels. The \textsc{mg-arepo} data are the same as in Fig.~\ref{fig:Pk_hmf_MGGLAMvsArepo}, while the \textsc{mg-gadget} data are from the two \textsc{lightcone} simulations, at higher ($L=768h^{-1}\mathrm{Mpc}$ and $N_{\rm p}^3=2048^3$) and lower ($L=1536h^{-1}\mathrm{Mpc}$ and $N_{\rm p}^3=2048^3$) resolutions, described in Ref.~\cite{Arnold:2018nmv}. The upper right panel shows the absolute values of $P(k)$ from \textsc{mg-glam} simulations with the four combinations of mass and force resolutions. In the lower right panel, the ratios of the power spectrum in each simulation to that of the highest resolution run ($N_{\rm p} = 2048$ and $N_{\rm g} = 4096$) are displayed, where the dark and light grey shaded regions denote respectively $\pm1\%$ and $\pm2\%$ differences from the benchmark. The vertical lines represent $k=1h\mathrm{Mpc}^{-1}$.
    }
    \label{fig:PkEnhancement_ResolutionTest_MGGLAMvsMGArepo}
\end{figure}

The left panel of Fig.~\ref{fig:PkEnhancement_ResolutionTest_MGGLAMvsMGArepo} presents the matter power spectrum enhancements at $z = 0$ from \textsc{mg-glam} and \textsc{mg-arepo}, as well as two \textsc{mg-gadget} simulations. We see that $\Delta P / P_{\rm GR}$ is relatively insensitive to the mass and force resolution variations considered here; this is consistent with previous experiences. However, increasing the mesh resolution from $0.25$ to $0.125h^{-1}\mathrm{Mpc}$ does improve the agreement between \textsc{mg-glam} and \textsc{mg-arepo}, by reducing $\Delta P / P_{\rm GR}$ (see, e.g., \cite{Li:2013MNRAS.428..743L} for a discussion of how a lower resolution simulation gives higher $\Delta P / P_{\rm GR}$). The highest resolution \textsc{mg-glam} run ($N_{\rm p} = 2048$ and $N_{\rm g} = 4096$) agrees with \textsc{mg-arepo} nearly perfectly down to $k \sim 1 \, h \,\mathrm{Mpc}^{-1}$, and the agreement is at the level of a couple percent down to $k\approx5h\mathrm{Mpc}^{-1}$ (ignoring the dip in $\Delta P / P_{\rm GR}$ at $k\approx4h\mathrm{Mpc}^{-1}$, which is apparently not physical). The slightly larger deviations at $k > 1 \, h \,\mathrm{Mpc}^{-1}$ can be still due to the lower force resolutions in the \textsc{glam} simulations, but we note that the agreement between the \textsc{mg-gadget} and \textsc{mg-arepo} runs (which have similar force resolutions) is at a comparable level, so the difference is likely also partly due to the different codes (or simulation realisations).

In the upper right panel of Fig.~\ref{fig:PkEnhancement_ResolutionTest_MGGLAMvsMGArepo}, we present the absolute matter power spectra from the \textsc{mg-glam} simulations at different resolutions. As expected, increasing the mesh/force resolution leads to a $P(k)$ curve that decays much more slowly at small scales (orange and green lines), while increasing the mass resolution (blue) gives little improvement. The lower right panel of Fig.~\ref{fig:PkEnhancement_ResolutionTest_MGGLAMvsMGArepo} shows the ratio of the matter spectrum in each simulation to that from the highest resolution run. The figures indicate $\approx1 \%$ convergence for $k \lesssim 1 \, h \,\mathrm{Mpc}^{-1}$, which is consistent with the convergence test of the original \textsc{glam} code presented in \citep{Klypin:2017iwu}.

\begin{figure}
    \centering 
    \includegraphics[width=\textwidth]{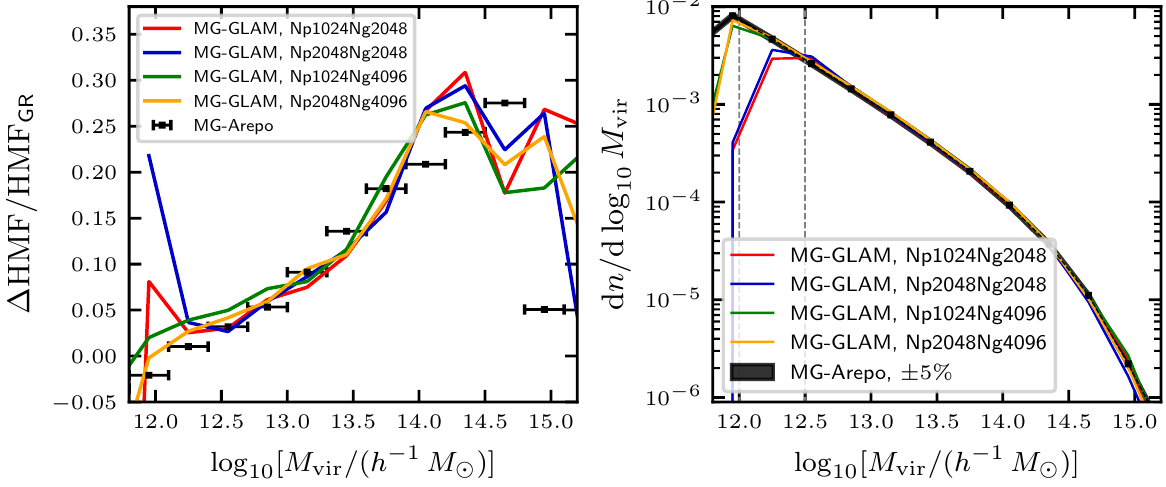}
    \caption{Comparison of halo mass functions of \textsc{mg-glam} and \textsc{mg-arepo} simulations at $z=0$.
    The relative enhancements with respect to $\Lambda$CDM and the absolute values of the HMFs are shown in the left and right panels, respectively. The two vertical lines in the right panel denote respectively the masses $10^{12}$ and $10^{12.5}h^{-1}M_\odot$. There is excellent agreement between \textsc{mg-glam}'s higher-resolution runs and \textsc{mg-arepo}, in both the HMF and its enhancement, down to $10^{12}h^{-1}M_\odot$.
    }
    \label{fig:dHMF_ResTest_MGGLAMvsArepo}
\end{figure}

The comparisons of halo mass functions are shown in Fig.~\ref{fig:dHMF_ResTest_MGGLAMvsArepo}, where note that we used different halo finders for the \textsc{mg-glam} and \textsc{mg-arepo} results, but the same halo mass definition, as described in Sect.~\ref{subsect:comparions}. The HMFs of \textsc{mg-glam} simulations are accurate in the range of $M_{\rm vir} \gtrsim 10^{12.5} \, h^{-1}M_{\odot}$ for $\Delta x = 0.25 \, h^{-1}\mathrm{Mpc}$ ($N_{\rm g} = 2048$), and $M_{\rm vir} \gtrsim 10^{12}\, h^{-1}M_{\odot}$ for $\Delta x = 0.125 \, h^{-1}\mathrm{Mpc}$ ($N_{\rm g} = 4096$). There is excellent agreement between \textsc{mg-glam}'s higher-resolution runs and \textsc{mg-arepo}, in both the HMF and its enhancement, down to $10^{12}h^{-1}M_\odot$.

\bibliographystyle{utphys}
\bibliography{MGGLAM_refs.bib}

\end{document}